\documentclass[pr,twocolumn,floatfix,notitlepage,nobibnotes,nofootinbib,showkeywords]{revtex4}
\usepackage{amsmath}
\usepackage{amssymb}
\usepackage{bm}
\usepackage{graphicx}
\usepackage{color}
\usepackage [cp1251]{inputenc}
\usepackage [english]{babel}

\usepackage{hyperref}

\newcommand{\sign}{\mathop{\rm sign}\nolimits}

\renewcommand{\Re}{\mathop{\rm Re}}

\newcommand{\hF}{\mathop{{}_2\rm{F}_1}}

\usepackage{graphicx} 

\def\be{\begin{equation}}  
\def\ee{\end{equation}}  
\def\ba{\begin{eqnarray}}  
\def\ea{\end{eqnarray}}  
\def\bc{\begin{center}}  
\def\ec{\end{center}}

\begin{document}
\title{High frequency electric field induced nonlinear effects in graphene (review)}

\author{M.M. Glazov}
\affiliation{Ioffe Physical-Technical Institute of the RAS, 194021
  St.-Petersburg, Russia}

\author{S.D. Ganichev}
\affiliation{Terahertz Center, University of Regensburg, 93040
  Regensburg, Germany} 

\newcount\timehh  \newcount\timemm
\timehh=\time \divide\timehh by 60
\timemm=\time
\count255=\timehh\multiply\count255 by -60 \advance\timemm by
\count255


\date{\today}

\begin{abstract}
The nonlinear optical and optoelectronic properties of graphene with the emphasis on the processes of harmonic generation, frequency mixing, photon drag and photogalvanic effects
as well as generation of photocurrents due to coherent interference effects, are reviewed. The article presents the 
state-of-the-art
of this subject, including both recent advances and well-established results. Various physical mechanisms controlling transport are described in depth including phenomenological description based on symmetry arguments, models visualizing physics of nonlinear responses,
and microscopic theory of individual effects. 
\end{abstract} 

\maketitle
\tableofcontents

\section{Introduction}\label{intro}

The discovery of graphene opened a new era in material science.  
Graphene is the first truly two-dimensional (2D) crystal consisting of just a single layer of carbon atoms arranged in a hexagonal lattice~
\cite{Novoselov04,Novoselov05,Zhang05,Geim07a,Neto09}.
The main consequence of such a crystal 
structure is the linear energy spectrum of electrons and holes, $\varepsilon_{\bm p} = \pm v |\bm p|$,
where $v\approx c/300$ is the effective speed, $c$ is the speed of light, $\bm p$ is
the charge carrier momentum and signs $\pm$ refer to the conduction and
valence bands, which merge at $\bm p=0$ point, at the edges of the 
Brillouin zone~\cite{Wallace47,McClure56,Slonczewski58}. 
Owing to a specific energy dispersion, 
graphene has revealed fascinating effects in a number of experiments.
In particular, the linear coupling of the charge carriers energy with their momentum leads to a peculiar modification
of the quantum Hall effect~\cite{Novoselov05,Novoselov07}
and plays an important role in phase-coherent phenomena such as, e.g., 
weak localization~\cite{Bib:McCann,Bib:Tikhonenko},
minimal electrical conductivity
\cite{Novoselov05,Zhang05,Katsnelson06,Nomura07,Tan07},  
Klein tunnelling \cite{Stander09,Young09}, etc., for reviews see \cite{Geim07a,Neto09,mccann2013}. 
The fact that the band structure resembles the dispersion relation of a
massless relativistic particle has created enormous excitement since
graphene provides an excellent model system for benchtop studies 
of quantum-electrodynamic effects \cite{Semenoff84,Haldane88} making
relativistic experiments in a solid state environment
feasible~\cite{Geim07a,Morozov:2008eng,Lozovik:2008eng}. 
Another important issue of this material is the presence of two valleys,
each containing a Dirac cone. This constitutes a two-state degree of freedom, which
was suggested to be used in valleytronics~\cite{Bib:Valley}. 
These and other specific features manifest themselves in a  
linear in electric field transport in graphene and have made it
attractive for fundamental research  
and numerous applications, for review see, e.g. Refs.~\cite{Falkovsky:2008eng,RevModPhys.82.2673,RevModPhys.83.407,Bonaccorso10}.

While linear in electric field phenomena in graphene are in focus
of the current research, nonlinear  
transport effects, where the response is proportional to the higher
powers of the electric field, are much less studied. 
In general, the such effects are caused by  
the redistribution of 
the charge carriers in the momentum and energy space induced by the  radiation incident on
the sample. 
The resulting nonequilibrium distribution can contain oscillating in time and space
components as well as steady-state and spatially homogeneous ones. Hence, the radiation may
cause both \emph{ac} and \emph{dc} current flows in a media, whose
magnitudes are nonlinear functions of the field amplitude and whose components 
are sensitive to the radiation polarization.
In conventional three- and two-dimensional semiconductors with
parabolic energy dispersion, as well as in metals and dielectric crystals, a large number of nonlinear effects was
observed and studied in great details.
Harmonic generation, frequency mixing, optical rectification, linear and circular photogalvanic effects,  
photon drag effect, photoconductivity,  coherently controlled ballistic charge currents,
etc. are the subjects of intense research and already found a number of applications~\cite{blombergen,Boyd,yariv,wegener,shen,sturmanBOOK,ganichev_book}. 
Moreover, these effects have been proven to be a very efficient tool to study 
nonequilibrium optical and electronic processes in semiconductors and provide information 
about their fundamental properties. For instance, they provide an access to the symmetry, 
peculiarities of the band structure, processes of electron momentum, energy/spin relaxation etc., 
as well as allow one to explore the processes of interaction of light with charge
carriers~(for review see, e.g.~\cite{blombergen,Boyd,yariv,wegener,shen,sturmanBOOK,ivchenkopikus,ganichev_book,ivchenko05a}).  Concerning the carbon based systems, so far the  nonlinear transport 
has  been extensively studied for carbon nanotubes and carbon films~\cite{ivch_spi,obraztsov:231112,ISI:000072756100009,PhysRevA.63.053808,Mele:2000oq,Kral:2000nx,mikheev:4854,doi:10.1021/nl203003p}, for review see, e.g. \cite{cnt:book}.

Naturally, nonlinear effects have attracted attention in graphene~\cite{Liu,Wang}, where a number of phenomena, including
 second~\cite{0295-5075-79-2-27002,10.1063/1.3483872,dean:261910,PhysRevB.82.125411,murzina} and third~\cite{10.1063/1.3483872,Hotopan11,2013arXiv1301.1697H,2013arXiv1301.1042K} 
harmonic generation, frequency mixing~\cite{Hendry10,10.1063/1.3483872,Hotopan11,2011rangel,GuT.:2012fk},
photon drag effects~\cite{karch2010,2010arXiv1002.1047K,entin10}, chiral edge photocurrents~\cite{edge},   ``bulk'' 
photogalvanic effects~\cite{PRB2010},  coherent
current injection~\cite{sun2010,Sun:2012ys,2012winzent}, time-resolved photocurrents~\cite{Prechtel:2012kx,Graham:2013uq}, photocurrents in graphene \textit{pn}-junctions~\cite{Mai:2011oq,doi:10.1021/nl8033812,PhysRevB.78.045407},
spatial self-phase modulation~\cite{doi:10.1021/nl2023405} and optical Kerr effect~\cite{2012chu,Yan},\footnote{Spatial self-phase modulation detected in colloidal
dispersion of graphene sheets in organic solvents as well as optical Kerr effect in this system are out of scope of the present review.
}
have been already addressed theoretically and experimentally.
These works demonstrated that the microscopic mechanisms of such
effects in graphene can be quite different from their counterparts in
ordinary semiconductor systems. Moreover, all the effects 
observed in graphene have a common feature:  they are strongly enhanced compared with their analogues in semiconductors.  
The reasons for this, on the first glance surprising, fact are the
high electron velocity and the linear dispersion in graphene. Indeed, the large velocity of electrons in
graphene, as
compared with typical semiconductor systems,  obviously implies  the efficient radiation -- electron motion
coupling. As for the electron dispersion, it crucially  affects the details of
optical transitions in the electron momentum space ($\bm k$-space). In particular, the gapless,
linear dispersion allows one to easily suppress some of the optical
excitation channels, e.g., leading to the resonant nonlinear response~\cite{entin10}.
Moreover, the nonlinearities and the nonlinear response can be
enhanced  via the excitation of the  plasmonic waves in graphene~\cite{PhysRevB.84.045432,novoselov_pl1,novoselov_pl2}.
Therefore, although being limited to a rather small amount of theoretical
and even less  experimental works, current research has already demonstrated that studying of 
nonlinear transport provides an access  
to various properties of graphene. Among others, these studies have proved that 
graphene, as a nonlinear element, is a promising material for a variety of different applications 
and may be used for the development of novel electronic devices 
for $\mbox{microwave-,}$ terahertz- and optoelectronic. Thus, the experimental and theoretical research in the 
field of 
nonlinear graphene optics and optoelectronics becomes an important task.

The paper is aimed to give an overview of the key properties 
of graphene as a nonlinear material, to outline the 
main theoretical and experimental results obtained in the nonlinear physics of graphene so far, and stimulate further studies of these effects in this material. 
We start with the brief introduction to the nonlinear phenomena in graphene. Then, we describe 
the second and third order effects. Each class of the effects is presented in a similar way:  
we start with the phenomenological analysis of different phenomena  based
on the symmetry arguments, provide theoretical background and, one by one, give an overview of the microscopic theory and
the main experimental results.
Finally, 
we summarize the results and discuss prospectives of 
future theoretical and experimental studies of the nonlinear
electromagnetic response of graphene. 


\section{General remarks}\label{remarks}
  
The standard way to treat the nonlinear effects without going into
microscopic details makes use of the symmetry arguments. This approach
allows one to conclude on the experimental geometry and conditions of observation of the
effect under consideration as well as  to describe its variation with change
of macroscopic parameters, such as intensity of the radiation, 
its polarization and angle of incidence without knowing of 
the microscopic origin.
In this way, the electron ensemble response to the external field 
can be most conveniently characterized by the coordinate- and
time-dependent electric current density $\bm j(\bm r,t)$. 
It is expanded in the power series in the external alternating
electric field $\bm E(\omega, \bm q)$ taken in the form of a plane wave 
\begin{equation} \label{plane}
{\bm E}({\bm r}, t)= {\bm E}(\omega, \bm q) {\rm e}^{- {\rm i} \omega t + {\rm i} {\bm q} {\bm r} }
+ {\bm E}^*(\omega, \bm q) {\rm e}^{{\rm i} \omega t - {\rm i} {\bm q} {\bm r}}\:,
\end{equation}
where $\omega$ is the radiation frequency and $\bm q$ is its wavevector. By that it has a form
\begin{multline}
\label{j:phen:gen}
j_\alpha (\bm r, t)= \left[\sigma_{\alpha \beta}^{(1)} E_\beta(\omega, \bm q)
\mathrm e^{- {\rm i} \omega t + {\rm i} {\bm q} {\bm r} }  + {\rm
  c.c.} \right]+ \\
 \left[ \sigma_{\alpha\beta\gamma}^{(2')} E_\beta(\omega, \bm q)
E_\gamma(\omega, \bm q) \mathrm e^{- 2{\rm i} \omega t + 2{\rm i} {\bm
    q} {\bm r} } + {\rm c.c.} \right] + \\
    \sigma_{\alpha\beta\gamma}^{(2)} E_\beta(\omega, \bm q)
E_\gamma^*(\omega, \bm q) +\ldots \ .
\end{multline}
Here Greek subscripts refer to the Cartesian coordinates, ${\rm c.c.}$
stands for the complex conjugate, and Eq.~\eqref{j:phen:gen}
is limited to the second order effects.  
While the first term in Eq.~(\ref{j:phen:gen}) describes the linear
transport, the other terms are the second order in electric field and include: (i) the contribution oscillating as $\exp{(-2\mathrm i \omega t)}$ responsible for the second harmonic
 generation (second term) and (ii) time-independent contribution yielding the  \emph{directed (dc)} current
generation (last term). 
These nonlinear processes are characterized by 
the nonlinear conductivities $\sigma_{\alpha\beta\gamma}^{(2')}$ and 
$\sigma_{\alpha\beta\gamma}^{(2)}$, respectively, whose specific form will be detailed below in Sec.~\ref{2nd}. 
The class of these phenomena can be extended by considering the nonlinear
polarization $\bm P$, which is described by the
equation similar to Eq.~\eqref{j:phen:gen} and leads 
to, e.g. the optical rectification effect. The higher order effects in
Eq.~\eqref{j:phen:gen} like third harmonic generation
are denoted by triple dot mark. The corresponding expressions and their
description will be given in Sec.~\ref{3rd}.

On a very general level, the enhanced nonlinear properties of graphene can be illustrated by considering the 
classical motion of the charge carrier under the action of the harmonic electric 
field $\bm E(t) = \bm E_0 \cos{\omega t}$, where $\bm E_0$ is the amplitude of the field, $\omega$ is
its frequency and taking into account the linear energy dispersion, $\varepsilon_{\bm p} = \pm v |\bm p|$.
The electron motion is described by the second Newton law
\[
\frac{d\bm p}{dt} = e\bm E_0 \cos{\omega t},
\]
where $e=-|e|$ is the electron charge.
It follows from this equation that electron momentum
exhibits harmonic oscillations $\bm p(t) = (e\bm E_0/\omega)
\sin{\omega t}$. In contrast to usual semiconductor systems with 
 parabolic or slightly nonparabolic dispersion, here the electron velocity, and,
hence, other observable quantities like, e.g. electric current, dipole moment 
or emitted radiation,  demonstrate strongly anharmonic temporal
behavior. Indeed, taking into account that for a massless particle the absolute value of the velocity is fixed, and its direction is determined by the direction of
the momentum, we have~\cite{0295-5075-79-2-27002,Lopez08}
\[
\bm v(t) = {\pm} v \frac{\bm p}{|\bm p|} = {\pm} v \frac{e\bm E_0}{|e\bm E_0|} \sign[\sin{\omega t}] = 
\]
\[
{\pm}  v
\frac{e\bm E_0}{|e\bm E_0|} \frac{4}{\pi} \left(\sin{\omega t}  +
  \frac{1}{3}\sin{3\omega t} +\frac{1}{5} \sin{5\omega t}  +  \ldots \right).
\]
Here $+$ and $-$ correspond to the electron in the conduction and valence band, respectively. In this simplified model, the nonlinear effects become
important even at very small fields: The coefficient at the third harmonic in the velocity is just $1/3$ of the first harmonic coefficient. 
For doped graphene with the typical Fermi energy $E_F \sim 100$~meV the estimations yield that the nonlinear response can already be observed at fields as low as
$10^2$ -- $10^3$ V/cm~\cite{PhysRevB.84.045432},
being several orders of magnitude smaller
than required for the same phenomena in other media.

Discussing various routes of nonlinearities in graphene, one should
consider the relation between photon and Fermi energies, which governs the nonlinear response 
of any material via microscopic mechanism of light-matter coupling.
Thus, before going in details of specific mechanism
we address different regimes of optical excitation in graphene.
The general description of radiation induced effects is based on 
the standard approach replacing the electron momentum $\bm p=(p_x,p_y)$ by $\bm p - e\bm A/c$ in Dirac Hamiltonian
\begin{equation}
\label{Dirac:Ham}
\hat{\mathcal H}(\bm p) = v(\hat{ \bm \sigma}\cdot \bm p),
\end{equation}
with $\bm A$ being vector potential of the 
electromagnetic field, and $\hat{\bm \sigma} = (\hat \sigma_x,\hat \sigma_y)$ is the vector 
composed of the Pauli matrices, which serve as basis matrices in the space of the 
electron states in the conduction and valence bands in the vicinity of
 Dirac point.\footnote{Such an approximation is suitable only for electrons 
 in a given valley of graphene in case where the interaction with other (distant) bands is neglected.}
The change of frequency, and/or the
Fermi energy, $E_F$, not only 
strongly influence the magnitude of the nonlinear phenomena, but may
change the microscopic picture 
of their formation or, at certain conditions, may give rise to resonance responses. 
Several regimes of light-matter interaction depending on the photon energy,
$\hbar \omega$, electron Fermi energy (in certain cases temperature) and its momentum
relaxation rate, $1/\tau$, are of importance. 
As a rule, physical problems of nonlinear transport are studied for graphene systems, for which the condition 
\begin{equation}
\label{free}
E_F\tau/\hbar \gg 1
\end{equation}
is fulfilled, allowing to consider electrons (holes) as free carriers. 
Moreover, taking into account 
that the energy distance from the Dirac point, $\varepsilon_{\bm p} =0$,  
to other bands in graphene is extremely large, exceeding
 $10$~eV~\cite{Bassani,PhysRevB.17.626}, one can also 
 disregard  direct optical transitions involving other bands.

\begin{figure}[t]
\includegraphics[width=\linewidth]{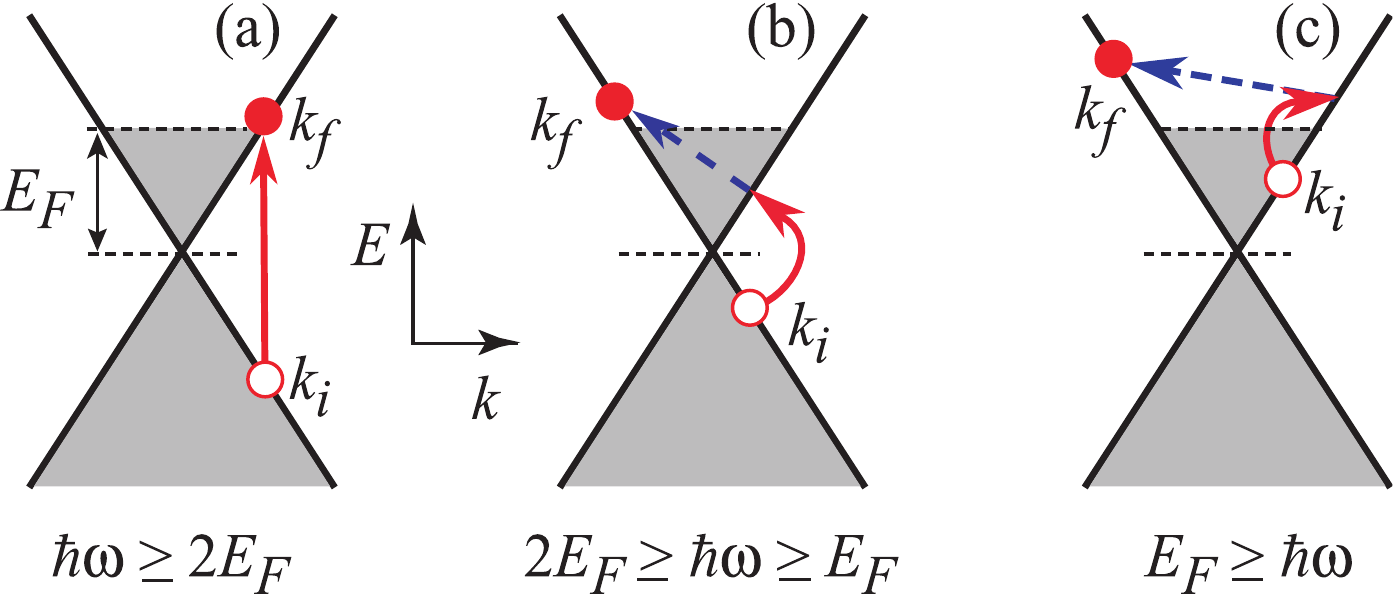}
\caption{Schematic illustration of the 
possible optical transitions: (a) direct \textit{inter}band transition,
(b) indirect \textit{inter}band transition,
(c) indirect \textit{intra}band transition. 
Solid and dashed arrows show electron-pho\textit{t}on interaction (solid arrows) and electron scattering caused by impurities or phonons (dashed arrows).
Initial and final states of a photoexcited 
carrier with wavevectors $k_i$ and $k_f$ are shown by open and 
solid circles, respectively. 
Inequalities define the corresponding photon energy ranges.
} \label{figt1}
\end{figure}

It follows then, that three regimes of radiation absorption are relevant to the discussed problem, namely, (i) direct \textit{inter}band transitions, (ii) indirect phonon or impurity assisted \textit{inter}band transitions and (iii) free-carrier absorption (Drude-like) due to indirect \textit{intra}band  transitions. These three processes are schematically shown in Fig.~\ref{figt1}(a), (b), (c), respectively. To obtain direct transitions, obviously, the condition
\begin{subequations}
\label{ranges}
\begin{equation}
\label{interband}
\hbar \omega \geqslant 2E_F,
\end{equation}
must be fulfilled. The indirect \textit{inter}band transitions become essential at
\begin{equation}
\label{indirect}
E_F \leqslant \hbar \omega \leqslant 2E_F.
\end{equation}
Finally, free carrier absorption caused by indirect \textit{intra}band transitions may contribute for any relation between $\hbar \omega$ and $E_F$. Similar to the case of conventional semiconductors its role increases for larger wavelengths, and for
\begin{equation}
\label{intra}
\hbar \omega < E_F,
\end{equation}
\textit{intra}band transitions become dominant absorption processes.\footnote{Note, that the impurity ionization, multiphoton and lattice absorption, while being possible, are out of the scope of the present review.}

Besides the microscopic origin of the radiation absorption we distinguish the classical and quantum regimes of light-matter
  interaction, which are conventionally given by the relation between $\hbar\omega$ and $E_F$. 
In the classical regime characterized by
\begin{equation}
\label{class}
\hbar \omega \ll E_F,
\end{equation}
\end{subequations}
the
  electron dynamics can be described by means of Newton equations
  of motion and Boltzmann equation for the distribution function
  $f(\bm p, \bm r,t)$, which depends on electron momentum, position,
  $\bm r$, and time. Such a description is valid for the arbitrary values of $\omega\tau$. We note also, that according to Eq.~\eqref{free} if $\omega\tau\ll 1$, then the condition $\hbar\omega \ll E_F$ is automatically fulfilled. 
For photon energy approaching the Fermi energy or for even higher photon energies, at which Eq.~\eqref{class} does not hold,
the   quantum-mechanical treatment of the radiation interaction with
electron system in graphene is required. In the intermediate frequency range, where $\hbar/\tau
  \ll \hbar \omega \ll E_F$, both classical and quantum-mechanical
  approaches merge.

All described regimes can easily be realized by variation of the 
photon energy and electron density in graphene. In the experiments reported so far, making use of the excitation with microwave/terahertz radiation and mid-infrared/visible light clearly corresponds to the classical or quantum mechanical regimes of the light-matter interaction, respectively. This is due to the fact that the nonlinear transport has been studied in ungated $n$- and $p$-type graphene samples, which have rather high carrier density of the order of several times of $10^{12}$~cm$^{-2}$ and, correspondingly, high Fermi energy ranging from  100 to 400~meV. 
Besides high Fermi energy, in all studied samples (exfoliated layers on SiO$_2$/Si substrate, epitaxial graphene on SiC or CVD graphene) 
the electron mobility is quite low, about several thousands cm$^2$/Vs at room temperature.
Such rather short scattering times $\tau$ ranging from units of $10^{-14}$ to $10^{-12}$~s 
enable the investigation of nonlinear effects for 
the parameter $\omega\tau$ about unity  giving rise to a number of
specific effects for THz/microwave frequencies. In particular, 
processes sensitive to the radiation helicity become important and may dominate the nonlinear response
for $\omega\tau \sim 1$, see Sec.~\ref{sec:micro}.

Before we begin the discussion of particular phenomena, we address one more important issue being crucial for all the nonlinear effect under study,
namely, the variation of  the radiation polarization state including degrees of 
linear and circular polarization. The controllable modification of the radiation 
polarization not only helps in the analysis of the mechanisms of the nonlinear 
response but also gives rise to new phenomena caused by transfer of the radiation 
angular momenta to the carriers in graphene. 
Below we show that the various contributions to the nonlinear response are proportional to the Stokes parameters.
Therefore in almost all experiments aimed to
nonlinear high frequency effects in graphene the polarization state 
of the radiation is controllably
modified by means of standard dichroic elements like, e.g., $\lambda/2$ and $\lambda/4$ 
plates or Fresnel rhombus. By that, assuming the radiation propagates along positive $z$ axis, 
the Stokes parameters~\cite{Saleh,BornWolf} are given by
\begin{subequations}
\label{S123}
\begin{equation}
\label{S12}
S_1 = \frac{|E_x|^2-|E_y|^2}{|E_x|^2+|E_y|^2}, \quad S_2 =
\frac{E_xE_y^*+E_x^*E_y}{|E_x|^2+|E_y|^2},
\end{equation}
\begin{equation}
\label{S3}
S_3 \equiv P_{\rm circ}= \mathrm i \frac{E_xE_y^*-E_x^*E_y}{|E_x|^2+|E_y|^2}.  
\end{equation}
\end{subequations}
Here $S_1$ and $S_2$ define the linear polarization of radiation in the $(xy)$
and rotated by $45^{\circ}$ 
coordinate frames, and $S_3$ describes the 
degree of circular polarization or helicity of
radiation. 
Rotation of the polarizer in respect to the polarization plane of the linearly 
polarized laser radiation with $\bm E_l\parallel x$ results in the 
variation of the $S_1$, $S_2$ and $S_3$. In particular, rotation of the half-wave plate 
results in the linearly polarized radiation with 
\begin{subequations}
\label{waveplates}
\begin{equation}
\label{S:alpha}
S_1 \propto \cos{2\alpha}, \, S_2 \propto \sin{2\alpha}, \, S_3=0,
\end{equation}
where $\alpha=2\beta$ defines the orientation of the polarization plane and $\beta$ is the angle between $\bm E_l$  the optical axis $c$. 
The radiation ellipticity, particularly, light helicity are conveniently varied by rotation of a quarter-wave
plate by angle $\varphi$, resulting in
\begin{equation}
\label{S:phi}
S_1 \propto \cos^2{2\varphi}, \, S_2 \propto \sin{4\varphi}, \, S_3
\propto \sin{2\varphi}.
\end{equation}
\end{subequations}
We note that at oblique incidence, crucially needed for some nonlinear effects 
in graphene, the functional behavior of nonlinear contributions in Eq.~\eqref{j:phen:gen} 
is also described by trigonometrical Eqs.~\eqref{waveplates}. This is in spite of the fact, 
that, strictly speaking, they are not directly given by the Stokes parameters 
$S_1$, $S_2$ $S_3$ in form of Eqs.~\eqref{S123}.

\section{Second order effects: Symmetry analysis}\label{2nd}

The class of the second order effects includes second harmonic generation, \emph{dc} 
photocurrent, and optical rectification effect. 
Obviously, the magnitude of the second-order in the electric field response is 
linear in the radiation intensity $I = c|\bm E(\bm q,\omega)|^2/2\pi$.\footnote{We emphasize that the magnitude of electric 
field acting on the charge carriers in graphene differs from that of an incident wave 
owing to the presence of substrate, finite conductivity of graphene itself and interactions. 
These effects require additional analysis and disregarded hereinafter.
}
The appearance and particular behavior of the effects upon variation of 
incidence angle and polarization state of the radiation are determined by the 
symmetry of the system. This is due to the fact that at a
spatial inversion the vector of electric current $\bm j$ changes its
sign while quadratic 
combinations $E_\alpha E_\beta$, $E_\alpha E_\beta^*$ in
Eq. (\ref{j:phen:gen}) do not. Hence,
the second order response is allowed if either (i) the spatial inversion is
incompatible with the symmetry of the structure under study, or (ii)
second-order conductivities $\sigma_{\alpha\beta\gamma}^{(2)}$ and
$\sigma_{\alpha\beta\gamma}^{(2')}$ change their signs at spatial
inversion. The latter is fulfilled, if components $\sigma_{\alpha\beta\gamma}$ are
proportional to the components of the radiation wavevector $\bm q$. 
This is because both photon wavevector $\bm q$ and
electric current $\bm j$ change their signs at the spatial inversion
and symmetry allows the linear coupling between the current and photon
wavevector in the second-order effects, $\bm j \propto \bm
q|E|^2$.
The sensitivity of the second-order phenomena to the spatial inversion 
reveals that peculiarities of the graphene structures, such as coupling with the substrate,
presence of adatoms, terraces, ripples, edges, etc. become crucial.
A further consequence is that these effects depend strongly on  the radiation polarization and the angle of incidence.
The addressed restrictions on the second-order
conductivities are given by the point-group operations and determine the
experimental geometry. They are analyzed in the first part of this
section. 
Afterwards, the existing experiments are introduced and discussed in the second part together with
microscopic models. 
This discussion is extended by a short account on microscopic theory
of some effects discussed in the literature but not observed so far.

Our analysis begins with \emph{dc} current generation, in order to 
demonstrate all important features of the nonlinear response, including 
an interplay between the spatial symmetry reduction  and wavevector 
induced effects and sensitivity to the radiation helicity. 
Further second order effects, such as generation of an \emph{ac} electric current 
giving rise to the harmonics generation or optical rectification, are discussed 
later on in Secs. \ref{2nd:phenom:sec}, \ref{rect}.
The effect of \emph{dc} current generation is given by the second term in
the right hand side of Eq.~(\ref{j:phen:gen}). As addressed above, the 
nonlinear conductivity ${\sigma}^{(2)}_{\alpha\beta\gamma}(\omega, {\bm  
q})$ has contributions due to both the reduced symmetry and radiation
wavevector $\bm q$. Therefore,  it can be conveniently decomposed in
the sum of two parts yielding the \emph{dc} current in the form
\begin{multline} \label{expansion}
{j_\alpha=\sigma^{(2)}_{\alpha\beta\gamma}(\omega, {\bm q})E_\beta E_\gamma^* =}\\
{\left[
\sigma^{(2)}_{\alpha \beta \gamma}(\omega, 0) 
+ \Phi_{\alpha \beta \gamma \mu}(\omega) q_{\mu}\right]E_\beta E_\gamma^*  \:,}
\end{multline}
where linear in the wavevector $\bm q$ terms are taken into
consideration, corresponding contribution is described by the fourth rank tensor $\Phi_{\alpha \beta \gamma \mu}(\omega)$. Such effects are related with the transfer of the photon momentum to the electrons. First we address the third-rank tensor
$\sigma^{(2)}_{\alpha\beta\gamma}(\omega, 0)$, which describes the class of 
phenomena known as  \emph{photogalvanic effects} 
(PGE)~\cite{sturmanBOOK,ivchenkopikus,ivchenko05a,ganichev_book} being present in
noncentrosymmetric systems only. Therefore, in ideal graphene all
photogalvanic effects are strictly forbidden by symmetry. However, in
most of real structures PGE
becomes possible, e.g. for the excitation in the vicinity of the edges, which
locally reduce the symmetry, in the samples with ripples, or if the
graphene layers are deposited on the substrate.  Two types of photogalvanic effects, linear PGE and
circular PGE, are known
and are already observed in graphene~\cite{PRB2010}. The linear PGE is sensitive to the
orientation of the radiation polarization plane, and is described by the
symmetrical with respect to the interchange of $\beta \leftrightarrow
\gamma$ part of $\sigma^{(2)}_{\alpha\beta\gamma}(\omega, 0)$. It is
given by
\[
j_\alpha \propto E_\beta E_\gamma^* + E_\gamma E_\beta^*.
\]
This symmetrized combination of electric field components is proportional to the linear combination of the Stokes parameters $S_1$ and $S_2$, see Eq.~\eqref{S12}.
By contrast, the circular PGE requires angular momentum of photons and,
correspondingly, given by the antisymmetric part of the tensor $\sigma^{(2)}_{\alpha\beta\gamma}(\omega,0)$,
\[
j_\alpha \propto E_\beta E_\gamma^* - E_\gamma E_\beta^* \propto
P_{\rm circ}.
\]
Here $P_{\rm circ}$ is the degree of circular polarization of the radiation given by the Stokes parameter $S_3$, see Eq.~\eqref{S3}.

While the photogalvanic effects are possible only in the systems 
lacking an inversion center, the \emph{dc} current generation proportional to the radiation wavevector $\bm q$ and described by second term in Eq.~\eqref{expansion} is allowed both in
centrosymmetric and noncentrosymmetric media and, consequently, can 
take place in any graphene system. 
 The fact, that the electric current can be caused by the
momentum transfer from photons to electrons was recognized as early as
in beginning of 1970s and the effect was named as a \emph{photon drag
  effect}~\cite{Ch7Yaroshetskii80p173,Ch7Gibson80p182}.  Even earlier, in 1954 Barlow derived
such a \emph{dc} current in 
 terms of \emph{ac} (dynamic) Hall effect: The joint action of electric,
 $\bm E$, and
 magnetic, $\bm B$, fields of the radiation causes a steady-state
 current in the form $\bm j \propto [\bm E \times \bm B]$~\cite{BARLOW1954}. These, at
 first glance, different mechanisms are related to the same phenomena,
 since for the plane wave in the form of Eq.~(\ref{plane}) the complex
amplitudes of electric and magnetic fields are coupled: 
\begin{equation}
\label{BE}
\bm B(\omega,\bm q) =\frac{1}{|\bm q|} [\bm q \times \bm
E(\omega,\bm q)],
\end{equation} 
and dynamic Hall contribution $\propto E_\beta B_\gamma^*$ can be written in form of photon drag effect, i.e. $\propto q_\delta E_\beta E_\gamma^*$. Therefore usually, the terminology choice between the photon drag and
dynamic Hall effects is determined by the microscopic treatment
in terms of the number of photons {absorbed} (quantum mechanical picture -- photon drag effect)
or the action of electromagnetic fields (classical picture -- dynamic Hall effect). While hereafter we
equally use both terms, for the phenomenological
consideration we prefer the term ``photon drag'' effect because the second term in Eq.~\eqref{expansion} is  
proportional to the wavevector $\bm q$. Similarly to the photogalvanic effect, the photon
drag effect may take place in response to both linearly and circularly polarized
radiation, which are described, respectively, by the  symmetric and
antisymmetric in 
$\beta\gamma \leftrightarrow \gamma\beta$ parts of the fourth-rank
tensor ${\Phi}_{\alpha\beta\gamma\mu}$. These effects are termed as linear and circular photon drag effects~\cite{Ivchenko1980,belinicher_cpde,Shalygin2007,PhysRevLett.103.103906}.

While point symmetry and, particularly, spatial inversion impose
restrictions on the conditions of observation, polarization and incidence angle dependence of the effects, another important symmetry operation, namely, time reversal places additional limitations affecting the 
frequency dependence of the response.
Electric current, $\bm j$, and
radiation wavevector, $\bm q$, are odd at time reversal $\bm j \to -\bm
j$, $\bm q \to - \bm q$ at $t\to
-t$. The bilinear combinations of the field related with the linear
polarization $E_{\beta} E_{\gamma}^*+E_{\beta}^*E_{\gamma}$ are
invariant under time reversal. Therefore, the symmetric part of the nonlinear conductivity $\sigma^{(2)}_{\alpha\beta\gamma}(\omega,0)$ 
describing \emph{linear photogalvanic effect}
is odd at time reversal,
i.e. it contains odd powers of scattering rate given by the reciprocal relaxation time $\tau^{-1}$. By contrast, the nonlinear conductivity responsible for
the \emph{linear photon drag effect}
is even at time reversal and may
contain even powers of dissipative constants. For current sensitive to the radiation helicity, i.e., circular photon drag and circular photogalvanic effects, the situation is just opposite. Now, the circular
polarization changes its sign at time reversal, therefore the constants
describing \emph{circular photogalvanic effect}  
are even at time reversal, while constants describing
\emph{circular drag effect} are odd. Owing to different properties under time reversal, the radiation frequency dependences of the linear and circular photocurrents, as well as of photon drag and photogalvanic effects, are distinct, see below for details.

\subsection{Photon drag effects in a single layer graphene}\label{drag:phenom:sec}

We shall start the consideration with the photon drag effect because this
mechanism of the \emph{dc} current generation does not imply a symmetry 
reduction and can be present in any graphene sample. Moreover, the photon drag effect makes it possible to illustrate all facets of phenomenological analysis, including dependence on the incidence angle and effects sensitive to the photon helicity.

\begin{figure}[t]
\includegraphics[width=\linewidth]{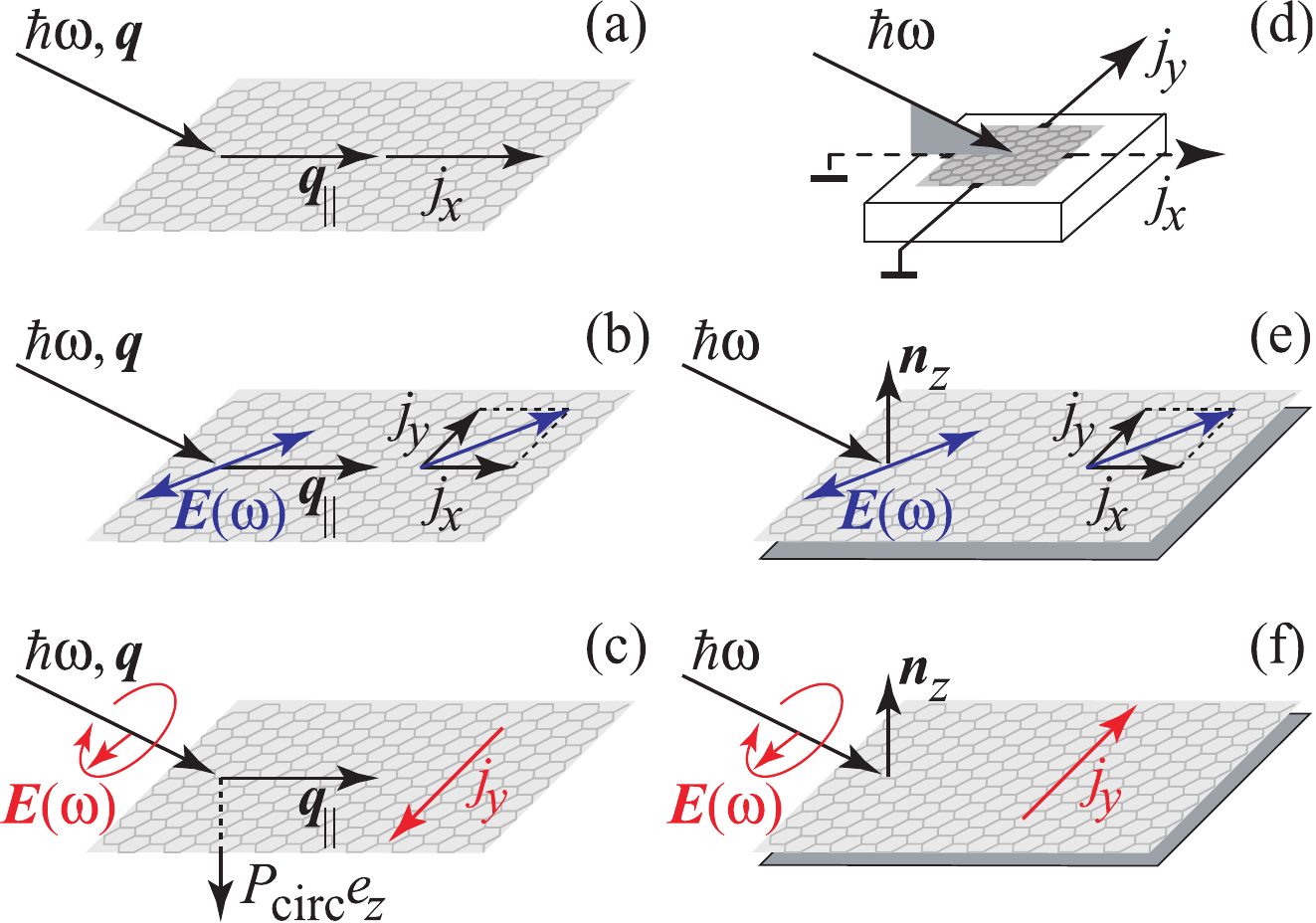}
\caption{Schematic illustration of the possible contributions
to the photon drag and photogalvanic effects.
Panels (a)-(b): linear and (c) circular
photon drag effects, respectively [see Eqs.~\eqref{j:x},~\eqref{j:y}]. 
Note that these figures are relevant both to ideal graphene and graphene on the substrate. Panels (e)-(f):   linear and circular photogalvanic effects 
allowed by symmetry in graphene samples with structure inversion asymmetry, i.e.
deposited on substrates [see Eqs.~\eqref{j:pge:x},~\eqref{j:pge:y}]. 
Panel (d) shows a relevant experimental geometry. 
After~\protect \cite{PRB2010}.
} \label{figt2}
\end{figure}

The consistent phenomenological theory of the photon drag effect in graphene layers
has been developed  
in Refs.~\cite{2010arXiv1002.1047K,PRB2010}. 
Disregarding the substrate, infinite homogeneous graphene layer is 
described by the centrosymmetric $D_{6\rm h}$ point group.
It follows that the tensor
${\Phi}_{\alpha\beta\gamma\mu}$ has five linearly independent
components, which give rise to corresponding contributions to the
photocurrent. However, two of them are related to normal to the
graphene layer component of electric field, $E_z$, or of the wavevector, $q_z$, and, in two-dimensional
system like graphene, are much weaker compared to the others. Hereafter, we disregard these contributions for all effects which can be induced without taking into account $E_z$ and $q_z$. Such a model will be named \emph{strictly} two-dimensional.
Hence, for ideal graphene layer, the photocurrent is given by:
\begin{subequations}
\label{drag}
\begin{equation}
\label{j:x}
j_x = T_1  q_x \frac{|E_x|^2 + |E_y|^2}{2}
+ T_2 q_x \frac{|E_x|^2 - |E_y|^2}{2}  ,
\end{equation}
\begin{equation}
\label{j:y}
j_y  = T_2 q_x \frac{E_xE_y^* + E_x^*E_y}{2}  - \tilde{T}_1  q_x
P_{\rm circ} \hat{e}_z (|E_x|^2+|E_y|^2). 
\end{equation}
\end{subequations}
Here $\hat{\bm e} = (\hat e_x,\hat e_y, \hat e_z) \equiv \bm q/|q|$ is a unit vector in the direction of light propagation, and we introduced the coordinate frame with axes $x$ and
$y$ chosen in the graphene plane, $z$ being the sample normal and
assume that radiation is incident in the $(xz)$ plane, therefore,
$q_y\equiv 0$, see Fig.~\ref{figt2}. Such a choice of the
coordinates is adjusted to convenient experimental geometry where the
current is investigated along and normal to the incidence plane. 
 Constants $T_1$ and $T_2$
describe linear photon drag effect. The specific feature of graphene
compared to three-dimensional cubic semiconductors and simple metals
is the presence of the circular photon drag effect given by the
constant $\tilde T_1$. A further peculiarity of the photon drag effect in graphene comes from its two-dimensional nature: Here the photon drag current is present under oblique incidence only and its direction changes 
 upon reversal of the incidence angle.\footnote{We note that in
  some systems with reduced symmetry due to e.g., 
ripples, strain  or terraces, the photon drag effect may also be allowed at
normal incidence. However, these contributions are expected to be very
small, as they are proportional to two small factors: the
photon wavevector and the degree of asymmetry. Hence, such effects are
out of scope of present review.} 
In the presence of substrate or adatoms deposited on one side of the sample, 
the symmetry of graphene reduces to
the noncentrosymmetric  group $C_{6\rm v}$. In such a case, the
equivalence of the  $z$ and $-z$ directions is removed.
Analysis shows that the form of Eqs.~\eqref{drag} remains the same for noncentrosymmetric graphene described by the $C_{6\rm v}$ point symmetry group. While the functional behavior does not change, the effect may originate from diverse microscopic mechanisms and, consequently, be characterized by different magnitudes of the corresponding constants in Eqs.~\eqref{drag}.

It follows from Eqs.~\eqref{j:x},~\eqref{j:y} that the photon drag current contains, in
general, three contributions illustrated in Fig.~\ref{figt2}, panels (a)--(c).
First one, schematically illustrated in Fig.~\ref{figt2}(a) results in the
 photocurrent, which flows  along the light incidence plane. 
 Two other effects are caused by the reduced symmetry of the system and exhibit a specific polarization dependence described by the combinations of electric
field components $E_x$, $E_y$ in Eqs.~(\ref{drag}). The terms
proportional to $T_2$ are sensitive to the linear polarization  
and yield photocurrent components (i) in the plane of incidence and (ii) 
perpendicular to the incidence plane, see Fig.~\ref{figt2}(b). By contrast, the
current proportional to $\tilde T_1$ is due to transfer of both light \textit{linear} and 
\textit{angular} momenta to electrons and reverses its  sign by changing
photon helicity. This is the
circular photon drag effect {or, as addressed in the previous section, }circular \textit{ac} Hall effect, which
appears in the transverse to the light propagation plane geometry,
see Fig.~\ref{figt2}(c).

In experiments described below, see Sec.~\ref{2nd:eandt}, the polarization 
state of incident radiation was controlled by half- or quarter-wave plates. 
In the former case of linearly polarized radiation, its helicity $P_{\rm circ} \equiv S_3 =0$ and, hence, only the components of 
the photocurrent proportional to $T_1$ and $T_2$ are excited, see Eqs.~\eqref{drag}. 
For not too large incidence  angles $\theta_0$, where $\sin{\theta_0} \approx \theta_0$ and $\cos{\theta_0}\approx 1$ (see Ref.~\cite{2010arXiv1002.1047K} for discussion of the 
arbitrary incidence angle), the component of the current in the radiation incidence plane, $j_x$, is
given by the sum of the polarization-independent contribution [$T_1$ term in Eq.~
\eqref{j:x}] and the contribution excited by polarized light being proportional to $S_1 \propto \cos{2\alpha}$ ($T_2$ term). 
The current component normal to the incidence plane, $j_y$, is excited by the polarized radiation only, it is proportional to the 
Stokes parameter $S_2$, which varies as $\sin{2\alpha}$. 
In the geometry with a quarter-wave plate one obtains elliptically polarized radiation and, in addition to the contributions described above,  one can generate current sensitive to the radiation helicity. 
Here the perpendicular to the incidence plane component of the 
photocurrent, $j_y$, can be presented as a superposition of the contributions excited by linearly and circularly polarized light, being described by the Stokes 
parameters $S_2$ and $S_3$, respectively. For the particular choice of the angles 
accepted by Eq.~\eqref{S:phi} it is described by
\begin{subequations}
\label{jABC}
\begin{equation}
\label{jAB}
j_y = j_{A} \sin{2\varphi} + j_{B} \sin{4\varphi} ,
\end{equation}
where $j_{A} = A |\bm E|^2 \theta_0$ and $j_{B} = B |\bm E|^2\theta_0$ are the 
magnitudes of circular and linear contributions, respectively. The current in the 
incidence plane is given by the superposition of terms $\propto S_1$ and $S_2$, namely, 
\begin{equation}
\label{jBC}
j_x = j_{B} \cos{4\varphi} + j_{C} ,
\end{equation}
\end{subequations} 
with $j_{C} = C |E|^2 \theta_0$.

\subsection{Photogalvanic effects in a single layer graphene}\label{PGE:symmetry}

Appearance of the photogalvanic effects implies that the spatial
inversion symmetry is broken. Such a situation for flat infinite sample is realized for the
graphene layer deposited on the substrate or if adatoms predominantly are present on one surface of the material. The symmetry of graphene on a substrate is $C_{6\rm v}$ and the PGE for the oblique incidence in the
$(xz)$ plane is described for not too large incidence angles by
\begin{subequations}
\label{pge}
\begin{equation}
\label{j:pge:x}
j_{x} = \chi_{l} \frac{E_x E_z^*+ E_x^*E_z}{2}  \:,
\end{equation}
\begin{equation}
\label{j:pge:y}
j_{y} = \chi_{l} \frac{E_y E_z^*+ E_y^*E_z}{2} +
\chi_{c} P_{\rm circ} \hat{ e}_x (|E_x|^2+|E_y|^2) \:,
\end{equation}
\end{subequations}
where two independent parameters, $\chi_l$ and $\chi_c$, describe linear
and circular PGE, respectively.
Similar to the photon drag effect, the PGE can be generated at
oblique incidence only, however, in this case it comes from the
necessity  to provide $z$-component of electric field, rather than in-plane photon wavevector $\bm q$, see
Eqs.~(\ref{pge}). Another specific feature of the PGE compared to the
photon drag effect is that it cannot be generated by unpolarized
radiation. Linear
and circular PGE currents given by $\chi_l$ and $\chi_c$ are schematically  shown in 
Fig.~\ref{figt2}(e) and (f), respectively. It follows from Eqs.~\eqref{j:pge:x},~\eqref{j:pge:y} that the
linear current flows along the projection of the electric field
onto the sample plane and, therefore, in general, it has both 
components along, $j_x$,  and normal, $j_y$, to the light incidence plane. 
By contrast, circular photocurrent flows perpendicularly to the
radiation incidence plane, i.e., along $y$ axis in the chosen
geometry. We note, that the dependence of the photogalvanic effect on the 
polarization state of light and, consequently on the wave plate orientation angles $\alpha$ and $\varphi$
is indistinguishable from that of the photon drag effect, see Eq.~\eqref{jABC}. In particular, for small incidence angles
\begin{subequations}
\label{jABC:pge}
\begin{equation}
\label{jBC:pge}
j_x = j_{B'} \cos{4\varphi} + j_{C'} ,
\end{equation}
\begin{equation}
\label{jAB:pge}
j_y = j_{A'} \sin{2\varphi} + j_{B'} \sin{4\varphi} ,
\end{equation}
\end{subequations} 
where $j_{A'} = A' |\bm E|^2 \theta_0$, $j_{B'} = B' |\bm E|^2\theta_0$ and $j_{C'} = C' |E|^2 \theta_0$ 
with constants $A'$, $B'$ and $C'$ are circular ($j_{A'}$) and linear ($j_{B'}, j_{C'}$) 
photocurrent components. It follows from Eqs.~\eqref{pge} the parameters $j_{B'}$ and $j_{C'}$ 
are interrelated according to $j_{C'}=-j_{B'}$ for $s$ polarization at $\varphi=0$, while $j_{C'}=3j_{B'}$ for $p$ polarization at $\varphi=0$~\cite{weber:GaN}. We emphasize,
however, that in the case of photogalvanic effects the microscopic sense of the 
parameters $A'$, $B'$ and $C'$ is distinct from that of the corresponding 
coefficients $A$, $B$ and $C$ for the photon 
drag effect, since PGE is related to the absence of an inversion center.
 
 The requirement of $z$-component of the field diminishes PGE in
graphene, since strictly two-dimensional carriers are almost unaffected by
$E_z$. This is in contrast to the conventional semiconductor
two-dimensional systems like, e.g., quantum wells and heterojunctions, where
in spite of the fact that there is no carrier motion in $z$ direction 
the wavefunction in extended over many atomic layers and can be easily
affected by an electric field. 
Due to the fact that the polarization behaviors of PGE and photon drag
are similar, the PGE is usually masked by the stronger drag effect.
Thus, the observation of PGE is most
likely in conditions where the photon drag effect is reduced, e.g. at high radiation frequencies, see Sec.~\ref{PGE} for details.

The situation changes, however, in graphene structures  of lower symmetry, which is  reduced, e.g., due 
to asymmetric ripples, curvatures, edges, terraces, etc. Here new contributions to
PGE appear, which do not require the action of $z$ component of
electric field on electrons. In particular, excitation of edges represents the natural route of the symmetry reduction. 
Disregarding the microscopic structure of the
  edge and presence of the substrate,
  we deal with the point symmetry $C_{2\rm v}$ having the two-fold
  rotation axis perpendicular to the edge and lying in the sample plane.
Corresponding additional to Eqs.~\eqref{pge} contributions to the photocurrent are given by
\begin{equation}
\label{phenom:edge}
{
j_y = R_l \frac{E_xE_y^* + E_x^*E_y}{2}+ R_c P_{\rm circ}
  \hat e_z (|E_x|^2+|E_y|^2)},
\end{equation}
where the edge is assumed to be along $y$ axis, and two
  constants, $R_l$ and $R_c$, describing the linear and circular edge
  PGE are introduced. Comparing Eqs.~\eqref{S123} and \eqref{phenom:edge} we see that the polarization dependences of these contributions are given by the Stokes parameters $S_2$ and $S_3$, respectively. We emphasize that edge photogalvanic
effect can be observed even for the normal incidence of
radiation where the photon drag is forbidden.  Obviously, it is sensitive to the quality and microstructure of the edge and provides an experimental access to this important parameters.

A further reduction of edge symmetry may come from the fact, that the edge orientation of the graphene layer is maintained
  with an atomic accuracy and its direction differs from
  high-symmetry ones. In this case, the point symmetry of the
  system lowers down to $C_{\rm s}$ (if the substrate is absent) or  further to
$C_1$ (with allowance for the substrate). In both cases even unpolarized radiation at normal incidence can cause the photocurrent flowing along the edge, and the direction of the current is determined by the microscopic structure of the edge.

Besides edge photogalvanic effects, the symmetry reduction compared to ideal graphene layer may also come
from the other factors both natural and produced on purpose, e.g., terraces, strain, ripples, artificial lateral superlattices etc. 
The symmetry of the system can be lowered 
depending on the specifics of the perturbation.
 In all these cases,
the current at normal incidence can be generated by linearly, circularly or even unpolarized light, its direction and particular polarization dependence
indicates the symmetry of perturbation (see e.g.~\cite{Kiselev11,Nalitov:2012vn}, where photogalvanic effects in perturbed graphene with lateral superlattice were addressed theoretically). 

\subsection{Photogalvanic and photon drag effects in multilayer graphene}\label{multilayer}

An important issue of graphene structures is the possibility
to arrange several atomic layers one on the top of the other. The
striking examples of these systems are graphene bilayers and
trilayers whose physical properties attract now a great 
interest~\cite{TaisukeOhta08182006,castro:216802,FengWang04112008,Mayorov12082011,Bao2011,Zhang2011,Lui2011}. 
In these kinds of systems the response to $z$ component of
electric field required for PGE current can be enhanced as compared with that in the single layer
graphene because each additional layer allows more freedom for electron to move along the sample normal. Therefore, it is expected to make the most pronounced impact on the photogalvanic effect by affecting the microscopic processes of the current formation.
 
Hence, in multilayer systems the the coefficients $\chi_l$ and $\chi_c$
in Eqs.~(\ref{pge}) describing linear and circular PGE under oblique
incidence may be strongly modified and enhanced. At the same time, the dominant contribution to the photon
drag effect in the multilayers is given by Eqs.~(\ref{drag}) and the
constants $T_1$, $T_2$ and $\tilde T_1$ differ from those in a single
layer due to the modification of electron energy spectrum and
scattering processes.

 Besides, the multilayer stacking may contribute to the symmetry reduction and may give rise to the novel 
 photogalvanic effects inherent in multilayer systems only.
The point symmetry
  of graphene $N$-layers 
  depends on the stacking type and on the layer number, $N$. Here we
  consider only two ``natural'' orderings: the rhombohedral stacking
  (ABCABC\ldots), described by the point symmetry
  group $D_{3\rm d}$, which contains an inversion
  center~\cite{PhysRevB.75.155424}, and the Bernal one  (ABAB\ldots).  In the
 latter case, the point 
symmetry is described by either $D_{3\rm d}$ group (for even $N$),
which contains an inversion center, or 
by $D_{3\rm h}$ group (for odd $N>1$)~\cite{PhysRevB.79.125426}. For
odd $N>1$ the CPGE is also symmetry forbidden for ideal
system, however, the linear photogalvanic current becomes possible even for normal 
incidence\footnote{{The electron states in the single valley 
$\bm K$ or $\bm K'$ of graphene monolayer possess $D_{3\rm h}$ 
point symmetry, hence, under the normal incidence the linearly 
polarized light can induce the valley-orbit currents described 
in Refs.~\cite{valleyLPGE,portnoi_book}.}} 
\begin{equation}
\label{D3d}
j_x=\chi_l' (|E_x|^2-|E_y|^2), \quad j_y =-\chi_l'(E_xE_y^*+E_yE_x^*),
\end{equation}
and described by a single parameter $\chi_l'$. Here $x$ axis is
chosen along one of $C_2$ axes in the sample plane.
 Finally, bulk graphite is described by $D_{6\rm  h}$ point symmetry 
group, which contains spatial inversion. Hence, in this material the photogalvanic effects are forbidden and only photon drag current is possible. Like in case of monolayers, the presence of the 
substrate or the top gate can reduce the symmetry of the multilayer 
graphene system and give rise to the photocurrents, which are forbidden in monolayers. For example, the symmetry of the bilayer deposited on the 
substrate reduces from $D_{3\rm d}$ to $C_{3\rm v}$ and the linear photocurrent 
described phenomenologically by Eqs.~\eqref{D3d} becomes possible.

\subsection{Second harmonic generation}\label{2nd:phenom:sec}

Phenomenological analysis of the second harmonic generation (SHG) 
in graphene and graphene-based systems is quite analogous to that 
of the \emph{dc} photocurrents, see Ref.~\cite{glazov:shg} for details. 
The important distinctions are as follows: (i) unlike \emph{dc} 
current generation where the current flows only in graphene plane, 
the \emph{ac} current associated with the harmonic generation can 
have normal to the graphene plane component, and (ii) the second 
harmonic is described by the quadratic
combinations $E_\beta E_\gamma$ in contrast to the \emph{dc}
current formation proportional to the $E_\beta
E_\gamma^*$. Consequently, the nonlinear conductivity
$\sigma^{(2')}_{\alpha\beta\gamma}$ is invariant under the permutation
of the last two subscripts. It follows that for unpolarized and linearly 
polarized radiation the symmetry description of the
SHG is similar to that of the linear
photon drag and photogalvanic effects: In the very same way, there are contributions to the
second harmonic related with the absence of an inversion center in
the medium and with the photon wavevector $\bm q$.
In \emph{strictly} two-dimensional model for an
ideal single-layer sample the current at a double frequency is described by two
linearly independent complex constants ${Q}_1$ and ${Q}_2$ as
\begin{subequations}
\label{phenom:2}
\begin{equation}
\label{phenom:2x}
j_x(2\omega, 2\bm q) = {Q}_1 q_x (E_x^2+ E_y^2) + {Q}_2[q_x(E_x^2-E_y^2)+2q_yE_xE_y],
\end{equation}
\begin{equation}
\label{phenom:2y}
j_y(2\omega, 2\bm q) = {Q}_1 q_y (E_x^2+ E_y^2) + {Q}_2[q_y(E_y^2-E_x^2)+2q_xE_xE_y].
\end{equation}
\end{subequations}
Comparing these expressions with Eqs.~(\ref{j:x}) and (\ref{j:y}) for linear photon drag effect we see that, as addressed above, the electric field and wavevector dependencies of these effects are just the same. 
Although second harmonic can be generated by unpolarized, linearly 
polarized or even circularly polarized radiation,  
no contribution sensitive to the radiation helicity to the second 
harmonic current is possible owing to the fact that the quadratic 
combinations $E_\beta E_\gamma$ in Eqs.~\eqref{phenom:2} are not 
sensitive to the radiation helicity. Therefore,  analogues of the 
helicity driven \emph{dc} current given by bilinear contributions 
$E_\beta E_\gamma^*-E_\beta^*E_\gamma$ in Eqs.~\eqref{drag}, 
\eqref{pge}, \eqref{phenom:edge} are absent for the SHG. 

\begin{figure}[t]
\includegraphics[width=\linewidth]{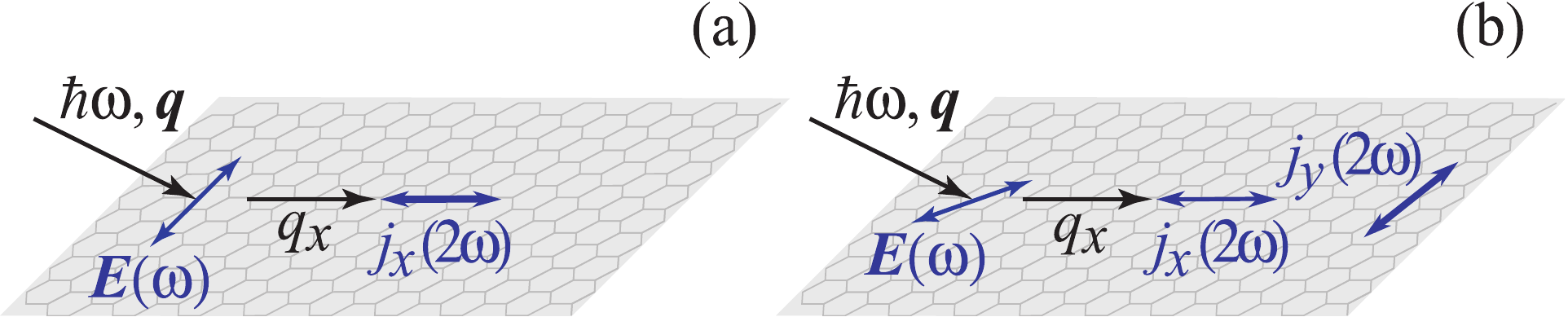}
\caption{Schematic illustration of the second harmonic generation described by Eqs.~\eqref{phenom:2}.  
Panel (a) shows current component in the incidence plane  at a double frequency, $j_x(2\omega)$ excited by 
fundamental frequency light polarized along $y$-axis, 
(b) shows both contributions to the second harmonic current parallel and perpendicular to the radiation incidence plane
for general direction of $\bm{E}(\omega)$.
  }\label{fig:directions} 
\end{figure}

Figure~\ref{fig:directions} schematically shows the geometry of the
second harmonic generation and the response at a double frequency,
$2\omega$. For ideal graphene sample the second harmonic is 
excited only at the oblique incidence of radiation and is caused by the photon wavevector $\bm q$. 
For instance, for $q_x\ne 0$, $q_y=0$, there is a component
of the in-plane oscillating current $\bm j(2\omega)$ parallel to the
light incidence plane described by $({Q}_1+{Q}_2)q_x
E_x^2+({Q}_1-{Q}_2)q_xE_y^2$ [Fig.~\ref{fig:directions}(a)]. Additionally,
there is a contribution, $2{Q}_2 q_x E_xE_y$, perpendicular to the incidence plane, see
Fig.~\ref{fig:directions}(b). 

Since the nonlinear conductivities $\sigma^{(2)}_{\alpha\beta\gamma}$ describing linear photon drag/photogalvanic effects and $\sigma^{(2')}_{\alpha\beta\gamma}$ responsible for the SHG, transform under symmetry operations in the same way, the phenomenological analysis of the effects of substrate, adatoms, ripples, edges, multilayer stacking, etc. on photocurrents, presented above, holds also for the second harmonic generation.

\subsection{Optical rectification}\label{rect}

Optical rectification complements the class of the discussed above second order 
effects resulting in \textit{dc} or \textit{ac} electric current.
It
refers to the formation of
the steady state dielectric polarization $\bm P$ in response to the radiation
propagating in the media~\cite{PhysRev.138.A534}. While the point symmetry
restrictions on the optical rectification effect are the same as for the \emph{dc} current
generation and described by Eqs.~(\ref{j:phen:gen}),~(\ref{expansion}), $\bm P$ does not
change its sign under the time reversal. 
Thus the constants responsible for the 
optical rectification and \emph{dc} current generation
have different properties at time reversal.
As a consequence, in contrast to photogalvanic or photon drag effects,
the optical rectification gives rise to the electric current during
the transient process only, when the illumination is turned on or off~\cite{ivchenkopikus,cote} 
\begin{equation}
\label{jP}
\bm
j = \frac{d\bm P}{dt}. 
\end{equation} 
Accordingly, microscopic mechanisms of the optical 
rectification, photon drag and photogalvanic effects are different. 
In particular, in contrast to the photon drag and photogalvanic 
effects, optical rectification does not require optical 
absorption and may take place in the transparency region.

To complete the picture, we note that the most general case of two incident 
waves with frequencies $\omega_1$ and $\omega_2$ 
(wavevectors $\bm q_1$ and $\bm q_2$) can also be considered. In this situation, 
the current or polarization 
response contains the nonlinear contributions corresponding 
to the sum and difference of the frequencies, $\omega_1 \pm \omega_2$, and wavevectors, $\bm q_1 \pm \bm q_2$, 
\begin{subequations}
\begin{equation}
\label{threewave}
j_\alpha \propto E_\beta(\omega_1, \bm q_1)E_\gamma(\omega_2, \bm q_2)\mathrm e^{-\mathrm i (\omega_1+\omega_2)t + \mathrm i (\bm q_1 + \bm q_2) \bm r},
\end{equation}
and
\begin{equation}
j_\alpha \propto E_\beta(\omega_1, \bm q_1)E^*_\gamma(\omega_2, \bm q_2)\mathrm e^{-\mathrm i (\omega_1-\omega_2)t + \mathrm i (\bm q_1 - \bm q_2) \bm r},
\end{equation}
\end{subequations}
respectively,
giving rise to the \textit{three}-wave mixing effects. Note that for $\omega_1=\omega_2$ in 
Eqs.~\eqref{threewave} we obtain second harmonic and \emph{dc} current generation described above. 
Phenomenological analysis of these effects can be carried out in a way similar to 
the description of the photon drag and photogalvanic effects.

\section{Second order effects: Theoretical background}\label{sec:micro}

Microscopic theory of second order effects in graphene was discussed in a number of works considering classical and quantum regimes of light-matter interaction~\cite{Mele:2000oq,PhysRevB.84.045432,entin10,karch2010,2010arXiv1002.1047K,glazov:shg,edge}. In order to illustrate the appearance of the second-order nonlinear effects in ideal graphene we first consider the classical range of radiation frequencies given by Eq.~\eqref{class} and describe the electron dynamics in the framework of the second Newtons law: 
\begin{equation} \label{Newton}
\frac{d \bm p}{d t} + \frac{\bm p}{\tau} = e
{\bm E}({\bm r}, t) +  \frac{e}{c} [{\bm v} \times {\bm B}({\bm r}, t)]
\:.
\end{equation}
The approach is a standard way widely used for other nonlinear media~\cite{legurevich67,perelpinskii73}, the specificity of graphene comes from the unusual relation between the momentum  $\bm p$ and velocity $\bm v$,
\begin{equation}
\label{v}
\bm v = v \bm p /p,
\end{equation}
and details of effective friction force, ${\bm p}/\tau$,  acting 
on the electron due to the scattering processes.
 Equation~\eqref{Newton} contains both the force acting from the electric 
 field of the radiation and the Lorentz force caused by the magnetic field of the wave.

\begin{figure}[t]
\includegraphics[width=\linewidth]{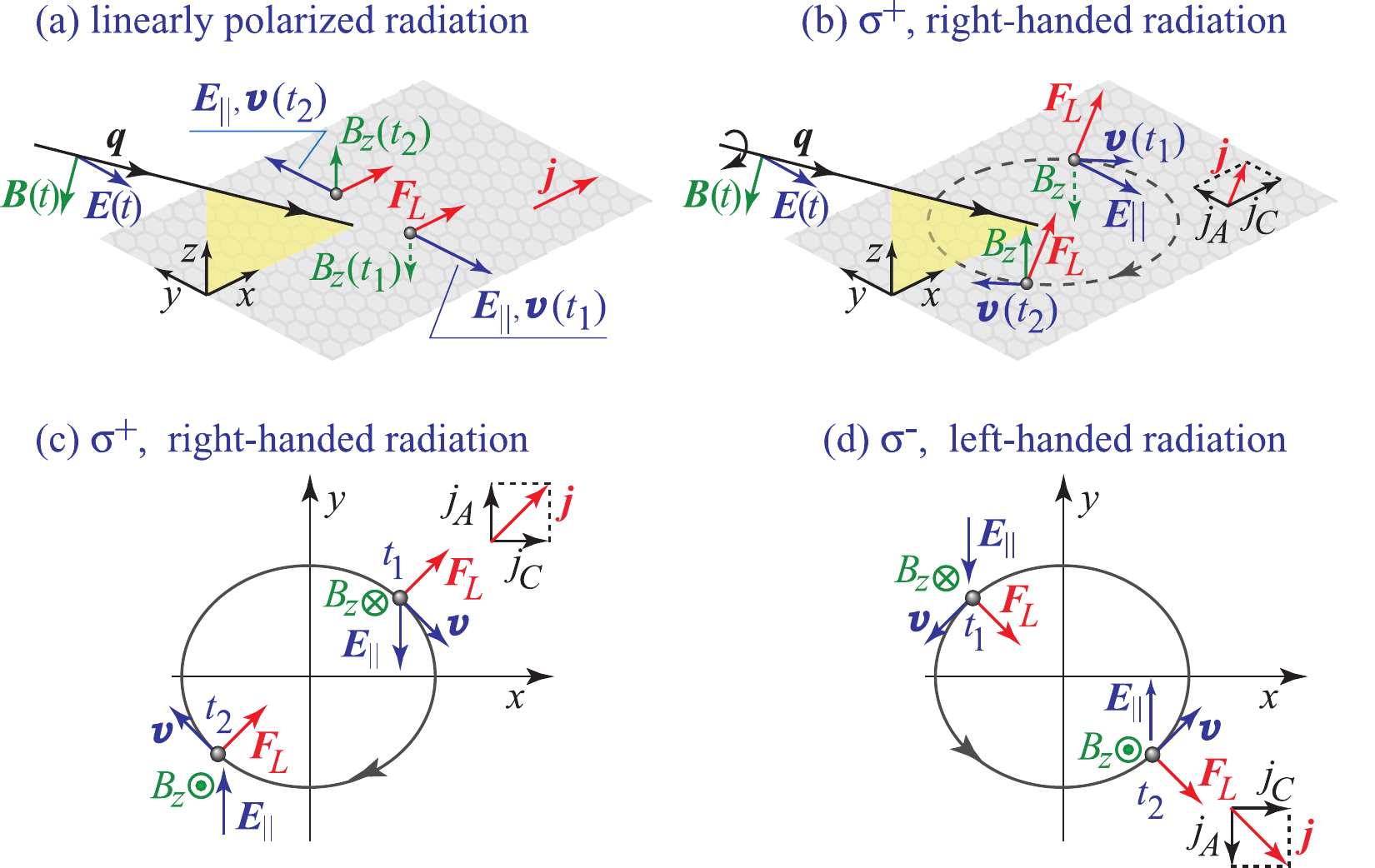}
\caption{
Schematic illustration of the dynamic Hall effect.  For
simplicity we assume positively charged carriers, i.e., holes. $\bm E_\parallel$, $B_z$ and $\bm v$ denote the in-plane components of electric field of the radiation, $z$-component of the magnetic field and the electron velocity, induced by electric field, respectively. These vectors are shown for two time moments, $t_1$ and $t_2$, corresponding to half-periods of the field oscillations. Microscopically, action of these fields results in $\bm F_L$ and, correspondingly, $\bm j$ are the Lorentz force and \emph{dc} current, respectively, see text for details. (a) Linear effect. (b) Circular effect, $\sigma^+$ radiation.
The
dashed orbit represents the hole elliptical trajectory caused by
the $ac$ $\bm E$-field.
(c) and (d) show top view of panel (b) for $\sigma^+$ and $\sigma^-$ radiation, respectively.
Data are given after~\protect \cite{karch2010}. 
} \label{fig2}
\end{figure}
 
Equation~\eqref{Newton} can be solved by iterations in the strength of electro-magnetic field. At a first stage we determine the linear response of electron on the oscillating electric field. The momentum oscillations are written as
\begin{equation}
\label{tildep}
\tilde{\bm p}(t) = \frac{e\tau \bm E_{\parallel}\mathrm e^{-\mathrm i \omega t}}{1-\mathrm i \omega \tau} + {\rm c.c.},
\end{equation} 
where $\bm E_{\parallel}$ is the field component in the plane of the graphene monolayer. The second stage of 
calculations is to determine the nonlinear response, induced by the forces in the right 
hand side of Eq.~\eqref{Newton}. It contains two contributions: One is related to the action of magnetic field, in which case the second-order correction to the 
electron momentum is caused by the Lorentz force. The other one results from the coordinate 
dependence of the electric field and does not require magnetic field at all. Below we 
consider both effects one by one and start with the response to the magnetic field. 
In this case, the steady state momentum is given by  
\begin{equation}
\label{hfh}
\bar{\bm p} = \overline{\frac{e\tau}{c} [\tilde{\bm v} \times {\bm B}( t)]}.
\end{equation}
Here the overline denotes the time-averaging, $\tilde{\bm v} $
is the oscillatory part of the velocity determined by Eqs.~\eqref{v} and~\eqref{tildep}. The coordinate dependence of the fields can be neglected. In the same way, the oscillations of the momentum at a double frequency (second harmonic generation) are given by the similar to Eq.~\eqref{hfh} expression:
\begin{equation}
\label{hfh2}
\tilde{\tilde{\bm p}} = \frac{e\tau}{(1-2\mathrm i \omega \tau)c} \widetilde{[\tilde{\bm v} \times {\bm B}( t)]}.
\end{equation}
Here wide tilde means taking the contribution oscillating at $2\omega$.

To illustrate the microscopic origin of the second-order responses we consider the photon drag effect as an example. 
The steady-state momentum $\bar{\bm p}$ in Eq.~\eqref{hfh} corresponds to the \emph{dc} current flow. Physically, it is related with the electron drift caused by the crossed electric and magnetic fields of the wave. Basic physics of this effect is illustrated in Fig.~\ref{fig2}(a). 
We assume the oblique incidence of the radiation in the $(xz)$ plane, and for the sake of illustration, consider the case of $s$-polarized radiation, where the electric field oscillates in the sample plane. At one of half-periods of oscillation, say, at time moment $t_1$, the radiation electric and magnetic fields acting on charge carrier  result in a Lorentz force and, consequently, electron drift in the direction of the light propagation (longitudinal current). At the second half of period $t_2$ both fields reverse, hence, the drift direction remains. The resulting current is so called dynamic or \emph{ac} Hall effect, which was considered by H.M.~Barlow~\cite{BARLOW1954}. The mechanism of the second-order response due to the joint action of electric and magnetic fields is named $EB$-mechanism. In quantum mechanical approach it corresponds to the magneto-dipole transitions.

While the longitudinal current is expected for unpolarized and even circularly polarized radiation, the appearance of the photon helicity dependent current is not obvious. However, as shown in Ref.~\cite{karch2010} such current indeed emerges if one takes into account the effect of retardation between the electric field $\bm E$ and the instant velocity of charge carrier $\bm v$, being most pronounced for $\omega\tau \sim 1$. The model picture of the circular \emph{ac} Hall effect is presented in Fig.~\ref{fig2}(b)-(d). For circularly polarized radiation, the electric field
rotates around the wavevector ${\bm q}$, sketched in
Fig.~\ref{fig2}(b) for $\sigma_+$ circularly polarized light.
Now, instead of linear oscillations, the carriers follow the elliptic orbit. At an instant of time
$t_{1}$, the carrier is accelerated by the
in-plane component ${\bm E}_{\parallel }$ of the $ac$ electric
field.
At the same time, the carrier with velocity ${\bm v}$ is subjected
to the out-of-plane magnetic field component ${\bm B}_z$.
Note, that the velocity ${\bm v}$ does not
instantaneously follow the actual ${\bm E}_{\vert \vert }$-field
direction due to retardation: There is a phase shift equal to
$\arctan(\omega \tau)$ between the electric field and the electron
velocity ${\bm v}$. The effect of retardation, well known in the Drude-Lorentz theory of high frequency conductivity,
results in an angle between the velocity ${\bm v}$ and the electric field direction ${\bm E}_{\parallel}$,
which depends on the value of $\omega \tau$.
The resulting Lorentz force $\bm F_L =  e /c [\bm v \times \bm B_z]$
generates a Hall current $\bm j$,
also shown in Fig.~\ref{fig2}.
Half a period later at $t_2 = t_1 + T/2$, both ${\bm v}$ and ${\bm B}_z$
get reversed so that the direction of $\bm F_L$
and, consequently, the current $\bm j$ stay the same. The oscillating
magnitude and direction of ${\bm B}_z$ along the trajectory
lead to a periodical modulation of the Lorentz force with nonzero average
causing a nonzero time-averaged Hall current with fixed
direction.
If, as shown in Fig.~\ref{fig2}(d), the light helicity is reversed,
the electric field rotates in the opposite direction and, thus, the charge
carrier reverses its direction. Hence, owing to retardation, the $y$-component of 
$\bm F_L$ at $t_1$ and $t_2$ is inverted. Consequently the polarity of
the transverse, time-averaged Hall current changes.  We stress that the origin of the circular {\emph{ac} Hall} effect is related with the retardation, which is very important if $\omega\tau \sim 1$. Such condition is readily realized for the state-of-the-art graphene samples at THz frequency range.

Now we turn to another mechanism of the second order response. As it follows from  Eq.~\eqref{Newton} this contribution comes from the fact that 
 that the momentum oscillations of an electron given by Eq.~\eqref{tildep} result in the oscillations of its coordinate, $\tilde{\bm r}(t)$. The electric force acting on electron depends on its position owing to the coordinate-dependence of $\bm E$, corresponding contribution has the form~\cite{perelpinskii73}
\[
{e \bm E_\parallel \mathrm e^{\mathrm i \bm q \bm r - \mathrm i \omega t}  \approx} e\mathrm i [\bm q \tilde{ \bm r}(t)] \bm E_\parallel\mathrm e^{-\mathrm i \omega t} + {\rm c.c.},
\]
and its time average results in the steady-state response, while its second temporal harmonic gives rise to the second harmonic generation. This mechanism named as $qE^2$ mechanism corresponds to the quadrupole transitions in a quantum approach.

The consistent theory of the second order response in the classical frequency range is developed in the framework of Boltzmann kinetic equation for the position $\bm r$, momentum $\bm p$ and time $t$ dependent electron distribution function:
\begin{equation}
 \label{kinetic:gen}
\frac{\partial f}{\partial t} + \bm v \frac{\partial f}{\partial \bm r} +
e\left(\bm E + \frac{1}{c}[\bm v \times \bm B]\right)
\frac{\partial f}{\partial \bm p} = Q\{f\}\:,
\end{equation}
where $Q\{f\}$ is the collision integral. Equation~\eqref{kinetic:gen} takes into account the action of electric and magnetic fields of radiation and is solved iteratively in the field amplitudes. The details of calculations are presented in Refs.~\cite{karch2010,2010arXiv1002.1047K,glazov:shg}. Corresponding results of calculations, comparison with available experimental data, and the extensions of treatment to cover quantum range of frequencies and to include the symmetry reduction owing to sample edges or substrate are reviewed below together with experimental results.

\section{Second order effects: Experiment and theory}\label{2nd:eandt}

\subsection{Second harmonic generation}

\subsubsection{Microscopic theory}
 
The SHG theory in graphene was presented in several  works~\cite{PhysRevB.84.045432,PhysRevB.82.125411,glazov:shg}. Reference \cite{PhysRevB.84.045432} deals with symmetry arguments and effects of radiation propagation in multilayer graphene-based systems. In the works \cite{PhysRevB.82.125411,glazov:shg} both quantum mechanical and classical regimes were discussed. The approaches of Refs.~\cite{PhysRevB.84.045432,glazov:shg} agree for the intermediate frequencies $\tau^{-1} \ll \omega \ll E_F/\hbar$. Here we follow Ref.~\cite{glazov:shg} and present the results of calculations for the constants ${Q}_1$ and ${Q}_2$
describing {two independent contributions to} the second harmonic current $\bm j(2\omega, 2\bm q) \propto Q_1, Q_2$, see in Eqs.~\eqref{phenom:2}. These calculations, based on Boltzmann equation and describe the classical frequency range, yield
\begin{subequations}
\label{S}
\begin{multline}
\label{S1}
 { {Q}_1 = -\frac{e^3v^4}{2\omega}\sum_{\bm k} \tau_{1,\omega}f_0' \times} \\ { \left[ \frac{\tau_{1,2\omega}}{\varepsilon_k}(3+\mathrm i
   \omega\tau_{2,\omega}) + (1-\mathrm i
   \omega\tau_{2,\omega})\frac{\mathrm d\tau_{1,2\omega}} {\mathrm
     d\varepsilon_k} \right],}
\end{multline}
\begin{multline}
\label{S2}
 { {Q}_2 = \frac{e^3v^4}{2\omega}\sum_{\bm k}
  \tau_{1,\omega}f_0'\times} \\
 {\left\{ \frac{\tau_{1,2\omega}}{\varepsilon_k}(1+4\mathrm i
   \omega\tau_{2,2\omega}) - \frac{\mathrm d} {\mathrm
     d\varepsilon_k}[\tau_{1,2\omega}(1-2\mathrm i
   \omega\tau_{2,2\omega})] \right\}.  }
\end{multline}
\end{subequations}
Here $f'_0=df_0/d\varepsilon$,
\[
\tau_{n,\omega} = \frac{\tau_n}{1-\mathrm i \omega \tau_n}, \quad (n=1,2),
\]
with $\tau_1$ and $\tau_2$ being the momentum and alignment relaxation
times, respectively and the condition $\hbar\omega \ll E_F$ is assumed.  Using the obtained \emph{ac} second harmonic current $\bm j(2\omega, 2\bm q)$ and Maxwell equation
\begin{equation}
\label{Maxwell}
\Delta \bm A(\bm r, t) + \frac{4\omega^2}{c^2}\bm A(2\omega) = -\frac{4\pi}{c} \mathrm e^{2\mathrm i \bm q_\parallel \bm \rho - 2\mathrm i\omega t}\delta(z) \bm j(2\omega) + {\rm c.c.},
\end{equation}
we obtain the vector potential 
$\bm A(\bm r, t)$ of emitted radiation. Here $\bm q_{\parallel}=(q_x,q_y)$ is the projection of radiation
wavevector onto the sample plane $z=0$. Note, that the current oscillating at a double frequency, $\bm j(2\omega)$, is proportional to the 
square of incident electric field, i.e. to the intensity of the fundamental harmonic. As a result, the intensity of the second harmonic is proportional to the fourth power of the incident electric field or the square of the fundamental harmonic intensity.

In the static limit, $\omega\to 0$, the coefficients ${Q}_1$ and ${Q}_2$
are real and diverge as $1/\omega$, but the net current $\bm j(2\omega, 2\bm q)$ remains
 finite due to factors $\propto \bm q$ in
 Eqs.~\eqref{phenom:2}. Coefficients ${Q}_1$ and ${Q}_2$ become, up to
 common factor, equal to the constants $T_1$ and $T_2$ describing
 linear photon drag effect, see Eqs.~\eqref{drag} and \eqref{classical} below, because at $\omega=0$ responses at
 zero and double frequencies are indistinguishable. At high
 frequencies, $\omega \tau_1 \gg 1$, $\omega \tau_2 \gg 1$, parameters
 ${Q}_1$ and ${Q}_2$ are proportional to $1/\omega^3$, hence, current
 density decays as $1/\omega^2$.

A remarkable feature of the SHG microscopic mechanism is the
fact, that for $\omega\tau \sim 1$ parameters ${Q}_1$ and ${Q}_2$ contain
real and imaginary parts, moreover, the phases of these quantities are
different. By that, excitation with linearly polarized radiation may cause  circularly polarized light at a double frequency. Indeed, if the incident radiation contains both
$x$ and $y$ components of $\bm E$, then the response at the double
frequency also contains $j_x$ and $j_y$, however, their oscillations
are phase-shifted. Thus, the second harmonic radiation becomes, in general, 
elliptically polarized. Calculation shows that the degree of circular
polarization of the emission can reach $90$~\%~\cite{glazov:shg}. Note, that this effect is not observed so far.

It is worth to mention that the response of graphene at a double
frequency due to the outlined mechanisms can be much higher than in conventional 
semiconductor systems, since electron velocity in graphene exceeds Fermi
velocity of electrons in semiconductor heterostructures. 
We compare the second order response in
  graphene with that of a two-dimensional centrosymmetric system with
  parabolic energy spectrum ($Q_1^{\rm parab}, Q_2^{\rm parab}$). In the high frequency limit
  ($\omega \tau \gg 1$, $\hbar\omega \ll E_F$)  Eqs.~\eqref{S} yield  ${Q}_2 = {Q}_1/2$ and we obtain the enhancement factor~\cite{PhysRevB.84.045432}
\[
\eta=\frac{{Q}_1^{\rm graphene}}{{Q}_1^{\rm parab}} = \frac{v^2}{2v_{F}^2},
\]
where $v=10^8$~cm/s, $v_{F} = \sqrt{2E_{F}/m}$ is the
Fermi velocity of electrons in the quantum well structure, $m$ is
their effective mass. For typical Fermi velocities on the order of
$v_{F} = 2\times 10^7$~cm/s one has the enhancement factor 
\[
\eta \sim 10.
\]
Hence, the second harmonic response in graphene may be about an order
of magnitude 
larger than that of other two-dimensional semiconductor systems. 
Moreover, it can further be enhanced due to the excitation of plasmons as 
suggested in Ref.~\cite{PhysRevB.84.045432}.

\subsubsection{Experiment}\label{2nd:exper}

Second harmonic generation has been first 
observed in single and multilayer graphene samples 
on SiO$_2$/Si substrates applying near-infrared radiation \cite{dean:261910,PhysRevB.82.125411}. 
The experiments on harmonic generation  reported so far applied 
linearly polarized pump beams and the linearly polarized response has been analyzed. 
The  nonlinear optical effects sensitive to the radiation helicity of 
the pump beam or resulting in generation of a circularly polarized light are 
still a challenge.  
In agreement with the phenomenological theory presented in Sec.~\ref{2nd:phenom:sec}, second harmonic has been observed applying radiation at oblique incidence. 
Either $p$- or $s$- polarized beam of a femtosecond Ti:Sapphire laser operating with pulse energy $\approx$0.06 nJ and duration about 150 fs  
in the wavelength range of $730-830$ nm have been used. The radiation 
falls on a graphene layer at an angle of incidence of $\theta =60^\circ$ and  is focused into an elliptical
spot size of approximately 7 --- 10 $\mu$m. 
 We note that in graphene, 
which is strictly two-dimensional system, the phase synchronization condition needed for 
harmonic generation in bulk materials is relaxed.
The signal at a double frequency is collected, optically filtered from the
fundamental light,  and detected using  a cooled photomultiplying tube and photon-counting
electronics.  It has been verified that the 
intensity of the {second harmonic} emission is proportional to the square of that for the incident radiation. The variations of the signal upon rotation of the radiation polarization vector as well as  
rotation the sample about the normal axis have been analyzed. Both methods allow the detailed 
characterization of the second harmonic and, together with averaging
over many rotations, improve the signal/noise ratio. The latter is of importance, since the second harmonic
intensity  from a small graphene sheet is very weak -- a few photons per second.
While all four combinations of $s$- and $p$-polarized fundamental and {second harmonic} light have been measured, the highest 
{second harmonic} intensity has been detected for $p$-polarization of both beams. 

\begin{figure}[t]
\includegraphics[width=0.65\linewidth]{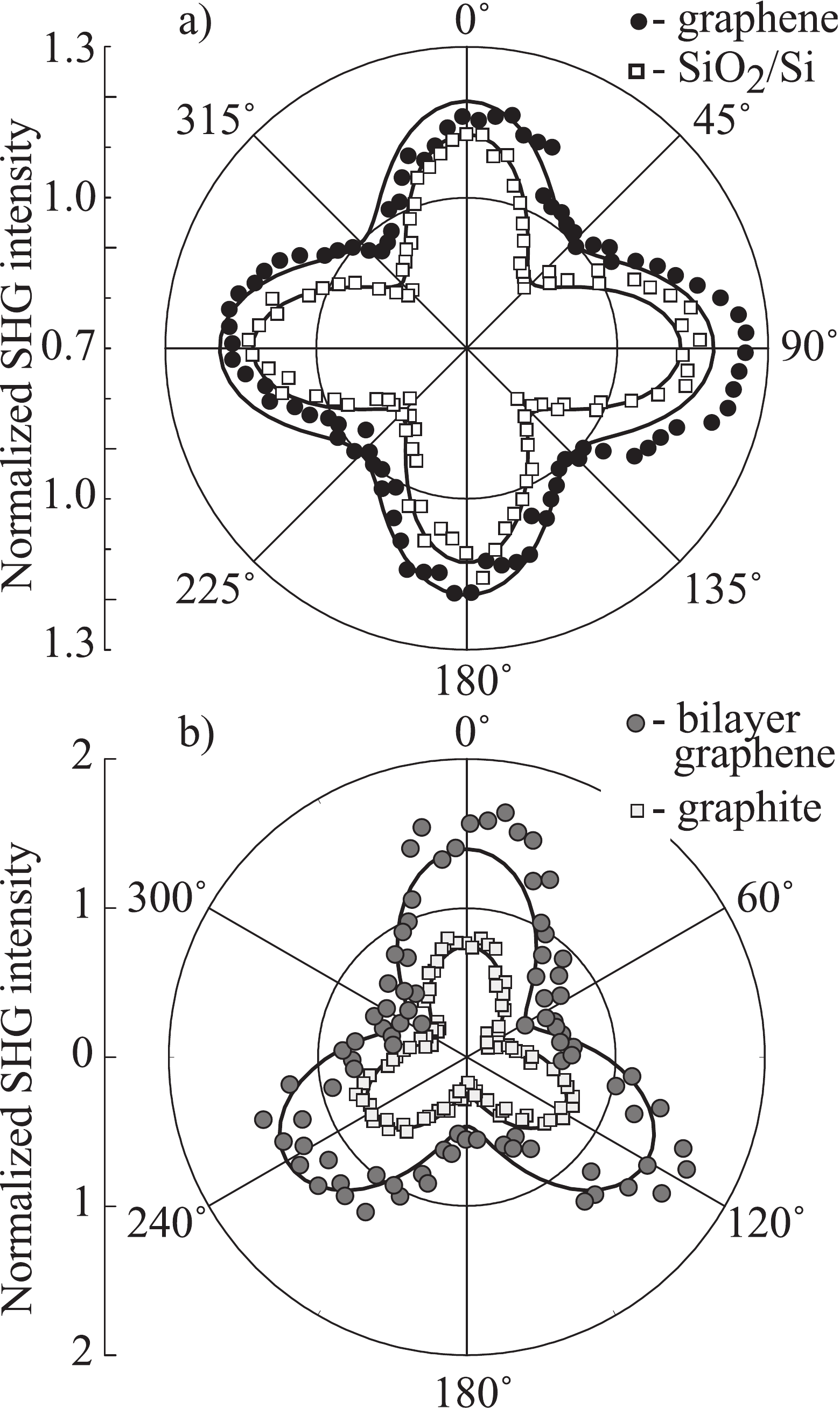}
\caption{
(a) Polarization dependence of the normalized second harmonic radiation 
intensity detected from graphene on SiO$_2$/Si substrate (filled circles)
and purely SiO$_2$/Si substrate (open squares) measured for the fundamental harmonic wavelength $\lambda=800$~nm at room temperature. 
Here {the polar angle $\gamma$} is the angle between the incidence plane and [100]-axis of the substrate (see text for details). Solid curves are fits according to Eq.~\eqref{SHG:Si:fit}. 
(b) Normalized second harmonic radiation intensity detected 
from bilayer graphene on SiO$_2$/Si (filled circles) and from 
bulk graphite (open squares) and a function of {$\gamma$}. Solid curves show the fits after Eq.~\eqref{SHG:bi:fit} and are
normalized such that the isotropic component of the second harmonic signal from silicon
would be unity. 
The absolute angle is arbitrary for both curves. 
After~\protect \cite{dean:261910}.
} \label{fig:SHG}
\end{figure}

The evidence of the SHG in the single layer samples requires careful analysis of the data, in particular, of the dependence on the incidence plane orientation characterized by  an angle {$\gamma$} between the incidence plane and $[100]$ axis of the substrate, see Fig.~\ref{fig:SHG}. As it follows from the phenomenological theory described above in Sec.~\ref{2nd:phenom:sec},
the second harmonic emission from graphene monolayer is isotropic, its intensity should not change upon variation of the angle $\gamma$. However, experimentally two contributions, the isotropic 
({$\gamma$}-independent) and quadrupolar ({$\propto \cos{4\gamma}$}) 
contributions are observed in single layers. As a result, the normalized SHG intensity 
as a function of angle $\gamma$ 
can be described by the following fitting equation:
\begin{equation}
\label{SHG:Si:fit}
I({\gamma}) = A_0 + A_4 \cos{(4{\gamma}+\delta)},
\end{equation}
where $A_0$ and $A_4$ are the amplitudes of the isotropic and quadrupolar components and $\delta \approx 0$ is the phase. The problem in analysis is that such a behavior is expected and indeed 
observed from the bare Si substrate~\cite{PhysRevB.82.125411,Tom:1983dq}. 
{However, contribution of graphene to the second harmonic manifests itself by reduction of anisotropy by about $30\%$ and an increase of intensity, compared with bare substrate}. This result is in agreement with phenomenological description, Eqs.~\eqref{phenom:2}, which demonstrates that the graphene response is isotropic: The intensity of the second order response is the same irrespective of the orientation of the incidence plane.

The situation drastically changes in multilayer graphene. Here, instead of four-fold, the symmetry of photoresponse becomes three-fold, which rules out the substrate contribution. It is shown in Fig. \ref{fig:SHG}, where the normalized {second harmonic} intensity is plotted as a function of {$\gamma$}. The experimental results now follow the phenomenological equation
\begin{equation}
\label{SHG:bi:fit}
I({\gamma}) = A_0' + A_3' \cos{(3{\gamma}+\delta')},
\end{equation}
where $A_0'$ and $A_3'$ are the amplitudes of isotropic (zeroth) and third angular harmonics, $\delta'$ is the initial phase. In this case the fourth angular harmonic is absent, indicating that the response is dominated by the multilayer graphene rather than by a substrate.
The data not only demonstrate a pure multilayer graphene response but also indicate the symmetry reduction to $C_{3\rm v}$ supporting the effect of the substrate induced structure inversion asymmetry, see Sec.~\ref{multilayer}. Based on this difference in Ref.~\cite{dean:261910} second harmonic generation effect was suggested for the diagnosing the layering structure of graphene samples. 
As recently shown in Ref.~\cite{murzina} the SHG can also be observed in flat graphene at the normal incidence, however, only if additionally an in-plane static field is applied to graphene sheet. These processes are already third order in electric field and will be discussed from the theoretical point of view in Sec.~\ref{3rd}.

\begin{figure}[t]
\includegraphics[width=0.8\linewidth]{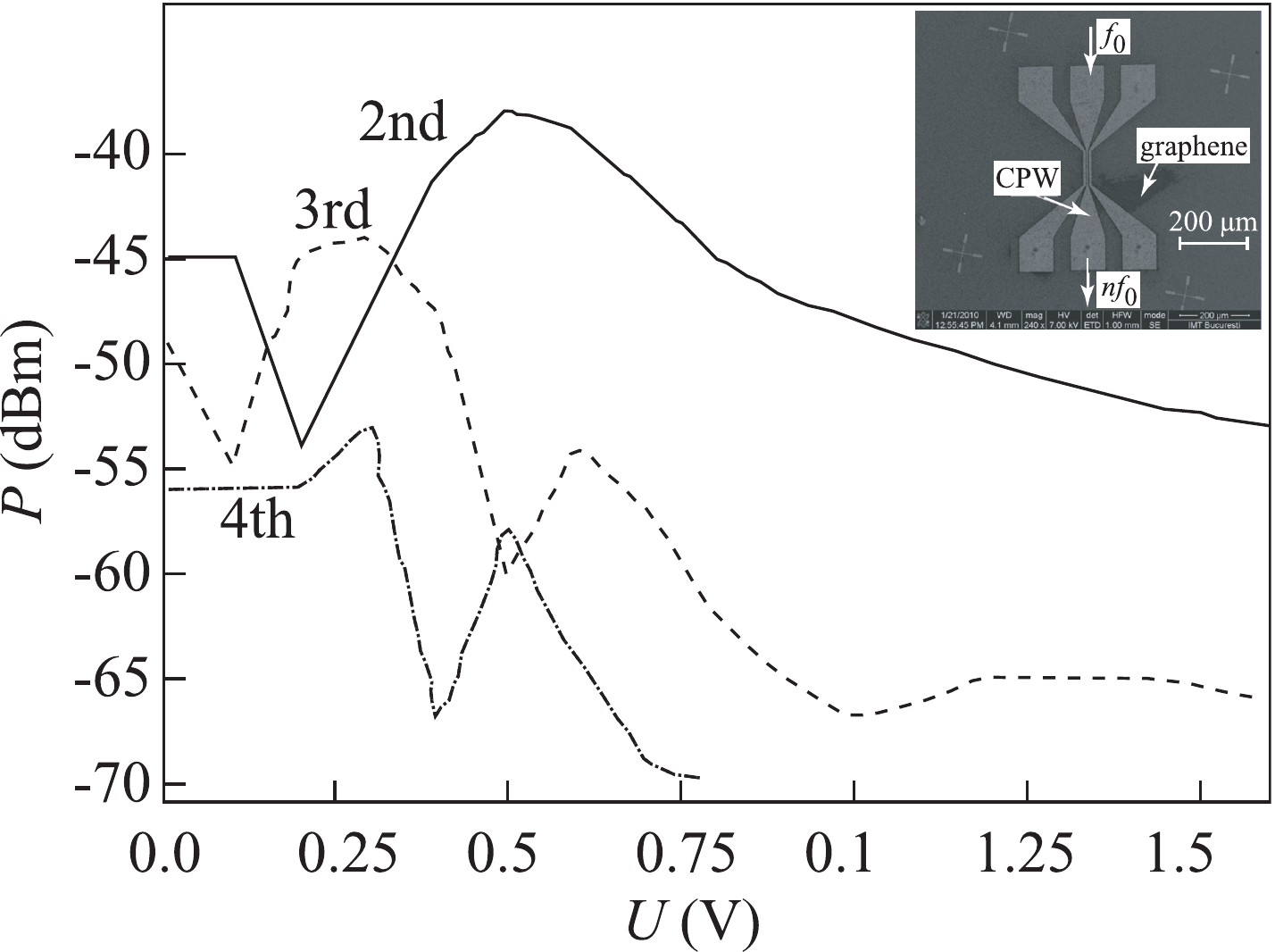}
\caption{ Output power $P$ as a function of 
\emph{dc} bias voltage $U$ for the second-order (solid line), third-order (dashed line), and fourth order
(dash-doted line) harmonics excited by radiation with frequency of $f_0 = 1$~GHz. 
Inset shows metallic coplanar line waveguide  
graphene multiplier device. After~\protect \cite{10.1063/1.3483872}. 
} \label{dragovan1}
\end{figure}

The discussed above experiments on the second harmonic generation apply infrared radiation with $\hbar\omega \gg E_F$ which corresponds to quantum mechanical regime. Moreover, the second and higher harmonic generation have been observed also for the gigahertz 
frequency range (wavelengths of the order of several millimiters) where the classical frequency range was realized~\cite{10.1063/1.3483872}.
For measurements of the microwave frequency
multiplication a  specific high-frequency structure, i.e., a metallic coplanar line
waveguide device,  was patterned directly on graphene, see inset in Fig.~\ref{dragovan1}. Importantly,
 the current-voltage characteristic of used device is linear, ruling out possible mechanisms 
 based on the nonlinear coupling between quasi-static field and current response. 
 Thus, the physical mechanism of frequency multiplication is related with strongly
  nonlinear electromagnetic response of Dirac fermions in
graphene. Figure \ref{dragovan1} shows the dependence of the signal at double 
frequency on the bias voltage, as well as the powers of the third and fourth harmonics measured on the same device. The second harmonic generation or second order 
nonlinear effect appears at zero bias and varies from $-60$ (at excitation with 10 GHz)
 to $-45$ dBm (for 1 GHz) as compared to the power of excitation frequency. In contrast
  to the measurements under infrared excitation~\cite{murzina}, here the static voltage does not lead to 
  an enhancement of the signal. Rather strong nonlinear response suggests that such
   graphene-based systems can be efficiently implemented as frequency multipliers in 
   GHz and may be even in THz ranges. Note that the latter is not yet realized.

{\subsection{Dynamic Hall  (photon drag) effect}\label{CaCHE1}}

\subsubsection{Microscopic theory}\label{theory:acHall}

The microscopic theory for the photon drag effect in graphene was developed 
in Refs.~\cite{karch2010,2010arXiv1002.1047K} for classical frequency range and 
in Ref.~\cite{PRB2010} for the quantum frequency range. Here we start with the 
presentation of the results of microscopic calculations based on the Boltzmann 
equation approach and considering the classical picture of the effects visualized 
in the model outlined in Sec.~\ref{sec:micro}. Amplitude and the sign of the resulting 
net \emph{dc} current $\bm j$ are given by the constants $T_1$, $T_2$, and $\tilde T_1$ 
in Eqs.~\eqref{drag}. Calculations carried out in Refs.~\cite{karch2010,2010arXiv1002.1047K} 
and taking into account both $E B$  and $qE^2$ contributions yield 
\begin{subequations}
\label{classical}
\begin{multline}
 T_1 = -\frac{2 e^3v^4}{\omega} \sum_{\bm k} \frac{\tau_1  f_0'}{1+\omega^2\tau_1^2} \times\\
  \left[2\left( \frac{d \tau_1}{d
 \varepsilon_k} + \frac{\tau_1}{\varepsilon_k} \right) -\frac{1-\omega^2\tau_1\tau_2}{1
 +\omega^2\tau_2^2}\left(\frac{d \tau_1}{d\varepsilon_k} - \frac{\tau_1}{\varepsilon_k}  \right)\right],
 \end{multline}
 \begin{equation}
 T_2 = -\frac{2 e^3v^4}{\omega} \sum_{\bm k} \frac{\tau_1  f_0'}{1+\omega^2\tau_1^2}
 \left(\frac{d \tau_1}{d\varepsilon_k} - \frac{\tau_1}{\varepsilon_k} \right),
\end{equation}
\begin{equation}
\label{classical:cde}
\tilde{T}_1 = e^3 v^4\sum_{\bm k}
\frac{\tau_1^2(1+\tau_2/\tau_1)f_0'}{[1+(\omega \tau_1)^2][1 + (\omega \tau_2)^2]}
\left( \frac{d \tau_1}{d \varepsilon_k} - \frac{\tau_1}{\varepsilon_k} \right)\:.
\end{equation}
\end{subequations}

Equations~\eqref{classical} show that the radiation frequency is an important issue for the current generation. The frequency dependence of linear and circular currents given by $T_1$ and $\tilde T_1$, respectively, is shown in the inset in Fig.~\ref{fig21} together with experimental data discussed in detail later, in Sec.~\ref{sexperiment_pd}. In line with qualitative model shown in Fig.~\ref{fig2} in the limit of $\omega\to 0$ the linear photocurrent is constant while circular one is zero. With the frequency increasing, i.e., for $\omega\tau \gg 1$ but $\hbar\omega \ll E_F$, the linear photocurrent decreases as
\begin{equation}
\label{hfBC}
 j_B, j_C \propto \frac{1}{\omega^2}.
\end{equation}
Moreover, due to an interplay of $EB$ and $qE^2$ contributions the linear photocurrent in the incidence plane not only decreases but may
change its sign as a function of the radiation frequency depending
on the dominant scattering mechanism~\cite{2010arXiv1002.1047K}.
By contrast, the circular photocurrent exhibits nonmonotonic frequency dependence: It rises with increasing frequency, reaches the maximum magnitude at $\omega\tau \sim 1$ and then drops down as ($\hbar/\tau \ll \hbar\omega \ll E_F$)
\begin{equation}
\label{hfA}
 j_A \propto \frac{1}{\omega^3\tau}.
\end{equation}
Although the drag effects are suppressed with an increase of frequency, they may still result in the observable signals, see below.
Such frequency dependence is in agreement with the phenomenological considerations. Indeed, the time reversal symmetry imposes restrictions on the constants $T_1$, $T_2$, and $\tilde T_1$ in Eqs.~\eqref{drag} and, hence, on the parameters $A$, $B$, and $C$ in Eqs.~\eqref{jABC} governing their frequency dependence. To illustrate these limitations, we consider the regime of low frequencies, $\hbar\omega \ll E_F$, where only \emph{intra}band transitions are possible. We note that the following quantities: $\bm j$, $\bm q$, $\omega\tau$, and $P_{\rm circ}$ are odd at the time reversal, while radiation intensity, $I$, is even at the time  reversal. Phenomenological Eqs.~\eqref{j:phen:gen} are invariant at time reversal. It follows from Eq.~\eqref{drag} that linear photocurrent is given by 
\[
j_\alpha \propto q_\beta \mathcal F_{d,l}(\omega\tau) I,
\]
where $\mathcal F_{d,l}(\omega\tau)$ is a function, forced to be even at time reversal. Hence, $\mathcal F_{d,l}(\omega\tau)$ contains only even powers of $\omega\tau$. By contrast, the circular photon drag effect given by Eq.~\eqref{j:y}
\[
j_\alpha \propto q_\beta P_{\rm circ} \mathcal F_{d,c}(\omega\tau) I,
\]
is described by the function $\mathcal F_{d,c}(\omega\tau)$ odd at time reversal, hence, containing only odd powers of $\omega\tau$. Similar relations are satisfied for the photogalvanic effect given by Eqs.~\eqref{pge}, in the latter case, however, since $\bm q$ does not enter the phenomenological expressions, function describing circular photon drag effect is even at time reversal, and vice versa.

\begin{figure}[hptb]
\includegraphics[width=\linewidth]{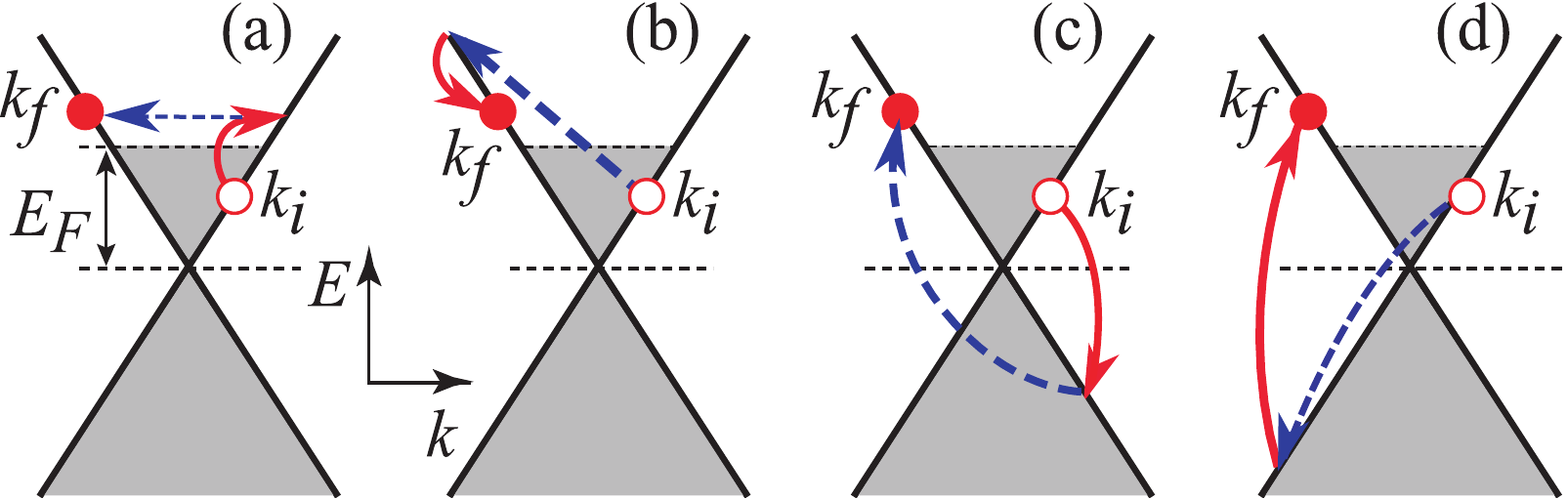}
\caption{Schematic illustration of the processes responsible for the
photon drag effect in the quantum frequency range under 
\textit{intra}band transitions ($\omega\tau \gg 1$, $\hbar\omega \leqslant E_F$). 
Panels (a)-(d) show phonon/impurity assisted 
indirect optical transitions via different intermediate virtual states. 
Arrows show electron-pho\textit{t}on interaction (solid arrows)
and 
electron scattering caused by phonons or impurities (dashed arrows).
Initial and final states of a photoexcited 
carrier with wavevectors $k_i$ and $k_f$ are shown by open and 
solid circles, respectively. }
\label{fig:inter}
\end{figure}

With further increase of the radiation frequency or decrease of the Fermi energy we turn to the quantum frequency range. We present the results for the case of  $\omega\tau\gg 1$ and $\hbar\omega \sim E_F$ studied in Ref.~\cite{PRB2010}. The absorption of the electromagnetic wave in the case of \textit{intra}band
transitions should be accompanied with the electron scattering, otherwise energy and
momentum conservation laws can not be satisfied. The corresponding processes are schematically depicted in Fig.~\ref{fig:inter}. 
As a result, one can express the coefficients $T_1$ and $T_2$ describing
 linear photocurrent in the following form ($\omega\tau \gg 1$)~\cite{PRB2010}
\begin{subequations}
\label{t:qnt}
\begin{equation}
 T_1 = -e^3v^4 \frac{32}{\hbar\omega^4} \sum_{\bm k_i} [f(\varepsilon_{k_i})
 - f(\varepsilon_{k_f})]\frac{\varepsilon_p }{(\varepsilon_{k_i} + \varepsilon_{k_f})^2}, \label{t1:qnt}
 \end{equation}
 \begin{equation}
T_2 = -e^3v^4 \frac{8}{\hbar\omega^4} \sum_{\bm k_i} [f(\varepsilon_{k_i})
- f(\varepsilon_{k_f})] \frac{\varepsilon_p^2 +\varepsilon_{k_i}^2 + (\hbar\omega)^2}{\varepsilon_{k_i}(\varepsilon_{k_i} + \varepsilon_{k_f})^2}.\label{t2:qnt}
 \end{equation}
\end{subequations}
Here $\varepsilon_{k_f} = \varepsilon_{k_i} + \hbar\omega$. It is noteworthy that
Eqs.~\eqref{t:qnt}  {are valid provided $\hbar \omega <E_F$}. We note that although the scattering rates are not
explicitly present in Eqs.~\eqref{t:qnt}, the scattering processes are crucial for the photocurrent formation. 

Note, that if the photon energy becomes much smaller as compared with the electron energies,
$\hbar\omega \ll \varepsilon_{k_i}, \varepsilon_{k_f}$, but $\omega\tau_1,\omega\tau_2 \gg 1$, the classical and quantum approaches merge. One can check that, in agreement with Eqs.~\eqref{hfBC}, Eqs.~\eqref{t:qnt} yield
\begin{equation}
\label{high-frq}
T_1 = 2T_2 = \frac{8 e^3v^4}{\omega^3} \sum_{\bm k} \frac{ f_0'}{\varepsilon_{k}}.
\end{equation}
{In this frequency range values of $T_1$ and $T_2$ are identical to those presented in~Eqs.~\eqref{classical}.}

\subsubsection{Resonant drag effect under \textit{inter}band transitions}

\begin{figure}[tb]
\includegraphics[width=0.6\linewidth]{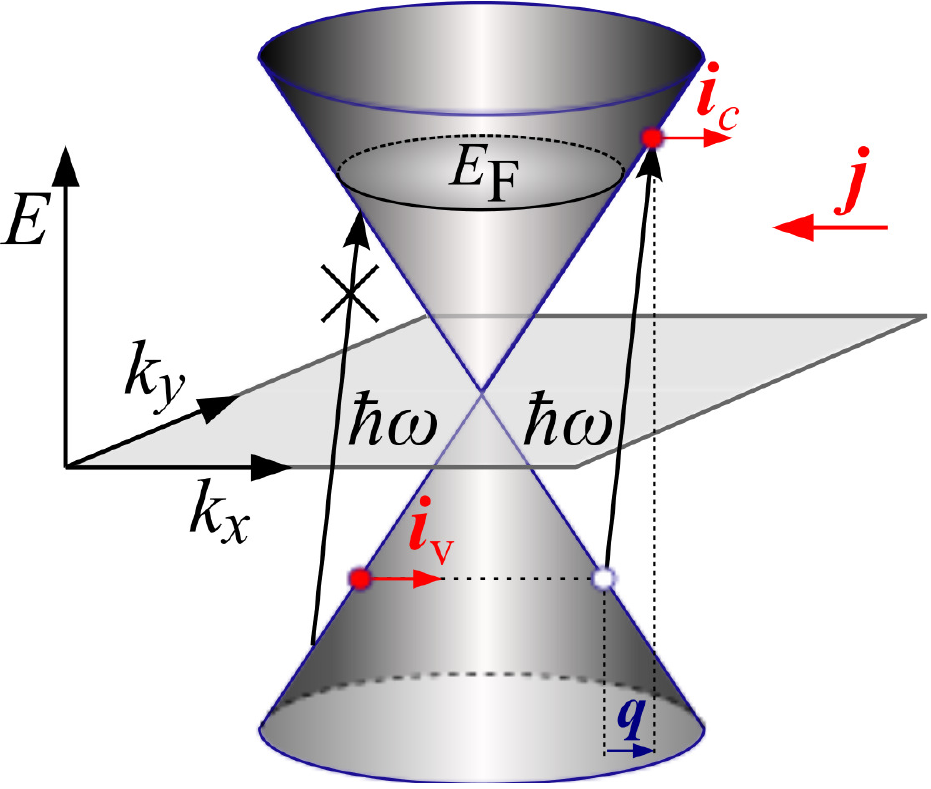}
\caption{Schematic illustration of the resonant
photon drag effect under \textit{inter}band optical transitions. 
The tilted arrows show optical \textit{inter}band transitions  inclined in the $\bm k$ space due 
to the transfer of photon momentum $\bm q$ to electrons. 
At resonance the final state of the transitions with negative $k_x$ lies below the Fermi energy and, therefore,
such transitions are forbidden. By contrast the transitions with positive $k_x$ are allowed. The optically induced imbalance of carriers in the $\bm k$-space results in fluxes of electrons  (filled circuits) in the conduction, $\bm i_c$, and valence, $\bm i_v$ bands.
Both fluxes contribute constructively to the electric current, $\bm j = e(\bm i_c + \bm i_v)$. After~\cite{entin10}. }
\label{fig:entin}
\end{figure}

A further increase of the radiation frequency or decrease of the Fermi energy opens another absorption channel, namely, if $\hbar\omega \geqslant 2E_F$ the direct \textit{inter}band transitions dominate the absorption of radiation. It gives rise to the novel regimes of
the photon drag effect as it is considered theoretically in
Ref.~\cite{entin10}. Schematics of the photocurrent generation is
illustrated in Fig.~\ref{fig:entin}. The microscopic origin of the
photocurrent generation in this frequency range is related with the
fact that the electron in the process of transition from the valence
band to conduction band shifts {in the $\bm k$ space} by $\bm q$, the photon wavevector. 

In the narrow frequency range
\begin{equation}
\label{rdrange}
|\hbar \omega - 2E_F | \leqslant \hbar v q,
\end{equation}
as it is seen from Fig.~\ref{fig:entin}, only transitions at
positive momenta are possible due to the final state filling effect. 
It results in the strong asymmetry of photoelectrons distribution, 
which gives rise to the resonant photocurrent. The interband absorption gives rise to the generation of electron-hole pairs. As a result, a photocurrent is contributed both by the photoelectron and photoholes. 
The hole contribution can be viewed as that of a valence band electron with an opposite wavevector. 
These fluxes of conduction and valence band electrons are shown by arrows in Fig.~\ref{fig:entin}. 
Since the velocity of quasiparticle is given by $\hbar^{-1}\mathrm d\varepsilon/\mathrm d{\bm k}$, the velocities for opposite wavevectors in the conduction and valence band are the same. Consequently, the fluxes in the conduction and valence bands are the same. Taking into account 
that the electron
generation rate is $ \pi \alpha I/(\hbar \omega)$, where $\pi
\alpha$ is the
monolayer graphene absorbance ($\alpha$ is the fine structure constant)~\cite{Nair06062008,Falkovsky:2008eng},
and all generated electrons contribute with velocity $v$ to the
drag current one has
\begin{equation}
\label{rde}
j \sim e v\tau \pi \alpha \frac{I}{\hbar\omega}.
\end{equation}
This effect, known as resonant drag effect, was suggested in
Ref.~\cite{entin10}. Although the magnitude of the current is
independent of the photon wavevector $\bm q$, the resonant effect
takes place in the narrow frequency range, Eq.~(\ref{rdrange}), the
smaller the smaller $q$. If the photon frequency is high enough,
 $\hbar \omega - 2E_F  > \hbar vq$, the resonant contribution 
 is absent and the ordinary (nonresonant) drag current is formed, 
 similarly to the case of semiconductor quantum well structures~\cite{PhysRevB.38.87}.

\subsubsection{Experiment}
\label{sexperiment_pd}

\begin{figure}[t]
\includegraphics[width=0.75\linewidth]{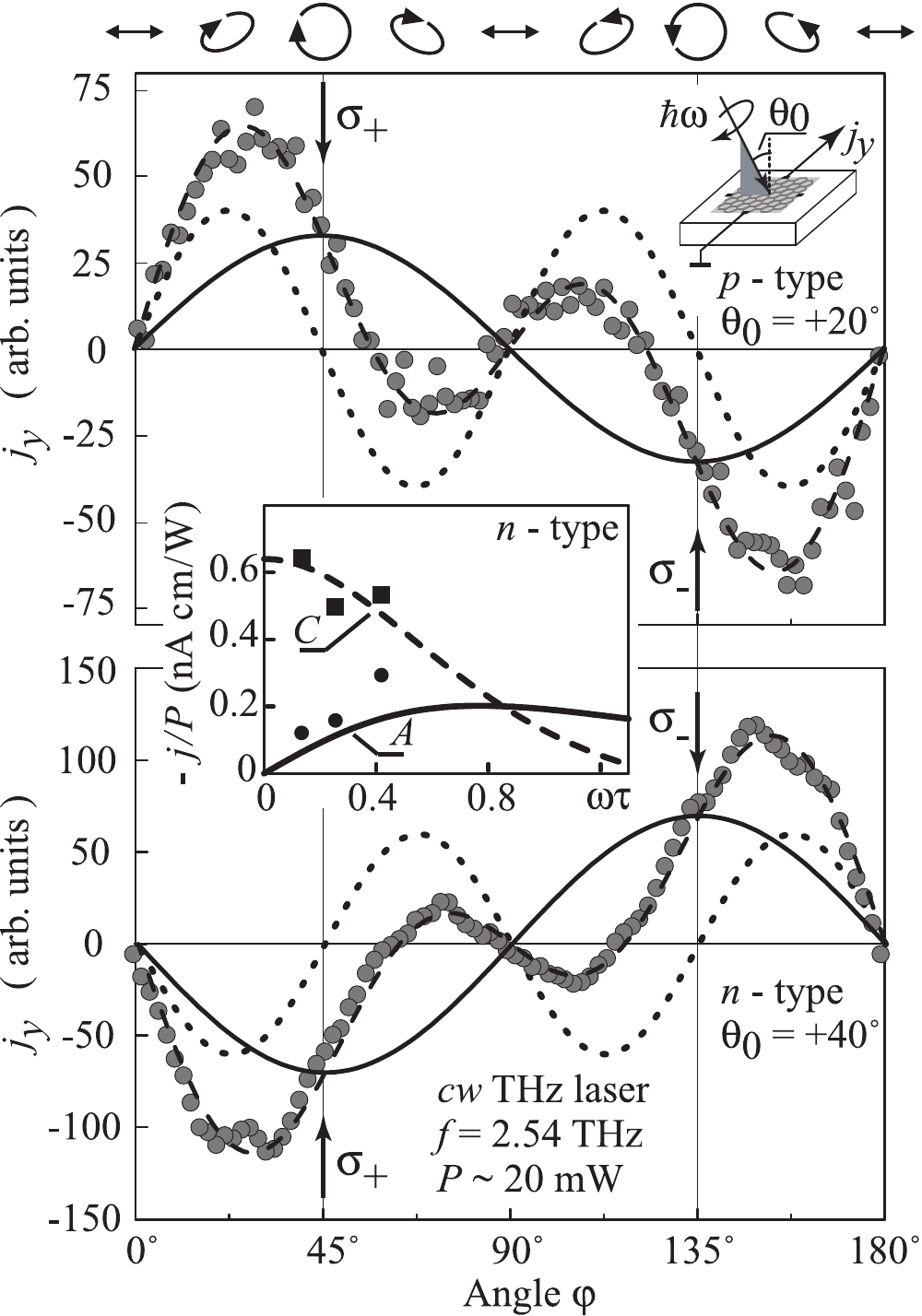}
\caption{Helicity dependence of the photocurrent, $j_{y}$, measured 
in the direction normal to the plane of incidence. 
The ellipses on top illustrate the polarization states 
for various angles $\varphi$. Dashed lines show 
fits to the calculated total current $j_A + j_B$ comprising the 
circular contribution $j_A$ (full line) and the linear contribution $j_B$ (dotted line), see Eq.~\eqref{jAB}.
Top and bottom panels correspond to $p$- and $n$- type samples, respectively 
and demonstrate that the current has opposite signs for opposite carrier polarity. 
Inset shows frequency dependence of the longitudinal linear, $j_C$, and circular, $j_A$, photocurrents. 
Circles and squares are experimental data, solid and dashed curves represent 
the results of calculation. The agreement is obtained without fitting parameters. After~\protect \cite{karch2010}.
} \label{fig21}
\end{figure}

Dynamic Hall and photon drag effects have been demonstrated applying THz and infrared 
laser radiation to unbiased graphene layers produced both by exfoliation 
and epitaxial techniques~\cite{karch2010,2010arXiv1002.1047K}.\footnote{While both types of samples showed the effect,
the micrometer sized exfoliated samples displayed an additional
edge contribution (discussed below in Sec.~\ref{edge}) as the
spot size of the terahertz laser of 1 mm$^2$ was larger
than the graphene flakes.}  In all experiments known so far, the limit $\hbar\omega < E_F$ was fulfilled.
To prevent high losses or electrical shunting by  conducting substrates high-resistivity Si or  
semi-insulating SiC substrates have been used. For some samples  nonconductive polymer films
were used for protection of graphene samples from the undesired 
doping in the ambient atmosphere~\cite{DrexlerC.:2013uq,suppllara2011}.
To measure photocurrents ohmic contacts were made at samples edges.  Details on the material growth
 and characterization
can be found in~\cite{karch2010,suppllara2011,Tzalenchuk2010,Emtsev2009}. 
For optical excitation  \textit{cw} and pulsed molecular optically  
pumped terahertz lasers or tunable CO$_2$ lasers were applied. 
In the measurements the spatial beam distribution 
has an almost Gaussian profile, independently measured by a pyroelectric 
camera~\cite{ch1Ziemann2000p3843}, and the laser spot is always centered between the contacts.
This arrangement prevents the temperature gradient between contacts necessary for the 
thermoelectric effect like that discussed for graphene in e.g. \cite{doi:10.1021/nl202318u}.
A pronounced signal is detected in a wide range of radiation
frequencies, from 0.6 THz ($\lambda \approx 500$~$\mu$m) up to
about 30 THz ($\lambda = 10~\mu$m), and intensities, from mW/cm$^2$ up to MW/cm$^2$. 
In agreement with theory presented above, the Hall photocurrent appears
 under oblique incidence.
Figure~\ref{fig21} shows results obtained on epitaxial single layer graphene excited by elliptically polarized light in transverse geometry. The polarization state of light was controlled by the rotation of the quarter-wave plate.
 This figure reveals that the photocurrent signal is a superposition of circular 
 and linear contributions of comparable strengths. We emphasize that the circular 
 contribution ($\bm j \propto P_{\rm circ} = \sin{2\varphi}$) manifests itself as a change of current direction for left- and 
 right- circularly polarized radiation. In accordance with the theory, Sec.~\ref{drag:phenom:sec}, the circular photocurrent is observed in the direction perpendicular to the incidence plane, while linear contribution is detected in the incidence plane
together with polarization independent current. Obviously, the latter effects can be and indeed have been observed for linearly polarized radiation. Functional behavior of the photocurrent components upon variation of the radiation polarization state, incidence angle and frequency is in a full agreement with that obtained theoretically in Sec. \ref{theory:acHall}. Moreover, the microscopic theory yields the absolute value of the photocurrent without fitting parameters with only assumption of the short-range scattering~\cite{karch2010}. It is worth to note, that in agreement with theoretical consideration the signal reverse its sign by change of carrier type from $p$ to $n$. Strikingly, due to the fact that the conduction- and valence-band are symmetric with respect to the Dirac point, the opposite polarities of the signal can be observed in the same sample just by changing the Fermi level position.

\subsection{Photogalvanic effect in the pristine graphene}\label{PGE}

\subsubsection{Microscopic theory}\label{theory:PGE}

Due to symmetry arguments addressed above, photogalvanic effect may emerge only in graphene systems where the inversion symmetry is broken. 
Moreover, from the same arguments  summarized
in Eqs.~(\ref{j:pge:x}), (\ref{j:pge:y}) it follows
that the photocurrent  in flat infinite graphene can be generated only with allowance for $z$-component of the incident
electric field. The latter condition hampers the photogalvanic effect formation. Indeed, for strictly two-dimensional model where only $\pi$-orbitals of
carbon atoms are taken into account, no response at $E_z$ is possible. However, taking into account other bands in electron energy
spectrum formed from the $\sigma$-orbitals of carbon atoms gives rise to the \emph{dc} current.

\begin{figure}[tb]
\includegraphics[width=0.75\linewidth]{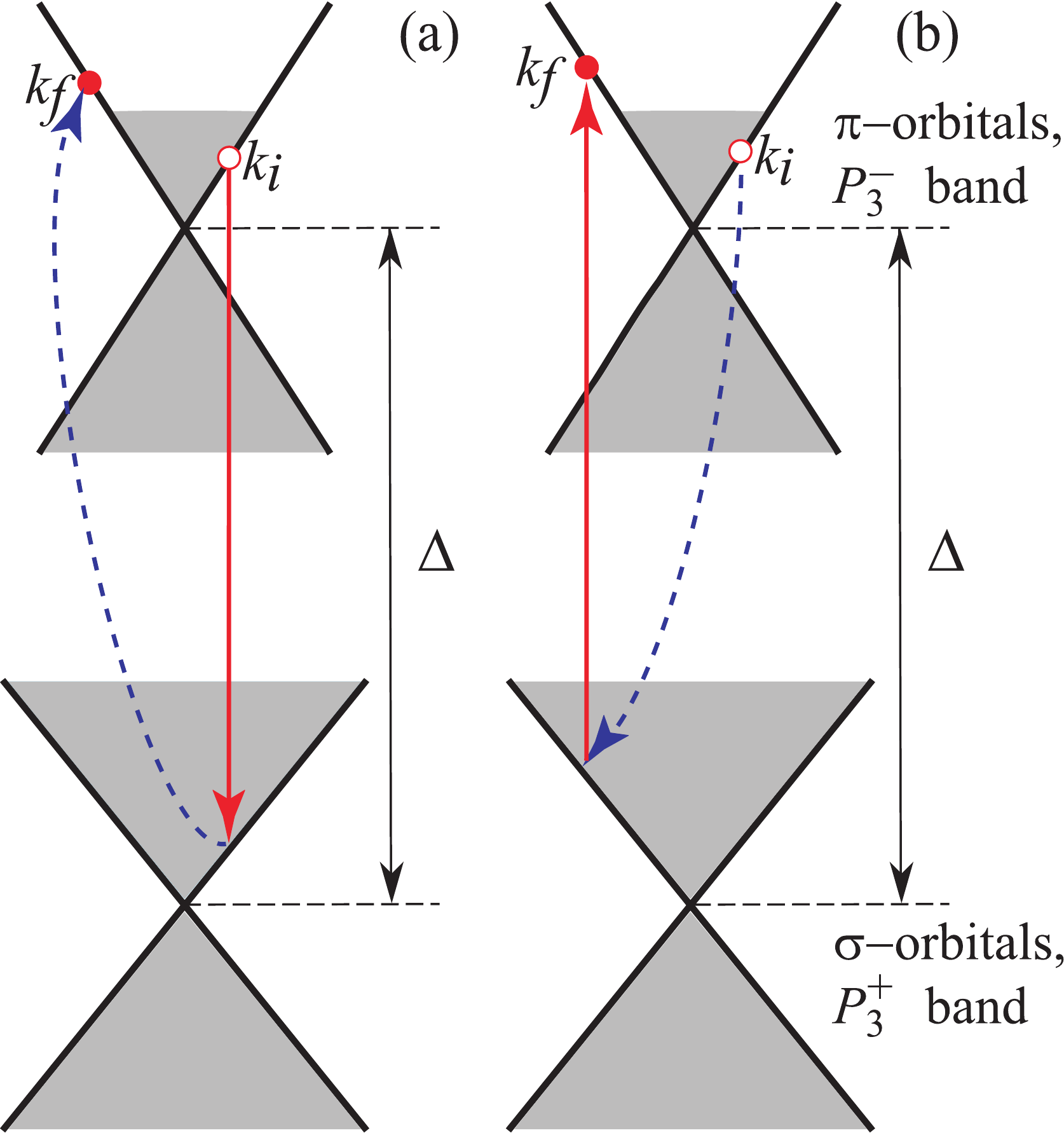}
\caption{Schematic illustration of indirect \textit{intra}band Drude transitions with intermediate states in $P_3^+$ band. These transitions together with those shown in Fig.~\ref{fig:inter} are important for photogalvanic effect, see text for details. After~\cite{PRB2010}.}
\label{fig:inter1}
\end{figure}

Microscopically, the  photogalvanic effect arises due to the
quantum interference of the Drude-like indirect optical transitions represented in
Fig.~\ref{fig:inter} (for $\bm q=0$) and the indirect \textit{intra}band
transitions with intermediate states in {distant} bands depicted in Fig.~\ref{fig:inter1},
similarly to the orbital mechanisms of the photogalvanic effects in
conventional semiconductor
nanostructures~\cite{Tarasenko2007,PhysRevB.79.121302,tarasenko11}. To illustrate the generation of the photocurrent 
we consider the circular photogalvanic effect. The current results from the anisotropic distribution of photoexcited electrons, which is caused the different 
dependence of the transition matrix elements on the wavevectors: The matrix element of the Drude-like 
transitions within one band (Fig.~\ref{fig:inter}) is linear in the wavevectors  
\[
M^{(1)}_{\bm k_f, \bm k_i} \propto \bm A\cdot (\bm k_i - \bm k_f),
\]
whereas the matrix element of the indirect optical transitions involving distant bands $M^{(2)}_{\bm k_f, \bm k_i}$ (Fig.~\ref{fig:inter1}) is almost $\bm k_i$ and $\bm k_f$ independent. The total transition rate calculated with allowance for the quantum mechanical interference is given by  
\begin{equation}
\label{rate}
W_{\bm k_f, \bm k_i} \propto |M^{(1)}_{\bm k_f, \bm k_i}+M^{(2)}_{\bm k_f, \bm k_i}|^2.
\end{equation}
As a result we obtain from Eq.~\eqref{rate} that besides $\bm k$-even contributions ($\propto |M^{(1)}_{\bm k_f, \bm k_i}|^2$, $|M^{(2)}_{\bm k_f, \bm k_i}|^2$), the transition probability contains $\bm k$-odd interference term:
\begin{equation}
\label{Minterference}
\propto 2\Re{[M^{(1)}_{\bm k_f, \bm k_i}M^{(2)^*}_{\bm k_f, \bm k_i} ]} \propto \bm A\cdot (\bm k_i - \bm k_f).
\end{equation}
It follows from Eq.~\eqref{Minterference} that the interference contribution is linear in the initial and final wavevectors, $\bm k_i$ and $\bm k_f$, hence, the distribution function of the photoexcited carriers is anisotropic in the $\bm k$-space. An imbalance of electron population in different regions of the $\bm k$-space results in the \emph{dc} current. We stress that the matrix element $M^{(1)}$ contains the in-plane components of the radiation vector potential, while the element $M^{(2)}$ is related with its $z$-component. Hence, the generated current is proportional to $\bm E_\parallel E_z^* \pm {\rm c.c.}$ in accordance with phenomenological analysis, see Eq.~\eqref{pge}.

As it follows from the above consideration, the optical transitions via distant bands, although providing a tiny fraction in the total absorption of graphene, are crucial for the current formation. Therefore, we consider them in more detail. Here, the distant bands, involved in the interference, are described by $P_3^+$ representation (even under the $z\to -z$ reflection), while the conduction and valence band states in graphene
transform according to the $P_3^-$ representation (odd under the $z\to -z$ reflection)~\cite{Bassani}. Microscopic
calculations performed within the basis of $2s$ and $2p$ atomic
orbitals~\cite{Bassani,PhysRevB.17.626} show that the distance from
the $P_3^-$ states forming conduction and valence bands and closest
deep valence bands $P_3^+$, $\Delta$, is about $10$~eV.  It is
noteworthy, that the electron dispersion in these bands has the form,
similar to that of conduction and valence bands: i.e. energy spectrum
near $K$ (or $K'$) point is linear, however, with different velocity,
as it is schematically illustrated in Fig.~\ref{fig:inter1}. Since matrix elements $M^{(1)}$ and $M^{(2)}$ have different parity under $z\to -z$ reflection, the  quantum interference is only possible in the case, where the graphene is deposited on the substrate/adatoms are present on one side of the sample, i.e. where the $z\to -z$ reflection symmetry is broken. 

In the further description we limit our consideration to the circular electric current only, $\bm j \propto \chi_c P_{\rm circ}$,  see Eqs.~\eqref{pge}. The calculations carried out in framework of the Fermi golden rule for $\omega\tau \gg 1$ and
$\hbar\omega< E_F$ yields~\cite{PRB2010}:
\begin{multline}
\label{jy}
\chi_c = -ev \frac{4\pi w}{\hbar}  \sum_{\bm k_i \bm k_f}
\frac{\tau_1(\varepsilon_{k_f}) \varepsilon_{ k_i}
  +\tau_1(\varepsilon_{ k_i})\varepsilon_{k_f}}{\varepsilon_{ k_i}
  + \varepsilon_{ k_f}} \times \\ [ f(\varepsilon_{k_i})
  - f(\varepsilon_{k_f})] \delta(\varepsilon_{k_f} - \varepsilon_{ k_i} - \hbar\omega),
\end{multline}
where
\[
w=\frac{2 {\pi} e^2 v p_0}{m_0 c\omega^2} \frac{\langle V_0V_1\rangle}{\Delta^2},
\]
$V_0$ and $V_1$ determine the electron-impurity or electron-phonon
scattering matrix elements within $\pi$-band and between $\sigma$- and
$\pi$-bands, respectively, $p_0$ is the \textit{inter}band optical matrix element, and $\langle V_0V_1 \rangle$ denotes the averaging of the product $V_0V_1$ over the disorder realizations.
 The treatment of the general case is given in Ref.~\cite{PRB2010}.

The direction of the current is determined by the
sign of the product $\langle V_0V_1\rangle$ and the radiation
helicity. The averaged
product $\langle V_0V_1\rangle$ has different signs for the same
impurities, but
positioned on top or bottom of graphene sheet. It is clearly seen that
the photogalvanic current vanishes in symmetric graphene-based
structures where $\langle V_0V_1\rangle=0$.

In the case of the degenerate electron gas with the Fermi energy $E_F$ and in the
limit of $\hbar \omega \ll E_F$ Eq.~\eqref{jy} can be recast as~\cite{PRB2010}
\begin{equation}
\label{jy1}
\chi_c = -8\frac{\alpha e d_0 }{\Delta}  \frac{\langle V_0V_1\rangle}{\langle V_0^2\rangle} \frac{E_F}{\hbar \omega},
\end{equation}
where we introduced effective dipole of \textit{inter}band transition
\[
e d_0 = \frac{e p_0 \hbar}{m_0 \Delta}.
\]
Equation~\eqref{jy1} allows us to evaluate the frequency dependence of the circular photogalvanic effect. Namely, at $\omega\tau\gg 1$, $\hbar\omega\ll E_F$, the circular photocurrent behaves as $1/\omega$,
 i.e. it is parametrically larger than the
circular drag (or circular \emph{ac} Hall) effect, which behaves as $1/\omega^3$, see
Eq.~\eqref{hfA}. This important property is related with the
time reversal symmetry: the coefficient $\chi_{c}$ describing
photogalvanic effect is even at time reversal, while $\tilde T_1$
describing circular drag effect is odd. Therefore, circular photocurrent formation
due to photogalvanic effect is possible at the moment of carriers
photogeneration. Since at $\omega\tau \gg 1$ for \emph{intra}band transitions the absorption rate is proportional to the electron scattering rate, $\tau^{-1}$, and current density is proportional to the electron scattering time, $\tau$, the circular photocurrent is independent of the scattering rate. Owing to different symmetry under time reversal the linear photogalvanic effect, by contrast, requires extra scattering, its description within the same model is presented for the classical frequency range in Ref.~\cite{PRB2010}. As a result for $\omega\tau \gg 1$ (but $\hbar\omega < E_F$) the following hierarchy of the current magnitudes is possible: (i) circular photogalvanic effect $\propto \omega^{-1}$, (ii) linear photogalvanic $\propto \omega^{-2}\tau^{-1}$ and photon drag effects $\propto \omega^{-2}$, (iii) circular photon drag effect $\propto \omega^{-3}\tau^{-1}$.

\subsubsection{Experiment}
\label{sexperiment_pg}

\begin{figure}[t]
\includegraphics[width=0.85\linewidth]{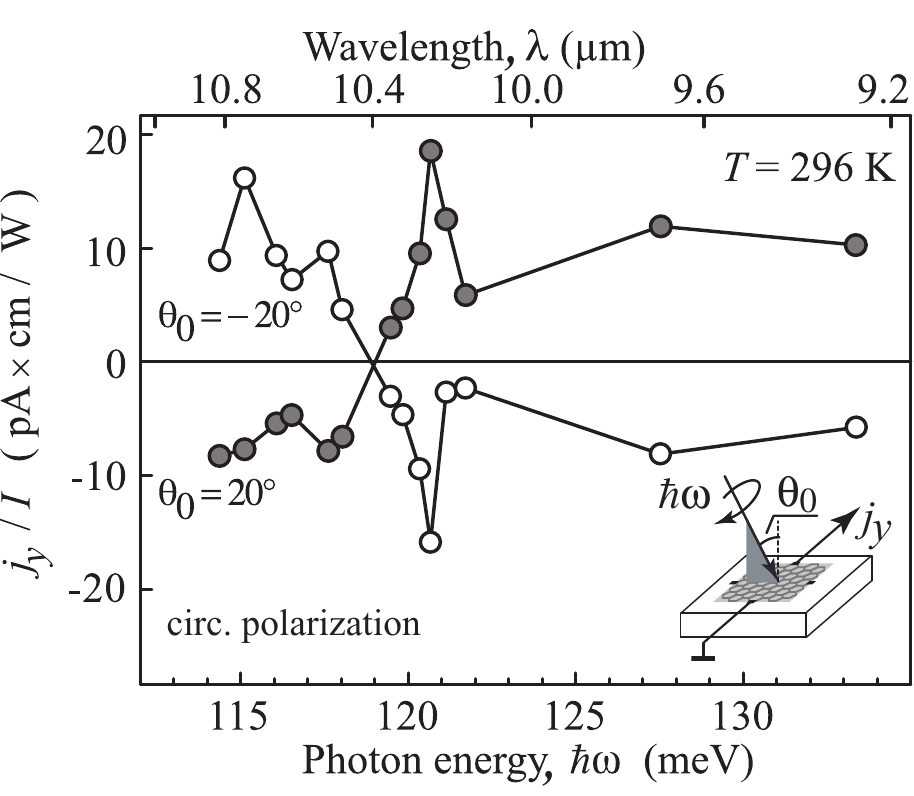}
\caption{Spectral dependence of the circular photocurrent $j_{y,A}$ measured in the direction normal to the plane of incidence. The data for epitaxial single layer graphene sample are obtained for circularly polarized infrared light ($\varphi = 45^\circ$) 
and two angles of incidence $\theta_0 = \pm 20^\circ$. The inset shows the experimental geometry. Data are given after~\protect \cite{PRB2010}.}
\label{fig2bis}
\end{figure}

As we emphasized above, see Sec.~\ref{PGE:symmetry}, the photogalvanic effects in pristine graphene involve $z$-component of electric field and may be observed only under special conditions, where the photon drag contribution is suppressed, in particular, in the quantum frequency range. Correspondingly, both the circular 
  and linear photogalvanic effects were observed in the mid-infrared range 
  of radiation frequencies (about 30~THz) on epitaxial graphene samples. The demonstration 
  of photogalvanic effects becomes possible due to two facts: On one hand, at 
  such a high frequencies the {photon drag} effect is suppressed, and on the other hand, 
  photogalvanic and {drag} effects appear to contribute to photocurrent with opposite signs.
   This interplay resulted in a change of sign of the photocurrent upon the variation of 
   radiation frequency, see Fig.~\ref{fig2bis}, providing an evidence for the existence 
   and substantial contribution of the photogalvanic effect \cite{PRB2010}. The value of 
   the circular photocurrent caused by the photogalvanic effect is close to the theoretical 
   estimate after Eq.~\eqref{jy1} for sufficiently strong asymmetry degree, $\langle
V_0V_1\rangle/{\langle V_0^2\rangle} \approx 0.5$.
We emphasize that the photogalvanic effect does exist only due to the structure 
inversion asymmetry. Therefore, no photogalvanic effect is expected in graphene
 with equivalent ``up'' and ``down'' surfaces, e.g. in free standing graphene. 
 It would be observable in such layers only for nonequal numbers of adatoms on
  the opposite sides of the graphene sheet. An experimental evidence for a large 
  structure inversion asymmetry due to adatoms and/or substrate has been given 
  most recently by observation and study of magnetic quantum ratchet effect in
   similar epitaxial samples~\cite{DrexlerC.:2013uq}.

\subsection{Edge photocurrents}\label{edge}

\subsubsection{Microscopic theory}\label{theory:edge}

\begin{figure}[t]
\includegraphics[width=0.9\linewidth]{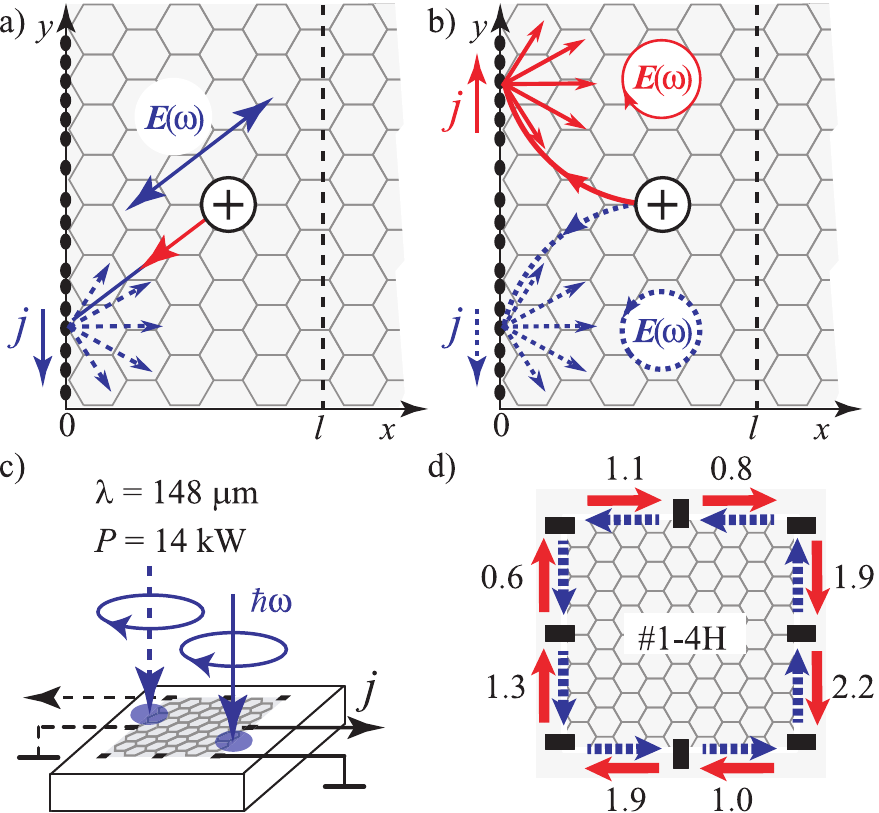}
\caption{(a) Schematic illustration of the linear edge photogalvanic effect. The oscillations of electric field $\bm E(\omega)$ are shown by double-sided arrow. The motion of a charge carrier (we consider holes for simplicity) towards the edge is shown by the solid red arrow, while the motion from the edge (after the diffusive scattering) is shown by multiple dashed arrows. The current is formed in the narrow stripe with the width of the mean free path $\ell$ near the edge.
(b) Schematic illustration of the helicity driven edge current generation.
The electric field of circularly polarized radiation rotates clockwise or counterclockwise
resulting in a circular motion of carriers, which is sketched by solid and dashed curved trajectories, respectively.  (c) Experimental geometry for the study of the edge photocurrents. (d) Edge photocurrent topology.
Solid and dashed arrows show the current direction for $\sigma^+$ and $\sigma^-$
polarizations, respectively. Numbers indicate the photocurrent amplitude
$J_{A}$ in microAmpers.}
\label{figure3}
\end{figure}

According to the symmetry analysis given in Sec.~\ref{PGE:symmetry}, the presence of sample edges breaks inversion symmetry of the system and, therefore, their illumination gives rise to the edge photocurrents, considered phenomenologically in Sec.~\ref{PGE:symmetry}. A microscopic process actuating the edge photocurrent generation is
illustrated in Fig.~\ref{figure3}(a) for the linearly polarized radiation where the semiinfinite sample occupying $x>0$ plane is shown together with the electron motion in the vicinity of the edge. The current formation involves the time dependent
motion of the charge carriers  under the action of the electric field  and the diffusive scattering at the sample edge. The electric current is formed in the narrow stripe with the width on the order of the mean free path $\ell$ in  the vicinity of the sample edge. It is contributed by the carries pushed towards the edge by the electric field in one half of a period, since for the diffusive scattering the electrons moving from the edge have random velocities along the boundary. 
We note that this mechanism is  similar to that of the surface photogalvanic effect
observed in bulk materials~\cite{PhysRevB.48.8307,Gurevich2000,magarill81,alperovich80eng}.
The above process results in the linear photogalvanic effect, given by the first term in phenomenological Eq.~\eqref{phenom:edge}, 
$j_y \propto E_xE_y^*+ E_yE_x^*$. The allowance for the trajectory winding under the action of circularly polarized radiation, shown in Fig.~\ref{figure3}(b), results in the contribution to the current sensitive to the radiation helicity reversing sign from $\sigma^+$ (solid) to $\sigma^-$ (dashed).
We note that the illumination of opposite edges of the sample results in the opposite sign of photocurrent (in a fixed frame of coordinates).

Edge photogalvanic effect may also result from the variation of the electron density in the vicinity of the edge due to the action of the field component perpendicular to the sample edge. To estimate the effect we use the continuity equation
\begin{equation}
\label{cont}
\frac{\partial \delta N}{\partial t} + \frac{\partial i_x}{\partial x} =0,
\end{equation}
which relates the variation of electron density $\delta N \equiv \delta
N(x,t) = N(x,t) - N_0$ with the electron flux density $\bm i = \bm 
j/e$, where $N_0$ is the unperturbed electron density, and the coordinate frame with axis $y$ parallel to the edge is used (see Sec.~\ref{PGE:symmetry}). The $x$ component of the flux contains  diffusive and drift contributions
\begin{equation}
\label{curr}
i_x = -D \frac{\partial \delta N}{\partial x} + \frac{\sigma(\omega)}{e} E_x,
\end{equation}
where $\sigma(\omega) = C(N_0) \tau/(1-\mathrm i \omega \tau)$ is the frequency-dependent conductivity,  $\tau$ is the momentum relaxation time, and $C(N_0) = e^2E_F/\pi \hbar^2$~\cite{Bolotin2008351,RevModPhys.83.407}. The electron gas is assumed to be degenerate,  $E_F = \hbar v
\sqrt{\pi N_0}$. The boundary conditions are as follows: at the sample edge $i_x =0$, while in the bulk of the sample the current is driven by the electric field only. As a result we have 
\begin{equation}
  \label{solution}
\delta N(x) = \delta N_0 \exp{\left(-\frac{1-\mathrm i}{l_{\rm eff}}x \right)},
\end{equation}
where $l_{\rm eff} = \sqrt{2D/\omega} = \ell/\sqrt{\omega \tau}$, 
$\ell=v\tau$ is the mean free path, $\delta N_0 =
\sigma(\omega) E_x l_{\rm eff}/[eD(\mathrm i -1)]$. This description holds for $l_{\rm eff} \gg \ell$. The electron density variation in the vicinity of the boundary is given by
\begin{equation}
  \label{integrated}
 \Delta N = \int_0^\infty \delta N \  dx 
  = 
 \frac{\sigma(\omega)E_x}{\mathrm i \omega e} \equiv  \delta N_0
 l_{\rm eff}/(1-\mathrm i).  
\end{equation}

The \emph{dc} edge photocurrent can be recast as a linear response to the $y$ component of electric field found with allowance for the $\Delta N$, the change of electron density induced by $E_x$ field component. The resulting expression for the total current $J_y = \int_0^\infty j_y(x) dx$ reads
\begin{multline}
\label{eq:edge}
J_y = 2\Re{\left\{\frac{\partial \sigma(0)}{\partial N_0} \Delta N E_y^*
  \right\}} \\
  = \frac{\tau^2}{e} \frac{\mathrm d C^2(N_0)}{\mathrm dN_0} 
\Re{\left\{ \frac{E_xE_y^*}{\mathrm i \omega(1-\mathrm i \omega \tau)}
    \right\}},  
\end{multline}
contains both linear and circular components of the photocurrent in agreement with phenomenological expression~\eqref{phenom:edge}.
The divergence of circular photocurrent present in Eq.~\eqref{eq:edge} for low frequencies, $\omega\tau \to 0$, results from the divergence of $l_{\rm eff}\propto (\omega\tau)^{-1/2}$, and may be removed taking into account the self-consistent field, finite size of the illuminated area and finite size of the contacts used to measure the photocurrent. 
We note that edge photocurrents have also been treated in the framework of Boltzmann Eq.~\eqref{kinetic:gen} in Ref.~\cite{edge}.

\subsubsection{Experiment}
\label{sexperiment_edge}

The photon drag and photogalvanic effects, described in Secs.~\ref{CaCHE1}, \ref{PGE} 
are induced in the ``bulk'' graphene 
layers applying THz/IR radiation at oblique incidence and vanish for normal incidence. 
By contrast, edge photocurrents require the illumination of sample borders 
and have a maximum at the normal incidence of radiation. 
Experiments on edge  photocurrents are challenging due to other types of photoresponses
 which may appear due to inhomogeneities, temperature gradients or illumination of contacts.
However, this  difficulty may be avoided by reduction of data analysis to the helicity 
dependent contribution, which changes its direction by switching the light 
polarization from right- to left-handed. Indeed,  all effects mentioned above
are unlikely to be sensitive to the  direction of electric field rotation. 
While photocurrents have been observed 
 in both large-area
and small-area samples~\cite{edge}, the analysis of the edge
 photocurrents is much easier in the large-area samples. Indeed, 
 in micrometer-sized exfoliated samples the radiation spot size is much 
 larger than the graphene flakes and the effects of different edges are superimposed complicating the separation of edge contributions from the data.
By contrast, in large area epitaxial samples, the illumination of a 
single edge by THz radiation could be realized enabling the accurate analyzis of the 
individual edge currents.

\begin{figure}[t]
\includegraphics[width=0.65\linewidth]{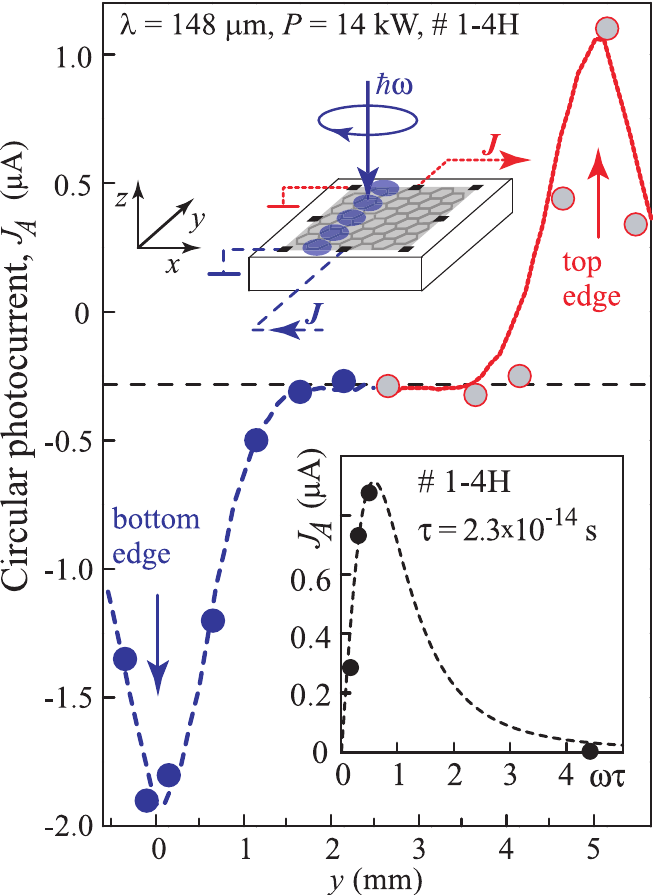}
\caption{
Circular photocurrent $J_{A}$ measured in epitaxial single layer  graphene 
sample as a function of the laser spot position.
The laser spot is scanned along $y$ and the current is
picked up from two contact pairs at the top
(open circles) or bottom (full circles) sample edges aligned along $x$  (see inset).
Lines represent the laser beam spatial distribution, which is measured by a
pyroelectric camera and scaled to the current maximum. Top inset shows the scanning geometry. 
Bottom inset shows the measured circular photocurrent $J_A(\omega \tau)$ 
at one of the edge segments of sample
(open circles) together with the fit after microscopic theory (dashed line) 
developed in the framework of the Boltzmann kinetic equation~\protect \cite{edge}. Data are given after~\protect \cite{edge}.}
\label{figedge}
\end{figure}

\begin{figure}[t]
\includegraphics[width=\linewidth]{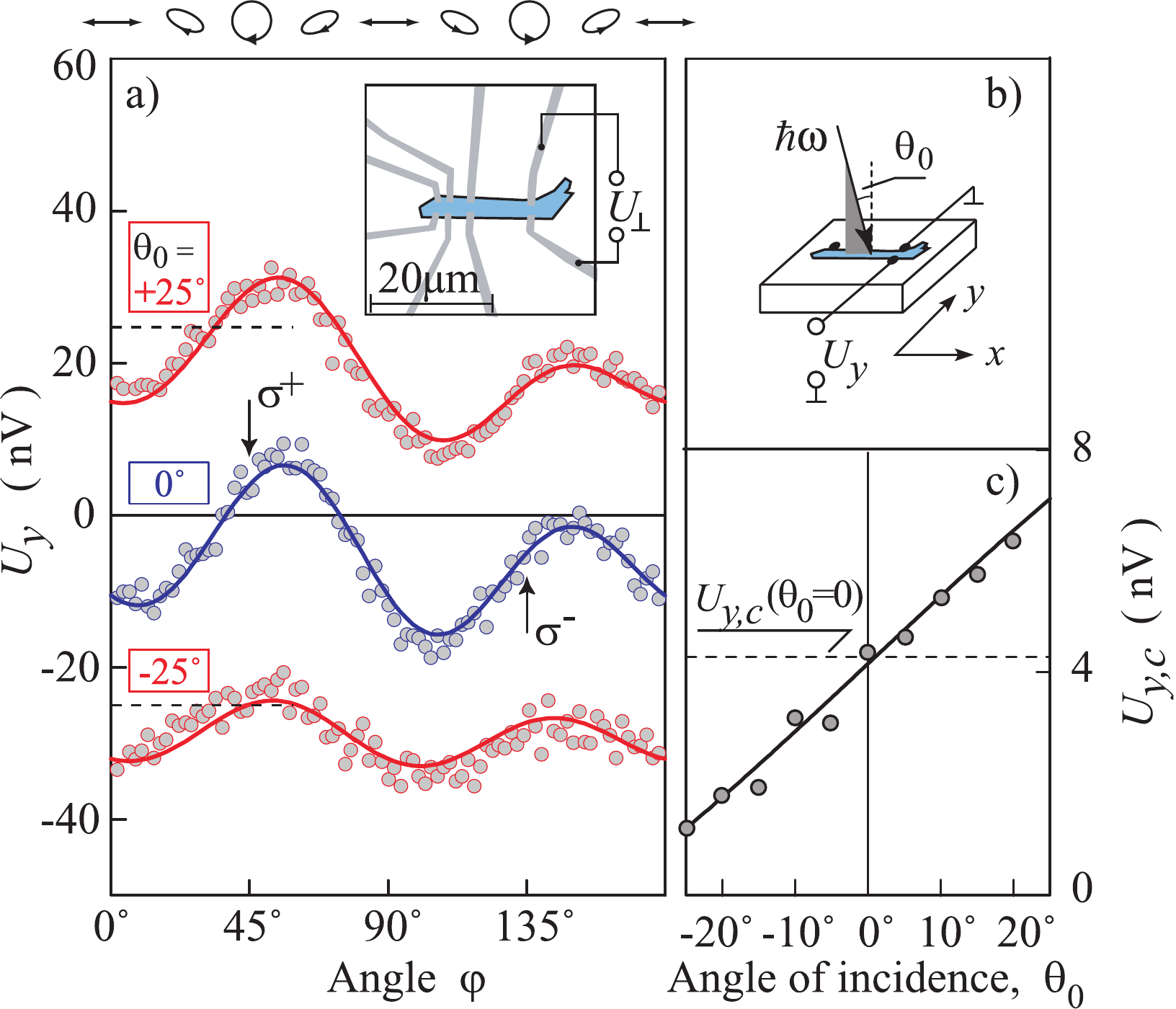}
\caption{
(a) Photosignal, $U_y \propto j_y$, in a single graphene sheet as a function of the angle $\varphi$., measured for various angles of incidence, $\theta_0$. The data are obtained applying radiation
 of the $cw$ THz laser with the photon energy $10.5$\,meV, power $\approx 20$\,mW and 
 a diameter of the laser spot about $1$\,mm. The data for $\theta_0 = \pm 25^\circ$ are shifted by
$\pm 25$~nV {for clarity}. The horizontal dashed lines show $x$-axes for the shifted data ($U_y=0$). 
Full lines are fits to {Eq.~\protect(\ref{fit:small})}. These fits can be obtained
by the superposition of the photon drag effect at oblique incidence given by Eq.~\eqref{jAB} and
the edge photogalvanic effect at normal incidence, Eq.~\eqref{phenom:edge}.
The inset shows the sample geometry. The ellipses on top of the left panel illustrate the states of polarization for various angles $\varphi$. Panel (b) illustrate the experimental configuration. 
(c) Signals due to circularly polarized radiation $U_{y,C} = [U(\sigma^+,\varphi = 45^\circ) - U(\sigma^-,\varphi = 135^\circ)]/2$ 
measured as a function of the incidence angle $\theta_0$.
Data are given after~\protect \cite{2010arXiv1002.1047K}. 
}
\label{figexfoliated}
\end{figure}

Figure~\ref{figedge} shows the circular edge photocurrent $J_A$ excited by THz radiation
for the laser spot scanned across the large-area epitaxial graphene sample along 
the $y$-axis. The signal is picked up from a pair of contacts at the sample top and bottom 
edges oriented along the $x$-axis. The current reaches its maximum for the laser spot 
centered at the edge and rapidly decays with the spot moving.
Comparison of the photocurrent with the laser spot cross-section (solid and dashed lines) 
shows that the signal just follows the Gaussian intensity profile showing that the current 
is due to illuminating the sample edges. The current direction for  $\sigma^+$ and  $\sigma^-$ circularly polarized radiation and the magnitude of $J_A$ for various contact pairs are shown in Fig.~\ref{figure3}(d).  
In these measurements the Gaussian laser spot is always centered between the contacts preventing 
the temperature gradient between the contacts, at which signal is picked-up.
The figure documents a remarkable behavior of the circular edge photocurrent: It forms a vortex winding around the edges of the square shaped samples, which reverses its 
direction upon switching from  $\sigma^+$ to  $\sigma^-$ light. The magnitude and frequency dependencies
of the circular edge current  shown in the inset to Fig.~\ref{figedge} are in agreement with the developed theory. The only 
adjustable parameter is a scattering time in the vicinity of the edge. The scattering times determined 
by this method for each sample edge are quite close to the average bulk scattering time, {the deviations} most 
likely reflect fluctuations of the local scattering time and hence inhomogeneities in the distribution 
of scatterers. Actually, measurements of chiral edge currents provide very sensitive method of mapping 
the scattering processes at the edges. Moreover, the sign of the current reflects the type of the 
charge carriers in the close vicinity of the edge. This feature allowed us to conclude, that the 
edges of the $n$-type epitaxial graphene are, in fact, $p$-type. 
The latter, at first glance, surprising result agrees with analysis of the spatially resolved Raman measurements 
indicating an enhanced density of $p$-type carriers at graphene 
edges \cite{doi:10.1021/nl8032697,10.1063/1.3474613},
transport measurements reporting on the the transition from $n$-to $p$-type 
of doping at the edges of graphene flakes on SiO$_2$~\cite{EJHLee}
and growth details of epitaxial graphene~\cite{edge,Tzalenchuk2010,Emtsev2009,erl}.
 
 The data reveals that the measurements of edge 
currents may serve as a local probe of edge properties of graphene even at the room temperature.

As addressed above chiral edge photogalvanic current has also been observed in small-area exfoliated graphene layers~\cite{2010arXiv1002.1047K}.
In this case  the spot size of the terahertz laser radiation of
1~mm$^2$ is much larger than the micron sized exfoliated flakes and
the current is caused by both edge photogalvanic and photon drag (dynamic Hall) effect.
Examples of the current helicity dependence are shown in Fig.~\ref{figexfoliated}(a).  
At normal incidence the data  can be well fitted by 
\begin{equation}
\label{fit:small}
{J = A_n\sin{2\varphi} + B_n \sin{4\varphi} + C_n\cos{4\varphi} + D_n,}
\end{equation}
{where $A_n$, $B_n$, $C_n$ and $D_n$ are coefficients.} For oblique incidence the 
functional behavior remains unchanged but the individual coefficients at the second and the fourth 
harmonics of the angle $\varphi$. The overall behavior at any angle of incidence is 
well described by the superposition of the edge photogalvanic and dynamic Hall effects of a comparable strength given by Eqs.~\eqref{jABC}, \eqref{phenom:edge}. 
The contributions can easily be distinguished by measuring the signal as a function of the angle of incidence.
This is illustrated in Fig.~\ref{figexfoliated}(b) for the circular photocurrent where its dependence on the angle of incidence is plotted. While the photosignal
generated at normal incidence is solely determined by the edge photogalvanic current $j \propto \cos{\theta_0}$, the dynamic Hall effect is given by $j \propto \sin{\theta_0}\cos{\theta_0}$, the latter is odd in the angle of incidence and shows up at larger values of $\theta_0$.

\section{Third order effects}\label{3rd}

\subsection{Phenomenological discussion}\label{3rd:phen}

We continue the discussion of nonlinear high-frequency radiation phenomena excited in graphene by turning to the effects, where induced electric
current is proportional to the third power of electromagnetic field. These phenomena are, in general, related to the class of
the four-wave mixing effects, where three waves of different
frequencies, $\omega_1$, $\omega_2$, and $\omega_3$, interact and give rise to the fourth
one~\cite{blombergen}. Such a situation is described by the general relation
\begin{subequations}
\label{3rd:general:eqs}
\begin{multline}
\label{j:3rd:gen}
j_\alpha(\bm r, t) = 
\sigma^{(3,{\rm g})}_{\alpha \beta\gamma\delta}(\omega_1,\omega_2,\omega_3) \times \\
E_\beta(\omega_1, \bm q_1)E_\gamma(\omega_2, \bm q_2)E_\delta(\omega_3, \bm q_3) \times \\ \mathrm e^{-\mathrm i (\omega_1+\omega_2+\omega_3)t+ \mathrm i (\bm q_1+ \bm q_2 + \bm q_3)\bm r} + {\rm c.c.},
\end{multline}
where $\bm q_1$, $\bm q_2$ and $\bm q_3$ are corresponding wavevectors of the waves and $\sigma^{(3,{\rm g})}_{\alpha \beta\gamma\delta}(\omega_1,\omega_2,\omega_3)$ is the general third order conductivity. In the field of nonlinear optics it is usual to write similar to Eq.~\eqref{j:3rd:gen} expression for the media polarization $\bm P(\bm r,t)$:
\begin{multline}
\label{j:3rd:gen:P}
P_\alpha(\bm r, t) = \chi^{(3,{\rm g})}_{\alpha \beta\gamma\delta}(\omega_1,\omega_2,\omega_3) \times \\ 
E_\beta(\omega_1, \bm q_1)E_\gamma(\omega_2, \bm q_2)E_\delta(\omega_3, \bm q_3) \times \\ 
\mathrm e^{-\mathrm i (\omega_1+\omega_2+\omega_3)t+ \mathrm i (\bm q_1+ \bm q_2 + \bm q_3)\bm r} + {\rm c.c.},
\end{multline}
\end{subequations}
where the third order susceptibility $\chi^{(3,{\rm g})}_{\alpha \beta\gamma\delta}(\omega_1,\omega_2,\omega_3)$ is introduced. Taking into account standard relation Eq.~\eqref{jP} between the current density and the polarization one obtains\footnote{This relation becomes ambiguous if the response is static, $\omega_1 + \omega_2 + \omega_3=0$. In this case current generation and dielectric polarization becomes independent, cf. Sec.~\ref{rect} where optical rectification was discussed.} 
\begin{equation}
\label{sigmachi}
\sigma^{(3,{\rm g})}_{\alpha \beta\gamma\delta}(\omega_1,\omega_2,\omega_3) = - \mathrm i (\omega_1 + \omega_2 + \omega_3) \chi^{(3,{\rm g})}_{\alpha \beta\gamma\delta}(\omega_1,\omega_2,\omega_3).
\end{equation}
 It is assumed in Eqs.~\eqref{3rd:general:eqs} that frequencies may take both positive and negative values, the corresponding fields being related through $\bm E^*(\omega, \bm q) = \bm E(-\omega,-\bm q)$, the wavevector dependence of $\sigma^{(3,{\rm g})}$ and $\chi^{(3,{\rm g})}$ is omitted to shorthand the notations.
It is worth
to mention, that under spatial inversion both current components,
$j_\alpha$, and cubic 
combinations, $E_\beta E_\gamma E_\delta$,  change their sign, therefore third order effects take
place in even centrosymmetric systems without allowance for the radiation
wavevector, $\bm q$. Moreover, as we addressed in Sec.~\ref{remarks} in graphene the third order response is possible for the normal incidence 
of radiation, where the field has only in-plane components:
 $E_x \ne 0$, $E_y \ne 0$, $E_z = 0$, and the current and/or polarization is induced in the plane of the structure.

\subsubsection{Effects of static and \emph{ac} fields}\label{ac+dc}

It is convenient to start the analysis of the third-order effects
from the case, where one of the fields is static, $\bm E(0,0)$.

One example of such effects is the electric field induced second 
harmonic generation, observed recently for monolayer graphene samples~\cite{murzina}. 
Symmetry analysis of this effect is  the same as that of the photon wavevector induced SHG~\cite{glazov:shg}, 
see Sec.~\ref{2nd:phenom:sec}, with the replacement of the wavevector 
components $q_\alpha$ by the components of the static field $E_\alpha(0,0)$. 
In particular, in the \emph{strictly} two-dimensional model, the phenomenological 
relations describing electric field induced second harmonic generation
are given by Eqs.~(\ref{phenom:2}) where the components of the photon 
wavevector $q_x,q_y$ should be replaced by the components of the static 
field $E_x(0,0)$, $E_y(0,0)$. 

\begin{figure}[tb]
\includegraphics[width=0.75\linewidth]{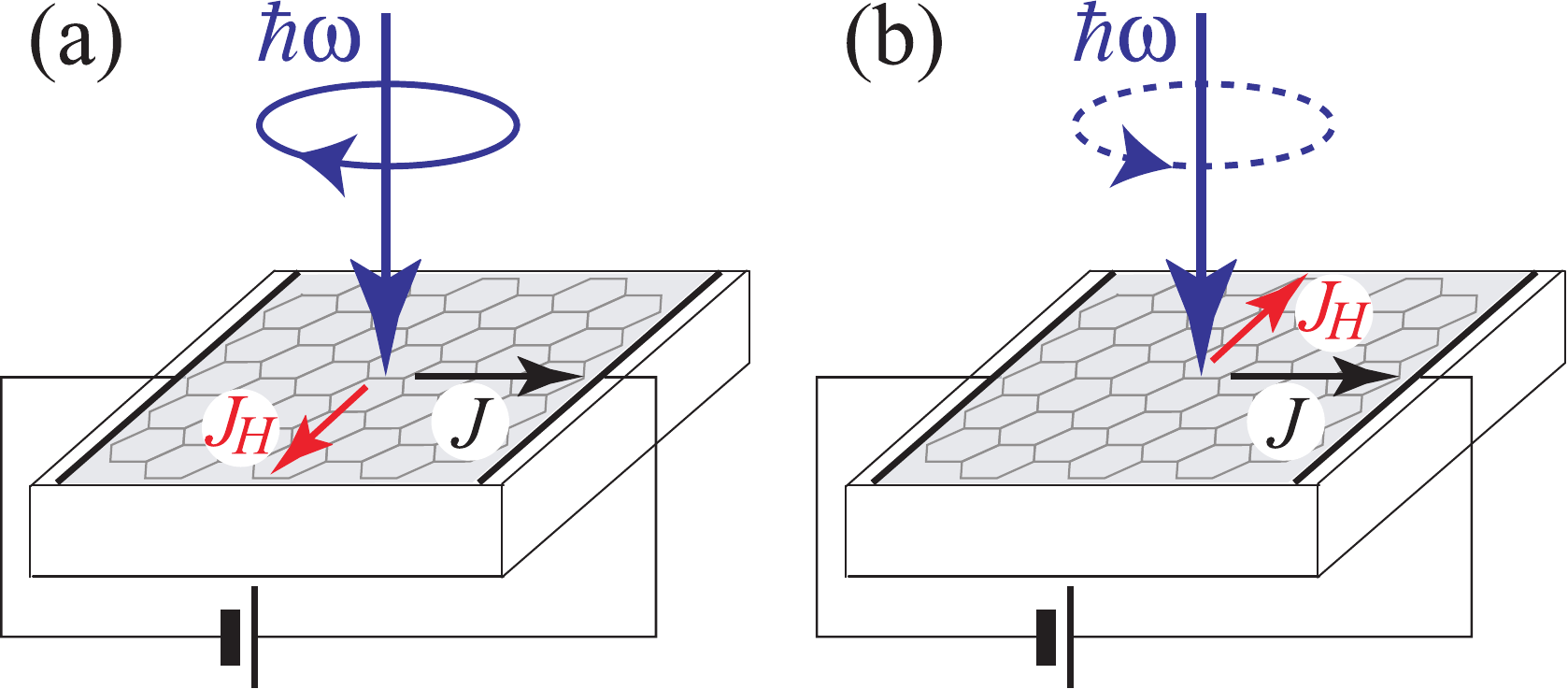}
\caption{Schematic illustration of the Hall effect 
excited by right (a) and left (b) circularly polarized 
radiation in a biased graphene sample. 
In the conducting layers the electric current $\bm J$ together with the electric field rotating at frequency $\omega$ the transverse component of the current, $\bm J_H$, appears, 
whose sign reverses with radiation helicity. After~\protect \cite{PhysRevB.79.081406}.
} \label{oka}
\end{figure}

Another particular example is the
\emph{photoconductivity phenomenon}, resulting in the \emph{dc} current
proportional to the intensity of the radiation at frequency $\omega$
and the static field $\bm E(0,0)$~\cite{sturmanBOOK}:
\begin{equation}
\label{photocond}
j_\alpha (\bm r, t)= \sigma_{\alpha \beta\gamma\delta }^{(3'')}
  E_\beta(\omega, \bm q) E_\gamma^*(\omega, \bm q) E_\delta(0,0),
\end{equation}
with $\sigma_{\alpha \beta\gamma\delta }^{(3'')} \equiv \sigma^{(3,{\rm g})}_{\alpha \beta\gamma\delta}(\omega,-\omega,0)$.
The
photoconductivity effects were studied in graphene theoretically and
experimentally in a number of
works~\cite{ISI:000275777700012,PhysRevB.77.195433,0295-5075-96-3-37006,0953-8984-21-44-445802,doi:10.1021/nl202318u,Sun2012}. 
Like photon drag effect, the photoconductivity is described by the fourth rank  tensor $\sigma_{\alpha
  \beta\gamma\delta }^{(3'')}$. It can be separated into the
symmetric and antisymmetric with respect to $\beta\gamma
\leftrightarrow \gamma\beta$ permutation parts giving rise to linear and circular photoconductivities, respectively. Anisotropic linear photoconductivity was discussed theoretically in detail in
Ref.~\cite{0295-5075-96-3-37006}. Circular photoconductivity effect also called photovoltaic Hall 
effect was predicted for graphene in Ref.~\cite{PhysRevB.79.081406}, see also
\cite{KibisPRB,PhysRevLett.107.216601,2011PhRvB..84w5108K,torres1,torres2,torres3}. It is schematically
depicted in Fig.~\ref{oka}. This effect results in the \emph{dc} current
flow perpendicularly to the static electric field under normal incidence
of radiation, $\hat {\bm e} \parallel z$:
\begin{equation}
\label{pvhe}
\bm j \propto [ \bm E(0,0) \times [\bm E(\omega, \bm q) \times \bm
E^*(\omega,\bm q)]] \propto  [ \bm E(0,0) \times P_{\rm circ} \hat {\bm e}].
\end{equation}
Equation~(\ref{pvhe}) demonstrates that the pseudovector of radiation
circular polarization $P_{\rm circ} \hat{ \bm e}$ plays a role of the
magnetic field in Hall effect, as illustrated in Fig.~\ref{oka}. The direction of the
transverse current, Eq.~(\ref{pvhe}), changes its sign if the
helicity of the radiation 
is reversed.

\subsubsection{Effects of \emph{ac} fields}\label{ac+ac}

The set of the third order effects under study further extends if all components of the incident field oscillate with time. First, let us consider a situation, where the sample is illuminated with the single monochromatic wave of a frequency $\omega$, Eq.~(\ref{plane}). In this case third order response can be written as
\begin{multline}
\label{j:3rd}
j_\alpha (\bm r, t)= 
 \sigma_{\alpha\beta\gamma\delta}^{(3')} E_\beta(\omega, \bm q)
E_\gamma(\omega, \bm q)E_\delta(\omega, \bm q) \mathrm e^{- 3{\rm i}
  \omega t + 3{\rm i} {\bm 
    q} {\bm r} }   + \\
    \sigma_{\alpha \beta\gamma\delta }^{(3)}
  E_\beta(\omega, \bm q) E_\gamma^*(\omega, \bm q) E_\delta(\omega, \bm q)
\mathrm e^{- {\rm i} \omega t + {\rm i} {\bm q} {\bm r} } + {\rm c.c.}\ .
\end{multline}
The first term described by the fourth order tensor $
\sigma_{\alpha\beta\gamma\delta}^{(3')}\equiv \sigma^{(3,{\rm g})}_{\alpha \beta\gamma\delta}(\omega,\omega,\omega) $ corresponds to the \emph{third harmonic
generation}, the effect already studied theoretically and observed experimentally
for graphene~\cite{doi:10.1021/nl2023405,2010arXiv1011.4841V,10.1063/1.3483872}.  
Corresponding fourth rank tensor, $\sigma^{(3')}_{\alpha\beta\gamma\delta}$, 
is symmetric with respect to $\gamma\delta \leftrightarrow \delta\gamma$ 
permutation. Hence, from the symmetry point of view its nonzero components are the same as for the 
corresponding part of tensor $\Phi_{\alpha\beta\gamma\delta}$ in 
Eq.~\eqref{expansion} describing linear photon drag effect.
The second term with the fourth rank tensor $ \sigma_{\alpha \beta\gamma\delta }^{(3)} \equiv \sigma^{(3,{\rm g})}_{\alpha \beta\gamma\delta}(\omega,-\omega,\omega)$
describes the current at the frequency $\omega$. This effect can be
seen as the correction to the ordinary high-frequency conductivity
$\sigma^{(1)}_{\alpha \beta}$ in Eq.~(\ref{j:phen:gen}), which is
proportional to the radiation intensity. Actually, it represents the \emph{two-photon absorption}~\cite{ivchenko05a}.

Now we turn to the situation where the frequencies of incident waves are different. An important example of such phenomena is the \emph{coherent injection of ballistic currents} also known as \emph{coherent photogalvanic effect}~\cite{shmelev,entin89}. In this case, the \emph{dc} current emerges under the illumination of sample with bi-harmonic field with frequencies $\omega$ and $2\omega$. As we show below in Sec.~\ref{3rd:theory}, the current is caused by the quantum mechanical interference of one and two photon absorption processes, in response to $\bm E(2\omega, 2\bm q)$ and $\bm E(\omega,\bm q)$, respectively, and has the following phenomenological form
\begin{equation}\label{jo2o}
j_\alpha = \bar \sigma^{(3)}_{\alpha\beta\gamma\delta}
E_\beta(2\omega,2\bm q) E_\gamma^*(\omega,\bm q) E_\delta^*(\omega,\bm
q) + {\rm c.c.} \:,
\end{equation}
where the field acting on the sample is given by:
\begin{equation} \label{plane2}
{\bm E}({\bm r}, t)= {\bm E}(\omega, \bm q) {\rm e}^{- {\rm i} \omega
  t + {\rm i} {\bm q} {\bm r} } + {\bm E}(2\omega, 2\bm q) {\rm e}^{-
  2{\rm i} \omega t + 2{\rm i} {\bm q} {\bm r} } + {\rm c.c.}\:,
\end{equation}
The corresponding nonlinear conductivity tensor, $\bar
\sigma^{(3)}_{\alpha\beta\gamma\delta}\equiv \sigma^{(3,{\rm g})}_{\alpha \beta\gamma\delta}(2\omega,-\omega,-\omega)$, is symmetric with respect to
the permutation of two last subscripts $\gamma\delta \leftrightarrow
\delta \gamma$, hence, the phenomenological description of this effect
in graphene is similar to that of linear photon drag effect and of the
second harmonic generation. In particular, in \emph{strictly} two-dimensional model, the coherent photogalvanic
effect is described by two  independent constants, ${M}_1$ and
${M}_2$ [cf. Eqs.~(\ref{phenom:2})]:
\begin{subequations}
\label{phenom:o2o}
\begin{multline}
\label{phenom:o2ox}
j_x = {M}_1 E_x(2\omega,2\bm q) \left[ (E_x^*(\omega,\bm q))^2+
  (E_y^*(\omega,\bm q))^2 \right] +\\ {M}_2E_x(2\omega,2\bm q)\left[(E_x^*(\omega,\bm q))^2-
  (E_y^*(\omega,\bm q))^2\right] + \\
  2{M}_2E_y(2\omega,2\bm q)
E_x^*(\omega,\bm q)E_y^*(\omega,\bm q) +{\rm c.c.}\ ,
\end{multline}
\begin{multline}
\label{phenom:o2o2y}
j_y = {M}_1 E_y(2\omega,2\bm q) \left[ (E_x^*(\omega,\bm q))^2+
  (E_y^*(\omega,\bm q))^2 \right] +\\ {M}_2E_y(2\omega,2\bm q)\left[(E_y^*(\omega,\bm q))^2-
  (E_x^*(\omega,\bm q))^2\right] +\\
  2{M}_2E_x(2\omega,2\bm q)
E_x^*(\omega,\bm q)E_y^*(\omega,\bm q) +{\rm c.c.}\ .
\end{multline}
\end{subequations}

If the static field $\bm E(0,0)$ in Eq.~\eqref{pvhe} is replaced by the linearly polarized \emph{ac} field $\bm E(\omega',\bm q')$, the transverse (Hall) current appears to be oscillating at the frequency $\omega'$. In such a case, the polarization plane of the \emph{ac} field reflected from or transmitted through the sample rotates, the direction of rotation is determined by the circular polarization of the field $\bm E(\omega,\bm q)$. This effect can be termed by \emph{optically induced Faraday/Kerr effect} similarly to the Faraday/Kerr rotation by 
optically induced spin polarization in semiconductors~\cite{opt_or_book}.

Below we briefly discuss theoretical approaches to calculate the third order effects and available experimental data.

\subsection{Theoretical background}\label{3rd:theory}

The microscopic mechanisms of the third order response are dominated
by the energy spectrum
nonparabolicity~\cite{ivchenkopikus,0295-5075-79-2-27002,0953-8984-20-38-384204}:
As already noted in Sec.~\ref{remarks}, the electron velocity in graphene $\bm v$ depends nonlinearly on the
electron momentum $\bm p$, see Eq.~\eqref{v},
hence, harmonic oscillations of $\bm p$ driven by external
electromagnetic field result in the anharmonic response in the
velocity and in the 
electric current, that is, in frequency
conversion~\cite{0953-8984-20-38-384204}. 

Since for the third-order effects neither the allowance for the
radiation wavevector nor the account of its magnetic field is needed,
its description is quite straightforward in the classical frequency
range, $\hbar\omega \ll E_F$. We employ the kinetic equation for
momentum and time dependent distribution function:
\begin{equation}
\label{kin3}
\frac{\partial f}{\partial t} + e\bm E(t) \frac{\partial f}{\partial
  \bm p} = - \frac{f(\bm p,t) - f_0({\bm p)}}{\tau},
\end{equation}
where the simplest form of the collision integral is taken, $f_0(\bm p)$ is the equilibrium distribution function. Its
solution, which takes into account electric field to all orders can be
written as~\cite{Bass1986237} 
\begin{multline}
\label{sol:exact}
f(\bm p,t) = f_0[\bm p - \bm p_0(t)] \mathrm e^{-t/\tau}
+ \\ \frac{1}{\tau} \int_{-\infty}^t \mathrm dt'
\mathrm e^{-\frac{t-t'}{\tau}} f_0[\bm p - \bm p_0(t) + \bm p_0(t')],
\end{multline}
where $\bm p_0(t) = \int^t_{-\infty} e\bm E(t) \mathrm dt$ is the electron
momentum acquired from the field and it is assumed that the field was
turned on at $t\to -\infty$. Equation~(\ref{sol:exact}) extends
the treatment developed in 
Refs.~\cite{0295-5075-79-2-27002,0953-8984-20-38-384204} for ballistic
electrons to allow for the scattering. Correspondingly, the induced
electric current at zero temperature for degenerate electrons in graphene with
density $n$ can be written at $t \gg \tau$ as
\begin{equation}
\label{current:exact}
\bm j = e n v \int_{-\infty}^t  \frac{\mathrm dt'}{\tau}
\mathrm e^{-\frac{t-t'}{\tau}} \frac{\bm P}{\sqrt{1+P^2}} \mathcal
G\left(\frac{2P}{1+P^2} \right),
\end{equation}
where $\bm P \equiv \bm P(t,t') = [\bm p_0(t) - \bm p_0(t')]/p_F$, $p_F$ is
the Fermi wavevector and function $\mathcal G(x)$ is related with
hypergeometric function as
\[
\mathcal G(x) = \hF{\left(\frac{1}{4},\frac{3}{4},2,x^2\right)}.
\]
Decomposing Eq.~(\ref{current:exact}) up to the third order of $\bm P$
we obtain the following expression for the nonlinear response:
\begin{equation}
\label{current:3}
\bm j = e N_0 v \int_{-\infty}^t  \frac{\mathrm dt'}{\tau}
\mathrm e^{-\frac{t-t'}{\tau}}  \left(\bm P - \frac{1}{8} \bm P P^2\right).
\end{equation}
Here $N_0$ is the electron density, the first term in parentheses describes linear response and second one
describes the third order effects.

For example, for the incident harmonic radiation $\bm E = \bm E_0 \mathrm
e^{-\mathrm i \omega t}$ the term $\bm P P^2$ is
oscillating at $3\omega$ with the result
\begin{equation}
\label{eq:3rd}
\bm j(3\omega) =  -e^4N_0v^4 \frac{3\bm E_0 E_0^2}{4E_F^3}  \tau_{\omega} \tau_{2\omega}\tau_{3\omega},
\end{equation}
with $\tau_\omega = \tau/(1-\mathrm i \omega\tau)$. In the limit
$\omega\tau \gg 1$ (but $\hbar\omega \ll E_F$) Eq.~(\ref{eq:3rd}) agrees with Eq.~(9) of \cite{0295-5075-79-2-27002}. 

As an example of the static field induced second harmonic generation we consider simplest situation where the static field ${\bm E(0,0) =}\bm E_0 \parallel x$, while the alternating (radiation) field $\bm E_1 \exp{(-\mathrm i \omega t)} + {\rm c.c.}$ is linearly polarized along $y$ axis, i.e. Stokes parameters of incident field are $S_1=-1$, $S_2 = S_3 =0$. Calculation shows that the 
current at a double frequency flows along $x$ axis and is given by
\begin{equation}
\label{j2:field}
\bm j(2\omega) =  e^4N_0v^4 \frac{\bm E_0 E_1^2}{4E_F^3}  \tau_{\omega}^2 \tau_{2\omega}^2\frac{2\omega^2\tau^2+6\mathrm i \omega \tau - 3}{\tau}.
\end{equation}
The microscopic theory of the field induced second harmonic generation for bilayer graphene was developed in Ref.~\cite{doi:10.1021/nl300084j} for the quantum frequency range. It was predicted that AB-stacked bilayer graphene can exhibit a giant and tunable second order nonlinear susceptibility if the in-plane electric field is applied. The susceptibility varies from 0 to $10^5$~pm/V depending on the magnitude of the static field and exceeds by 3 orders of magnitude that of conventional nonlinear crystal AgGaSe$_2$. Such a high values of the electric field induced response is related to the specifics of the bilayer band structure, and its detailed consideration is out of the scope of this review.

\begin{figure}[tb]
\includegraphics[width=0.75\linewidth]{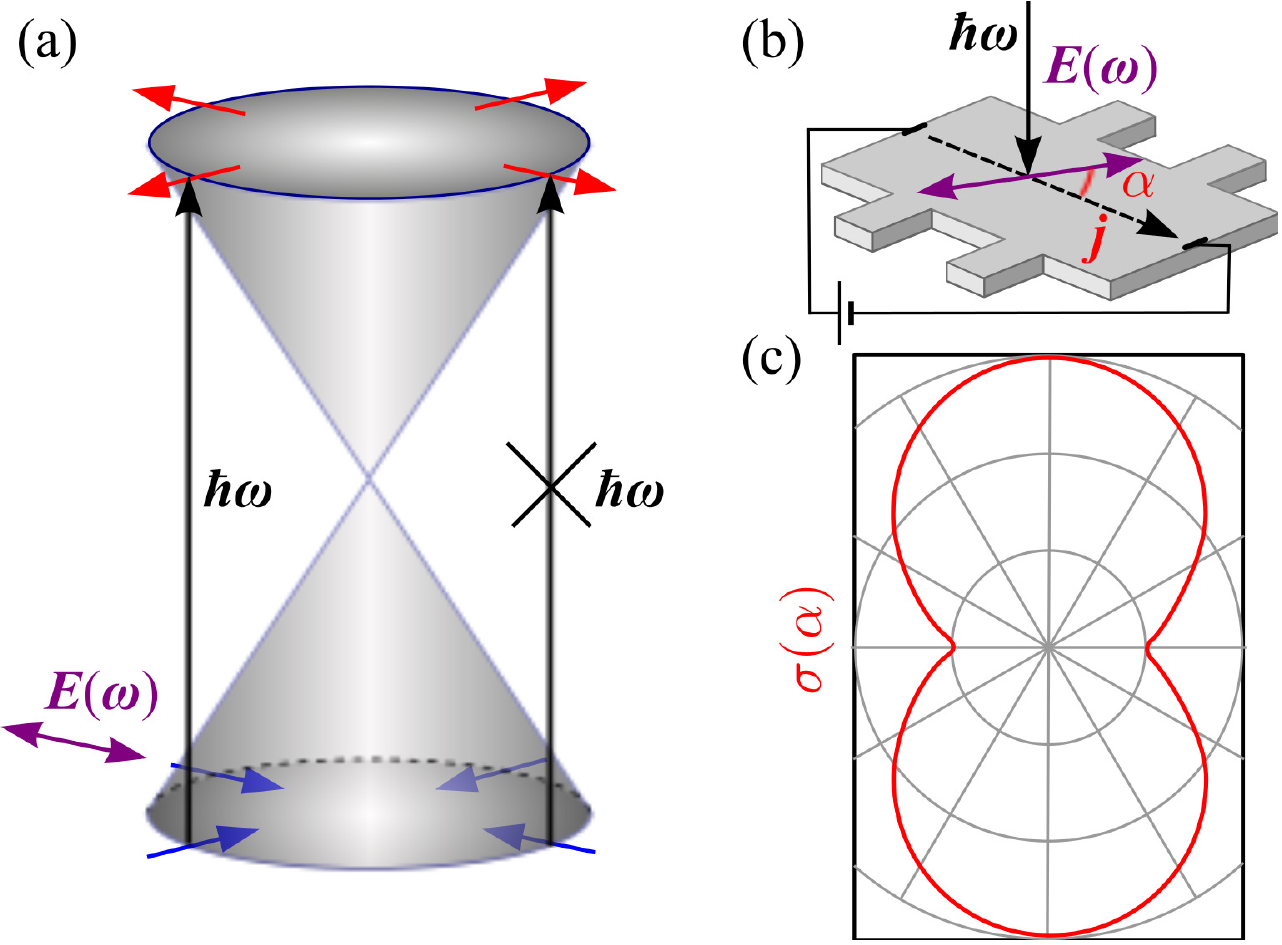}
\caption{(a) Scheme of \textit{inter}band optical transitions excited by 
linearly polarized irradiation with $\hbar \omega \geq 2 E_F$. Due to selection 
rules the transitions are forbidden for $\bm k$ parallel to the linear polarization 
plane of light. (b)  Graphene Hall bar sample irradiated with linearly polarized 
electromagnetic wave. Application of a bias voltage leads to an electrical current 
$\bm j$, whose magnitude depends on the in-plane orientation of polarization plane of the linear polarized light given by the azimuthal angle $\alpha$. (c) Photoconductivity, $\sigma (\alpha)$, as a function of $\alpha$, an angle 
between the current and linear polarization plane of radiation.
 After~\protect \cite{0295-5075-96-3-37006}. 
} \label{trushin}
\end{figure}

\begin{figure}[t]
\includegraphics[width=0.85\linewidth]{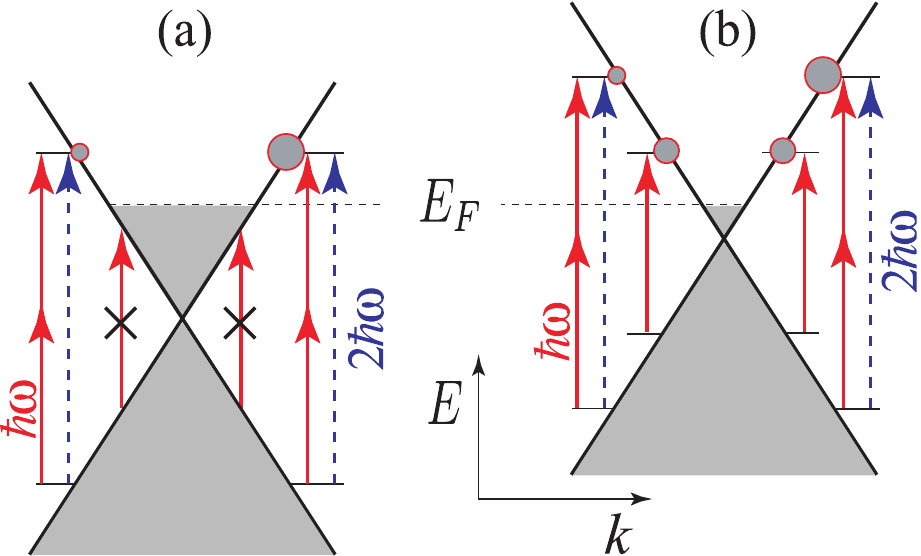}
\caption{Optical \textit{inter}band optical transition for (a) $E_F<\hbar\omega<2E_F$
and (b) $\hbar\omega>2E_F$. The dotted horizontal lines indicate the Fermi 
level. Solid arrows denote transitions caused by the beam of frequency 
$\omega$ and dashed arrows show transitions caused by the $2\omega$ beam. 
Filled circles of various diameters sketch asymmetric electron populations at $\pm k$ caused by the quantum interference of a single and two-photon absorption processes. 
The photoinduced imbalance in $\bm k$-space causes $dc$ current generation.  After~\protect \cite{sun2010}. 
} \label{sipe0}
\end{figure}

A detailed theory of linear photoconductivity in graphene for the case of \emph{inter}band optical transitions was developed in Ref.~\cite{0295-5075-96-3-37006}. Following this work we note, that due to the optical selection rules
the excitation with linearly 
polarized light generates the distribution of photocarriers containing
second angular harmonic (momentum alignment) whose orientation is
determined by the polarization plane of the radiation, see
Fig.~\ref{trushin} and Refs.~\cite{0295-5075-96-3-37006,portnoi_book}
for details. Indeed, the \textit{inter}band transitions are forbidden for electron momentum  $\bm p$
being parallel to the linear polarization plane of radiation, since the perturbation 
due to electromagnetic radiation $\propto v(\bm \sigma \cdot \bm A)$, where $\bm A$ 
is the vector potential of radiation does not mix eigenstates of the Dirac 
Hamiltonian, Eq.~\eqref{Dirac:Ham}, with ${\bm p} \parallel \bm A$. {The matrix element of the \textit{inter}band transition has a form
\begin{equation}
\label{Minter}
M \propto p_x A_y - p_y A_x.
\end{equation}
The distribution function of photoelectrons $\delta f$ is determined by the transition rate $\propto |M|^2$, namely,
\begin{equation}
\label{dfE2}
\delta f \propto p_x^2 |E_x|^2 + p_y^2|E_y|^2 - p_xp_y (E_xE_y^* + E_x^*E_y),
\end{equation}
where $E_x$ and $E_y$ are the incident field components. Apart from the isotropic part $\propto (|E_x|^2+|E_y|^2)$ the photoelectrons distribution contains second angular harmonics of electron momentum $\bm p$: $\cos{2\varphi_{\bm p}} \propto (|E_x|^2-|E_y|^2)$ and $\sin{2\varphi_{\bm p}} \propto (E_xE_y^* + E_x^*E_y)$, where $\varphi_{\bm p}$ is the angle between $\bm p$ and $x$ axis.
As a result, 
the magnitude of the current 
of photoelectrons driven by external bias
depends strongly on the mutual orientation of the polarization plane
of radiation and external electric field. For the classical frequency range the second-order in the \emph{ac} field correction to the distribution function assumes the same form of Eq.~\eqref{dfE2} giving rise to the anisotropic photoconductivity.

The description of the coherent photogalvanic  and frequency
mixing phenomena can be carried out along the same lines for the
classical range of frequencies. As a particular example we consider
bichromatic field in the form
\begin{equation}
\label{bi:inc}
\bm E = \bm E_1 \cos{\omega t} + \bm E_2 \cos{(2\omega t+ \delta)},
\end{equation}
incident on the sample. The parameter $\delta$ describes the phase
shift between $\omega$ and $2\omega$ fields.  In the geometry $\bm
E_1 \parallel \bm E_2 \parallel x$ the $x$-component of the \emph{dc} current described by phenomenological parameter ${M}_1$ in Eqs.~\eqref{phenom:o2o} yields
\begin{subequations}
\label{j:dc:o2o:gen}
\begin{equation}
\label{j:dc:o2o}
j_x = -e^4N_0v^4 \frac{9E_1^2 E_2\cos{\delta}}{16E_F^3} \frac{\tau^3}{1+5\omega^2\tau^2+4\omega^4\tau^4}.
\end{equation}
If, by contrast, $\bm
E_1 \parallel y \perp \bm E_2 \parallel x$, then the \emph{dc} photocurrent described by phenomenological parameter ${M}_2$ in  Eqs.~\eqref{phenom:o2o}  has form
\begin{equation}
\label{j:dc:o2o:y}
j_x = -e^4N_0v^4 \frac{3E_1^2 E_2\cos{\delta}}{16E_F^3} \frac{\tau^3}{1+5\omega^2\tau^2+4\omega^4\tau^4}.
\end{equation}
\end{subequations}
Note that, similar to Eq.~\eqref{j:dc:o2o} expression was derived in Ref.~\cite{AlekseevErement} (see
also~\cite{Bass1986237}) for the semiconductor system with
nonparabolic energy dispersion. It follows from Eq.~(\ref{j:dc:o2o})
that the coherent photocurrent is extremely sensitive to the phase
relation between two waves: The current is proportional to the cosine
of the phaseshift. It is worth to mention that in the ballistic case
($\tau \to \infty$) the first term in Eq.~(\ref{sol:exact}) also gives rise
to the current $\propto \omega^{-3}\sin{\delta}$, see Ref.~\cite{AlekseevErement} for the semiconductor system with nonparabolic dispersion and Ref.~\cite{doi:10.1021/nl204283q} for ``mini-gapped'' graphene on a substrate. The mechanism of
the coherent photogalvanic effect in the systems with parabolic
dispersion is presented in Ref.~\cite{entin89}.

In the quantum frequency range, $\omega\tau \gg 1$, $\hbar\omega \sim E_F$ (or even $\hbar\omega \gg E_F$) the description of the third order phenomena can be carried out in a similar fashion. Instead of applying Boltzmann equation~\eqref{kin3} one may use similar equation for the density matrix where the collision integral is absent. Such a treatment is outlined in Refs.~\cite{Hendry10, sun2010,PhysRevB.83.195406,Wright09,Mele:2000oq}. In the case of \textit{intra}band transitions, where the double photon energy exceeds $2E_F$ and two-photon transition becomes possible, see Fig.~\ref{sipe0}, the coherent photogalvanic effect can be understood in terms of quantum interference of single and two photon processes~\cite{sun2010}. These processes are schematically shown in Fig.~\ref{sipe0}. To begin with, consider the case where the direct absorption of a single photon with the frequency $\omega$ is forbidden, as illustrated in Fig.~\ref{sipe0}(a). The matrix element describing the electron transition from the valence to the conduction band caused by the absorption of one photon with the frequency $2\omega$ is linear in the electron wavevector $\bm k$ and has a form
\begin{equation}
M^{(1)}_{2\omega} \propto k_x A_y(2\omega) -  k_y A_x(2\omega).
\end{equation}
Here $\bm A(2\omega) = [A_x(2\omega), A_y(2\omega)]$ is the vector potential of the field oscillating at $2\omega$.
Due to the condition $\hbar\omega<2E_F$ the direct \emph{inter}band absorption of the radiation with the frequency $\omega$ is possible only via the two-photon absorption. Such a second-order process takes place via the intermediate states in the same band, yielding the matrix element of the two-photon process in the form
\begin{equation}
M^{(2)}_{\omega} \propto [k_x A_y(\omega) -  k_y A_x(\omega)][k_xA_x(\omega) + k_yA_y(\omega)],
\end{equation}
with $\bm A(\omega) = [A_x(\omega), A_y(\omega)]$ being the vector potential of $\omega$-oscillating field. As both, one $2\omega$ photon absorption and two $\omega$ photon absorption, processes mix the same states they interfere. The total absorption rate is proportional to the $|M^{(1)}_{2\omega} + M^{(2)}_{\omega}|^2$ with the interference contribution in the form 
\begin{equation}
\label{o2o:interference}
\propto 2\Re{[M^{(1)}_{2\omega}M^{(2)^*}_{\omega} ]},
\end{equation}
which results in the anisotropic distribution of photoelectrons, shown by filled circles of different sizes in Fig.~\ref{sipe0}(a) and, correspondingly, in the electric current. The magnitude and direction of electric current are controlled by the orientation of $\bm A(\omega)$, $\bm A(2\omega)$ and their phase difference. Similar situation occurs if $\hbar\omega>2E_F$, i.e. where single photon absorption is also possible, see Fig.~\ref{sipe0}(b). While the interference of one and two-photon absorption processes is possible and gives rise to the electric current, the absorption of single photon with the frequency $\omega$, although being possible, does not result in the asymmetry of electron distribution and does not lead to current generation.}

To conclude this Section, we present the results of analytical calculations of the optically induced Faraday/Kerr effect introduced in Sec.~\ref{ac+ac}. Theoretical estimate of this effect for the classical frequency range can be obtained considering the incident radiation in a form
\begin{equation}
\label{circ:lin}
E_x(t) = E_1\cos{\omega t} + E_2\cos{\omega t}, \quad E_y(t) = \mp E_1 \sin{\omega t},
\end{equation}
corresponding to the combination of the circularly polarized wave with the amplitude $E_1$ and linearly polarized wave with the amplitude $E_2\ll E_1$. Signs $\mp$ in expression for $E_y$ correspond to right and left circular polarizations for the wave propagating along negative $z$ axis. It follows from Eq.~\eqref{current:3} that the transverse component of the current in the classical frequency range is
\begin{multline}
\label{j:farad:y}
j_y = \mp e^4N_0v^4 \frac{3E_1^2 E_2}{8E_F^3} \frac{\tau^3}{1+5\omega^2\tau^2+4\omega^4\tau^4} \times \\
 \left(2\omega\tau\cos{\omega t} -\sin{\omega t} \right).
\end{multline}
The appearance of $j_y \ne 0$ is responsible for the Faraday rotation of the polarization plane of the transmitted (and Kerr rotation of reflected) probe beam $E_2$ incident on the excited by circularly polarized beam $E_1$ graphene.

\subsection{Third and higher harmonic generation and frequency mixing: Experiment}\label{THG}

\begin{figure}[t]
\includegraphics[width=0.75\linewidth]{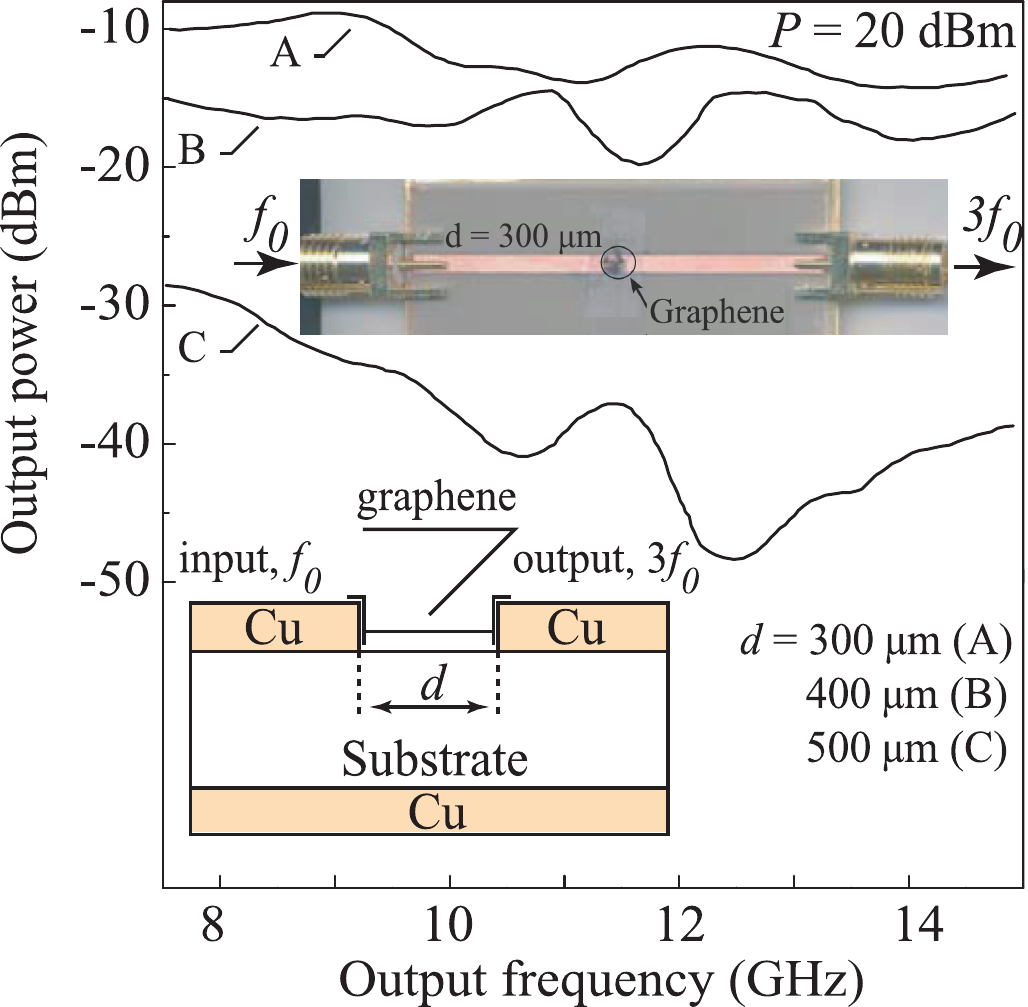}
\caption{
Radiation output power as a function of output frequency, $3 f_0$, for a graphene based frequency tripler. Curves $A$, $B$ and $C$ present the characteristics of three devices with different gap lengths of 300~$\mu$, 400~$\mu$m, and 500~$\mu$m. Middle inset shows photograph of manufactured device and bottom inset demonstrates the device cross section.  After~\cite{doi:10.1163/156939311798072090}.}
 \label{camblor}
\end{figure}

The generation of third harmonic and higher orders nonlinearities (up to seventh order harmonic) were reported first for 
millimeter waves in Ref.~\cite{10.1063/1.3483872} (this experiment on a monolayer graphene is already described in detail
in Sec.~\ref{2nd:exper}),
and in Refs.~\cite{Hotopan11,doi:10.1163/156939311798072090}. 
In the latter work a graphene based frequency tripler was manufactured. The sketch of the setup and photograph of the device is
shown in Fig.~\ref{camblor}. 
The nonlinear component of the device consists of a microstrip line with a small gap covered by a 
few layer graphene film.  A standard microwave set-up consisting of 
a generator tunable in the 2.5 --- 5 GHz range and a spectrum analyzer was used.
Output frequencies in the range between 8 and 15 GHz have 
been obtained with a received output power up to $-10$~dBm as shown in the main panel of Fig.~\ref{camblor}. Almost flat frequency 
behavior can be obtained in the whole output frequency range.

\begin{figure}[t]
\includegraphics[width=0.75\linewidth]{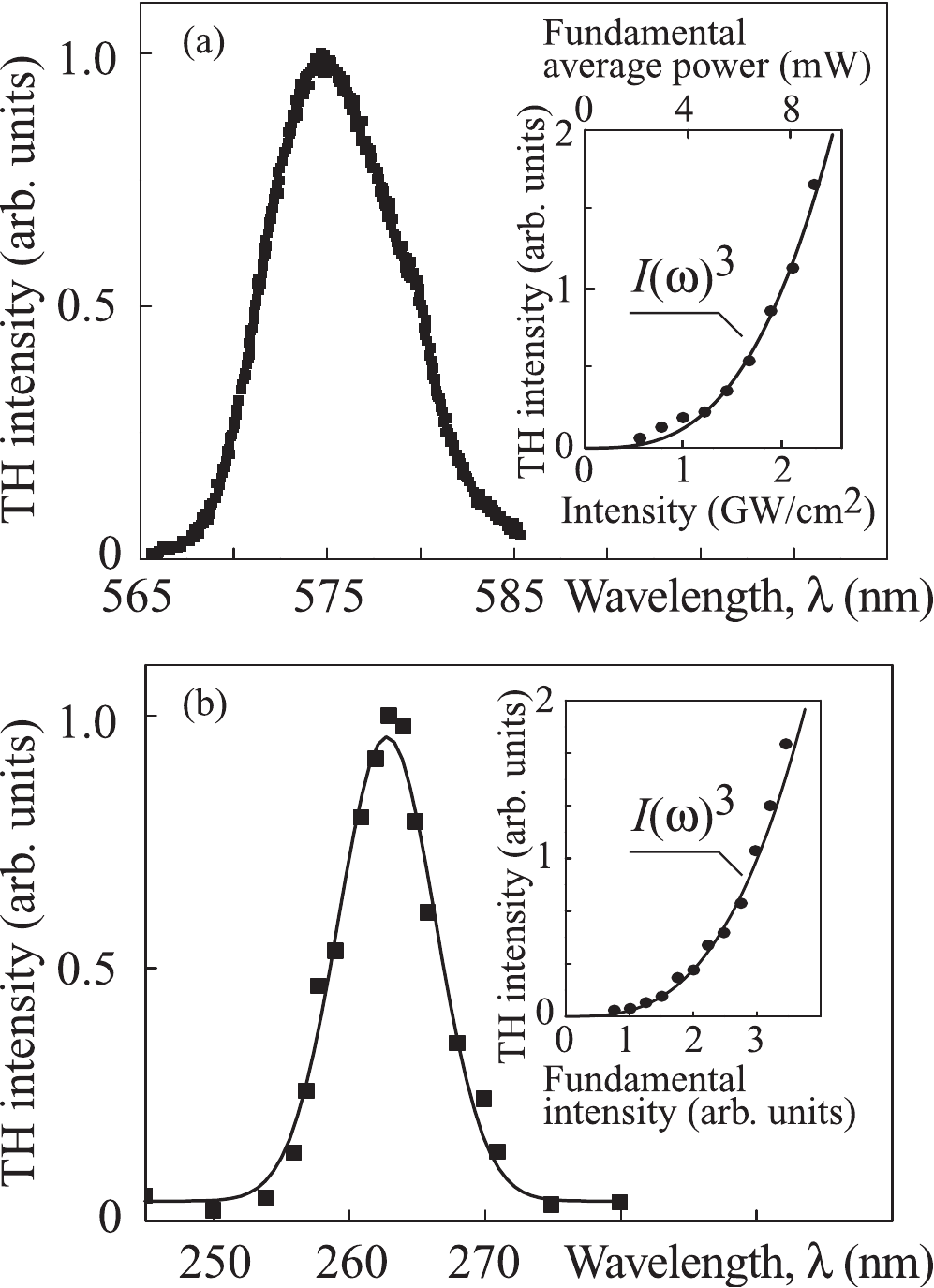}
\caption{Spectra of third harmonic generation measured in Ref.~\cite{2013arXiv1301.1042K}, panel (a), and Ref.~\cite{2013arXiv1301.1697H}, panel (b). Insets show the measured third harmonic (TH) power as a function of the fundamental beam intensity (points). Solid lines show the cubic law fit. After~\protect \cite{2013arXiv1301.1042K,2013arXiv1301.1697H}. 
} \label{triple}
\end{figure}

Third harmonic was most recently observed in graphene for the fundamental frequency $\omega$
 in the near infrared range in two works~\cite{2013arXiv1301.1042K} (exfoliated graphene)
 and \cite{2013arXiv1301.1697H} (CVD graphene). In Ref.~\cite{2013arXiv1301.1042K} the fundamental 
 wavelength is $\approx 1.72$~$\mu$m (third harmonic wavelength is $\approx 0.575$~$\mu$m), 
 while in Ref.~\cite{2013arXiv1301.1697H} the fundamental wavelength was somewhat shorter 
 $\approx 0.8$~$\mu$m (third harmonic corresponds to $\approx 0.265$~$\mu$m). In the latter 
 case the third harmonic was close to resonance with the optical transition in the $M$ point 
 of the Brillouin zone making it possible to enhance the signal. Figure~\ref{triple} 
 demonstrates the spectra of the third harmonic measured in Ref.~\cite{2013arXiv1301.1042K}, 
 panel (a), and in Ref.~\cite{2013arXiv1301.1697H}, panel (b). 
 The insets demonstrate that the third harmonic intensity indeed scales as cube of fundamental 
 harmonic intensity. According to Ref.~\cite{2013arXiv1301.1042K} 
 the third order susceptibility of graphene for such near-IR frequencies is on the order of $10^{-8}$~esu (electrostatic units) 
 and is by several order of magnitude larger than in transparent materials. 

Besides the third harmonic generation, several other effects caused by the third order nonlinearity  have been reported for near-infrared, optical, and UV frequencies. In particular, the $\chi^{(3)}$ have been studied for graphene in \textit{solutions} by means of the time-resolved pump-probe techniques~\cite{2012chu,doi:10.1021/nl2023405}. In Ref.~\cite{doi:10.1021/nl2023405} a purely coherent nonlinear optical response of
 high-quality graphene sheets functionalized by alkylamine has been demonstrated. These graphene sheets has been investigated, using near-infrared, visible, and ultraviolet continuos wave and ultrafast laser beams, and spatial self-phase modulation has been  observed in the solution dispersions.
The ultrafast third-order nonlinear optical properties of graphene in both suspension and film state were studied using femtosecond time resolved optical Kerr gate technique in Ref.~\cite{2012chu}. The third-order nonlinear optical susceptibility of about $4 \times 10^{-14}$ esu was observed for solution of 0.010 mg/ml. While huge nonlinear response has been detected, the signal may result from the superposition of the nonlinear  response of graphene itself and the effect of reorientation and alignment of graphene sheets in \textit{solutions} induced by the electromagnetic field which is similar to the case of liquid crystals~\cite{doi:10.1021/nl2023405}. Thus, the detailed discussion of these interesting and important for application results are out of scope of the present review aimed to pristine graphene and graphene on substrates.

\begin{figure}[tb]
\includegraphics[width=\linewidth]{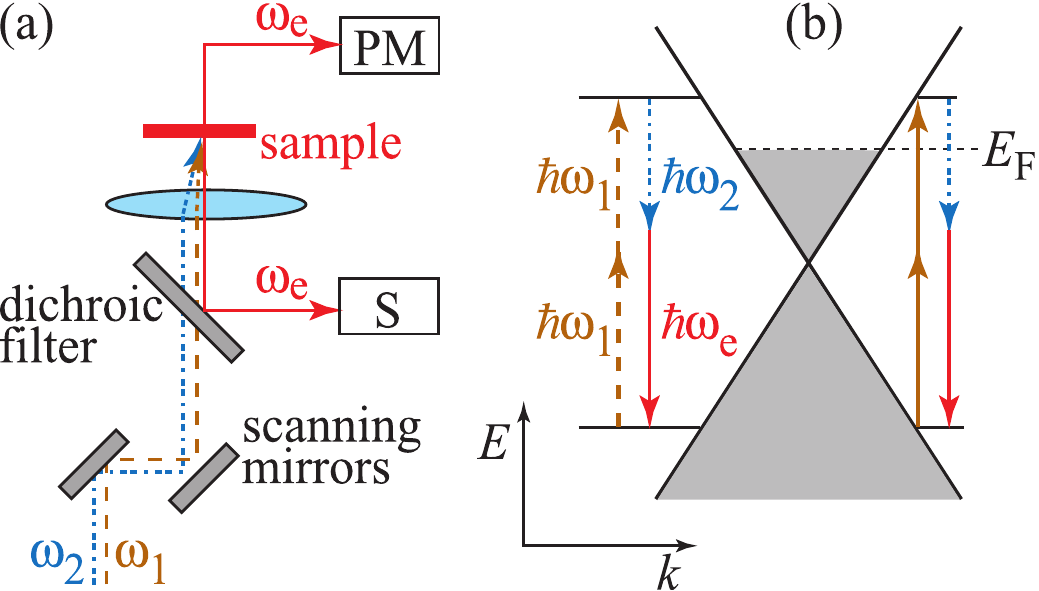}
\caption{(a) Schematic illustration of the experimental setup used in the frequency mixing experiments.  Emission beam with frequency $\omega_e$ caused by mixing of the beams with frequencies $\omega_1$ and $\omega_2$ is detected by a photomultiplier, PM, and spectrometer,
S. (b) Sketch of the frequency mixing effect in graphene with the three resonant photon energies (arrows) involved in the process. After~\protect \cite{Hendry10}.
} \label{savchenko1}
\end{figure}

\begin{figure}[tb]
\includegraphics[width=0.7\linewidth]{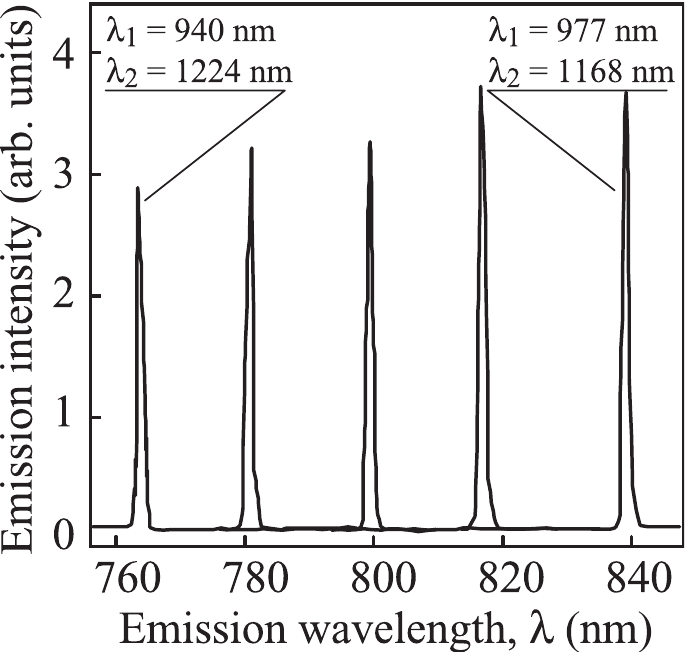}
\caption{Emission spectra of an exfoliated graphene flake
excited with pump pulses of different wavelengths, ($\lambda_1$, 
$\lambda_2$): (940 nm, 1224 nm), (950 nm, 1210 nm), (958 nm, 1196 nm),
(967 nm, 1183 nm), and (977 nm, 1168 nm) from left to right, respectively.
After~\protect \cite{Hendry10}.
} \label{savchenko2}
\end{figure}

\begin{figure}[tb]
\includegraphics[width=0.8\linewidth]{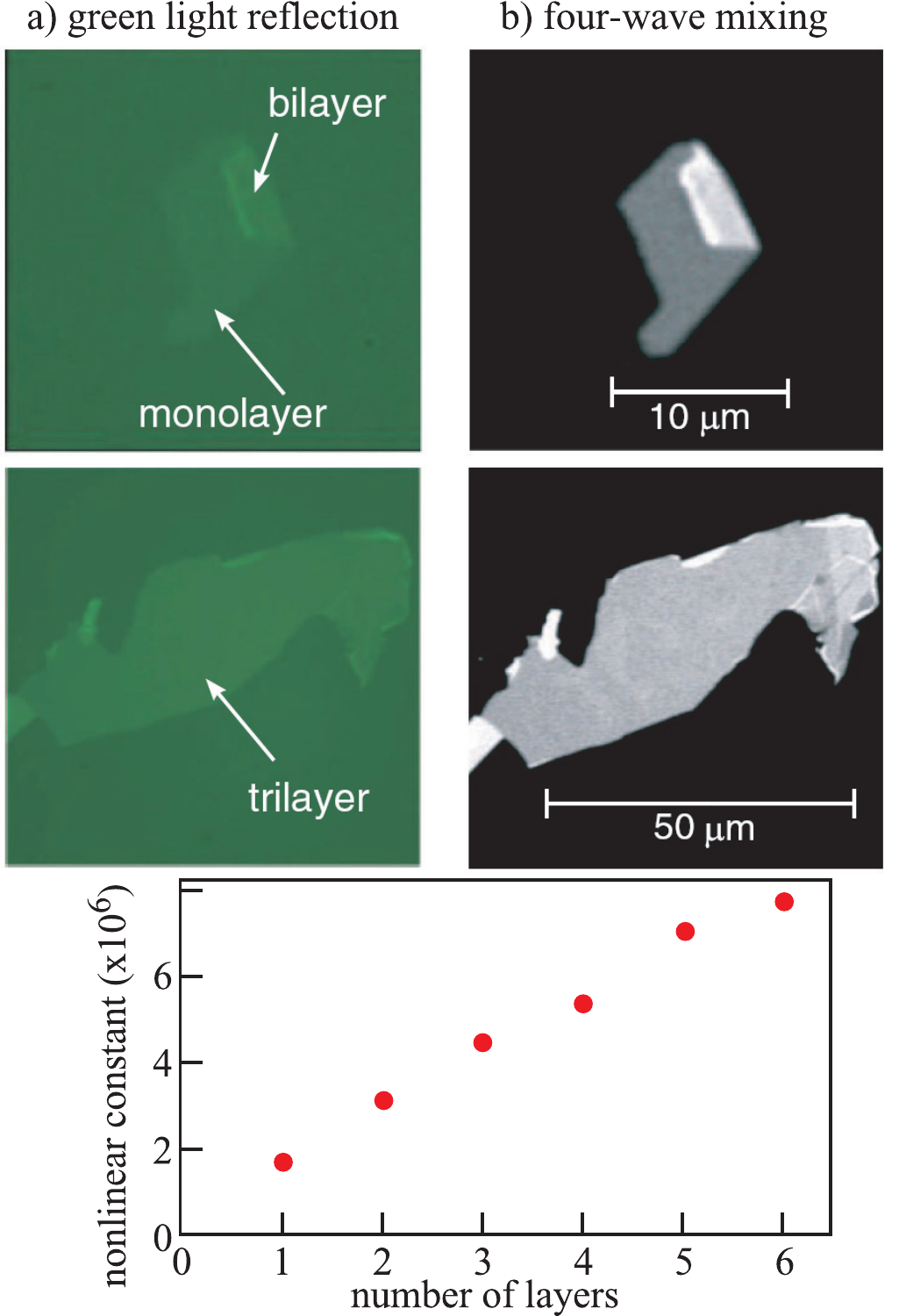}
\caption{(a) Green light (550 nm) reflection
images of two exfoliated graphene flakes. (b) Nonlinear optical
images measured with pump wavelengths of 969 nm and
1179 nm. Image acquisition times are approximately 0.6 s.
(c) The contrast in four-wave mixing images as a function of
the number of graphene layers.
After~\protect \cite{Hendry10}.
} \label{savchenko3}
\end{figure}

Another experimental manifestation of the third order nonlinearity is a frequency mixing, recently 
demonstrated for infrared/red light and radiation of THz and GHz frequency 
ranges~\cite{Hendry10,Shareef:12,Hotopan11}. Figure \ref{savchenko1}(a) shows the  setup used in Ref.~\cite{Hendry10} for the four-wave mixing experiments, which involves the generation
of  optical frequency harmonics 2$\omega_1- \omega_2$ under irradiation by two monochromatic waves with the
frequencies $\omega_1$ and $\omega_2$ as depicted in Fig.~\ref{savchenko1}(b).  Two
incident pump laser beams with wavelengths $\lambda_1$ (tunable from 670 to 980 nm) 
and $\lambda_2$ (1130 to 1450 nm) duration about 6~ps are focused 
collinearly onto a sample and mix together to generate a third, coherent beam of wavelength $\lambda_e$. 
The incident pump pulses are focused onto the sample using a water 
immersion objective with a numerical aperture of 1.2, giving rise to 
a spot size $< 1$~$\mu$m and time averaged and peak excitation powers at the 
sample of about 1~mW and 10~W, respectively. Note that in these experiments the peak beam power 
is much higher than in other experiments. This 
fact indicates that the graphene samples are  robust and rather higher power 
can be used without damaging samples. 
The nonlinear signal is presented in Fig.  \ref{savchenko2} for different combinations of incident wavelengths $\lambda_1$ and $\lambda_2$. The results of Ref.~\cite{Hendry10} evidence that the graphene has an exceptionally
high nonlinear response, with the effective
nonlinear susceptibility $\chi^{(3)}=10^{-7}$ esu being by about an order of magnitude larger than that obtained by third harmonic generation in Ref.~\cite{2013arXiv1301.1042K} and  by several orders of magnitude larger than that for, e.g., gold or glass. Moreover, this nonlinearity is shown to be almost dispersionless
in a wide range of emission wavelengths (from 760 to 840 nm). Interestingly, a high third order nonlinearity yields an enormously large contrast between the responses of the sample and substrate as well as between the samples with different number of graphene layers, the latter is due to the linear increase of the $\chi^{(3)}$ with the number of layers in the sample, see Fig.  \ref{savchenko3}(c). This results in much better microscopic images of graphene compared to those obtained in a normal optical reflection as demonstrated in Fig. \ref{savchenko3}(a) and (b). Further application of graphene signal mixing has been addressed in Ref. \cite{Hotopan11} where signals with MHz frequencies were mixed in different combinations by three layer graphene device with linear current-voltage characteristic.

\subsection{Coherent injection of ballistic photocurrents: Experiment}

Ballistic photocurrents related to the third order nonlinearity have been experimentally 
demonstrated for multilayer epitaxial graphene film produced
on the C-terminated face of single-crystal 4H-SiC~\cite{sun2010,Sun:2012ys}. 
In these multilayer epitaxial graphene films the first few layers were heavily doped (10$^{13}$ cm$^{-2}$)
with the doping decreasing rapidly by four orders of magnitude.
Interestingly, samples used in these experiments have from 9 to 63
graphene atomic layers; but have been shown by independent studies to
have the graphene-like linear band structure and be distinct
from bulk graphite~\cite{sun2010,13,24,25,26,27,28,kop}.  
In order to investigate the coherent photocurrent arising on subpicosecond 
timescale the THz radiation emitted by the current pulse has been measured.
 Application of short current pulses for THz radiation generation (Auston switch) has 
 been developed in early 90's \cite{15,17} and is currently widely used for generation of THz 
 radiation and time-domain THz spectroscopy~\cite{ganichev_book,Sakai,Lee}.
  In this case 
 the generated electric field 
\begin{equation}
\label{THz:det}
\bm E(t) \propto d \bm j(t)/d t \sim \bm j/\tau_p,
\end{equation}
where $\bm j(t)$ is the current pulse density and $\tau_p$ is its duration. Since typical 
current pulse durations correspond to picosecond timescale, the emitted field corresponds 
to THz frequency range. Consequently, the dynamics of the emitted THz field reflects 
behavior of the generated current. 

\begin{figure}[ht]
\includegraphics[width=\linewidth]{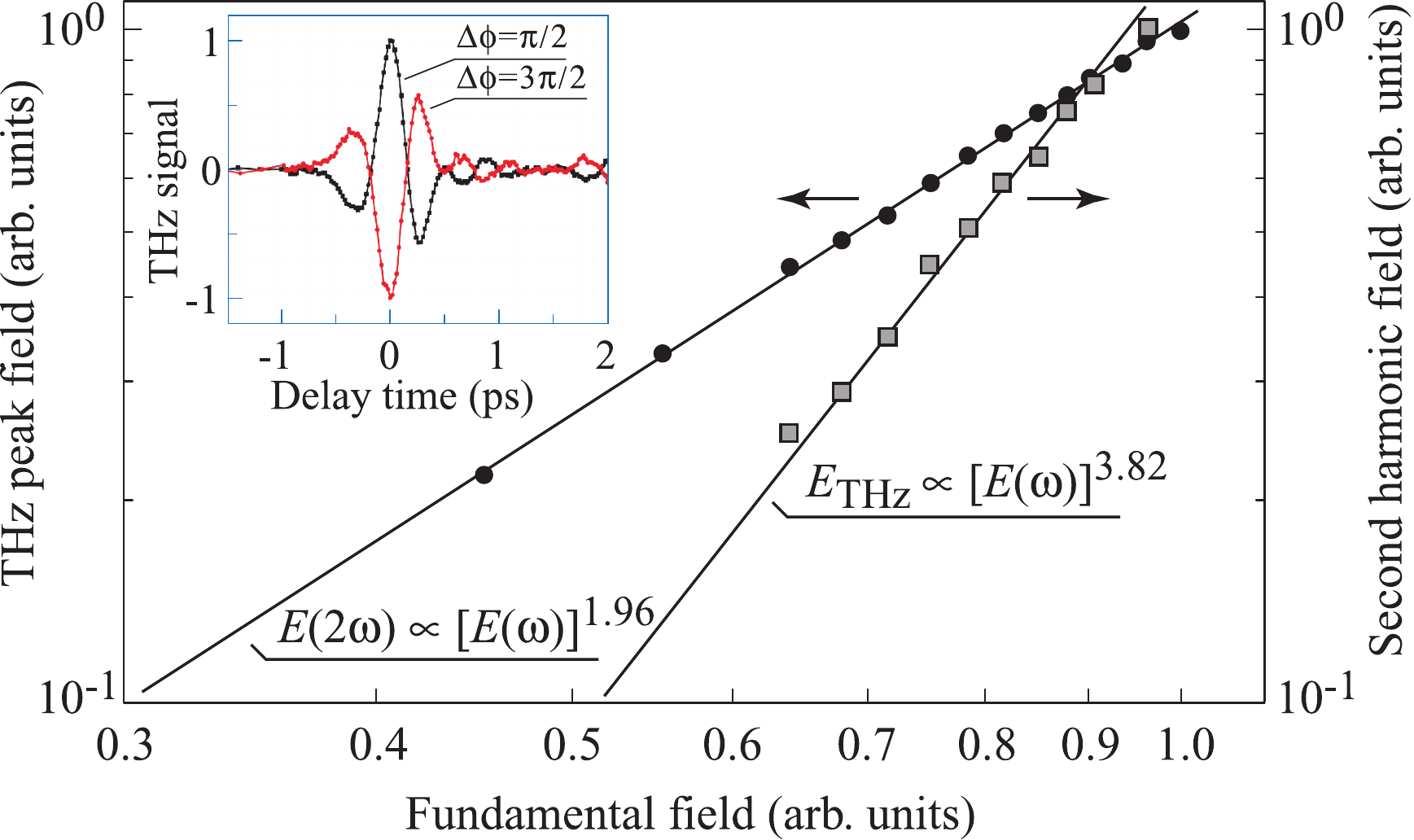}
\caption{Fundamental beam power dependence of the field $E(2\omega)$ (black circles) and THz field proportional to the electric current generated in the sample (grey squares). Solid lines are corresponding power-law fits.
Inset shows THz signal from the multilayer graphene sample as a function of time for two values of relative phases of the first and second harmonics $\Delta \phi$.
The fundamental beam wavelength used in this experiment is 4.8~$\mu$m. 
After~\protect \cite{sun2010}. 
} \label{sipe}
\end{figure}

\begin{figure}[ht]
\includegraphics[width=\linewidth]{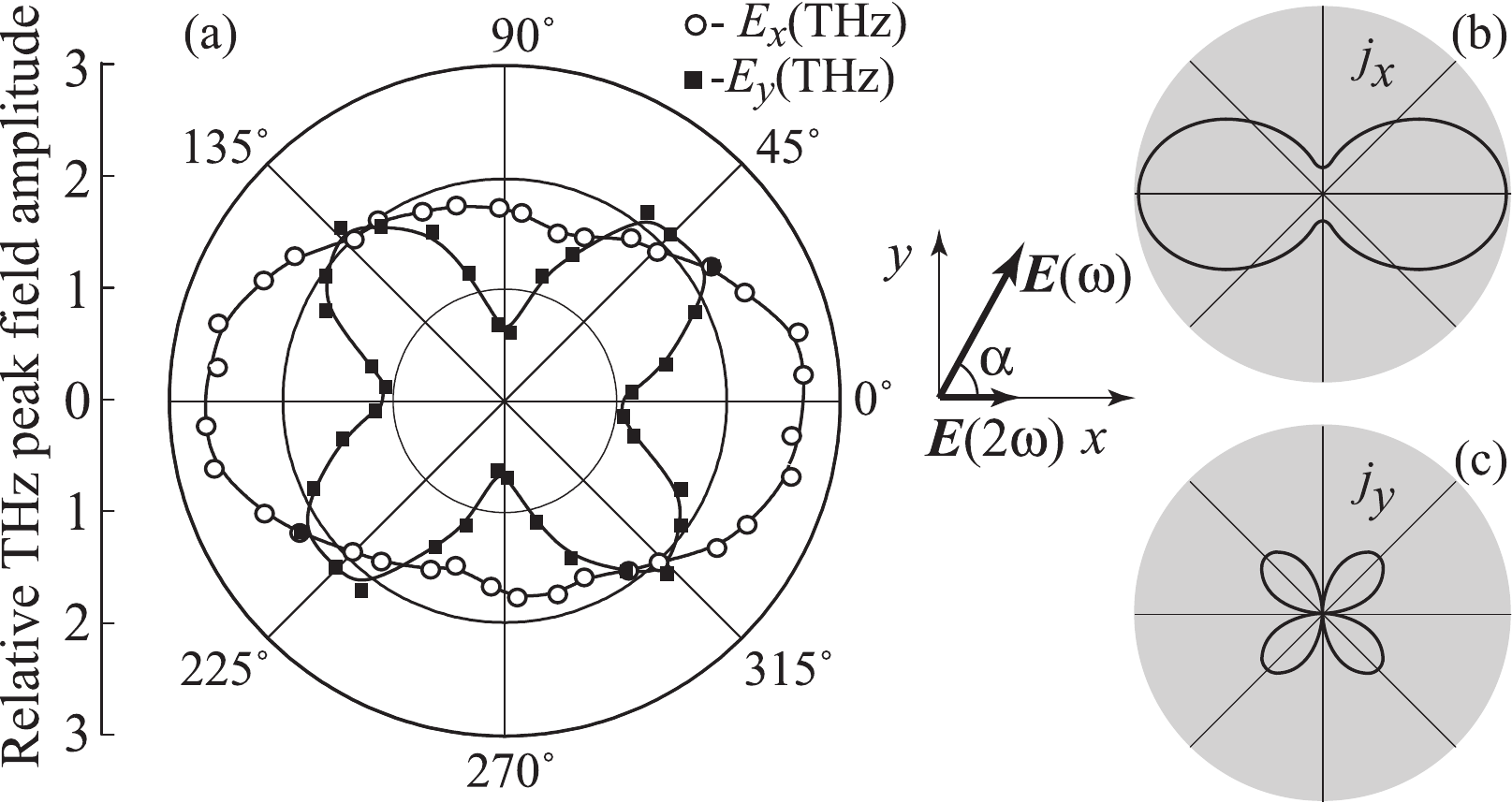}
\caption{(a) Experimentally measured $x$ and $y$ components of relative peak THz amplitude proportional to corresponding components of the photocurrent in the sample, Eq.~\eqref{THz:det}, as a function of the polarization angle between $\omega$ and $2\omega$ pulses. The field geometry is illustrated by the coordinate frame on the right, $\bm E(2\omega)\parallel x$. Theoretical dependence of photocurrent $j_x$ [panel (b)] and $j_y$ components [panel (c)] calculated for a 70~\% uncoupled-layer and 30~\% coupled-layer. The boundary of the shaded circle represent unit amplitude.
After~\protect \cite{Sun:2012ys}. 
} \label{sipe1}
\end{figure}

The  system used in these experiments consists of pulsed Ti:Sapphire laser,  optical parametric amplifier,  and a differential frequency generator yielding 200~fs infrared pulses
with intensity of GW/cm$^2$ range. In order to generate the coherent current, which requires two coherent beams at frequencies 
$\omega$ and $2\omega$, the beam of the laser operating at a fundamental frequency is 
split into two beams. One of those is directed to the sample, whereas the second is frequency 
doubled by the second harmonic generation process in the nonlinear crystal. As a result, the fields at the 
frequency $\omega$ and $2\omega$ are coherent and scale with radiation power 
as $E_{\omega} \propto P_\omega^{1/2}$ and $E_{2\omega} \propto P_{2\omega}^{1/2} \propto P_\omega$. As described above in Sec.~\ref{3rd:theory}, due to the interference of the two-photon transition with the frequency $\omega$ and a single photon transition with the frequency $2\omega$, the current is generated. This current induces the radiation of THz range, see Eq.~\eqref{THz:det}. 
An example of the power dependence of the emitted THz radiation for the fundamental beam wavelength 
of the 4.8~$\mu$m is shown in Fig. \ref{sipe}. As it is seen from Eq.~\eqref{THz:det}  terahertz radiation signal detected by the method of electro-optical sampling is associated with the two
color current injection process and scales with the pump power 
as $|E_{THz}| \propto |j| \propto P_\omega P_{2\omega}^{1/2} \propto P_\omega^2$. The experiment data
in Fig. \ref{sipe} support the expected power dependence and
are consistent with a third order optical process. Another proof 
of the third order optical process comes from the studies of polarization 
dependence of the relative THz peaks amplitudes carried out in Ref.~\cite{Sun:2012ys}. 
This is shown in Fig. \ref{sipe1}(a) where the dependence of the THz amplitudes, and, 
correspondingly, amplitudes of the photocurrent $\bm j$ on the angle between the polarization directions of $\omega$ and $2\omega$ pulses, see Eq.~\eqref{THz:det}. 
The data show that neither model of single layer graphene, nor that of a bilayer graphene describes experimental data. The 
results are in agreement with theoretical model~\cite{Sun:2012ys} where the mixture of 70\% of 
uncoupled layers and 30\% of bilayers was assumed, demonstrating that the interlayer coupling 
modifies the polarization dependence of coherently controlled currents, as shown in Fig.~\ref{sipe1}(b), (c). 
This work demonstrates that (i) nonlinear electric transport can be studied on  a femtosecond time scale and (ii) the photocurrents can be studied without 
necessarily to fabricate contacts to the graphene layer. Both advantages 
provide a unique access to dynamic of the nonlinear phenomena as well as 
allows one to characterize graphene layers in a 
contactless way.

\section{Conclusions and outlook}\label{concl}

The physics of nonlinear phenomena in graphene, although being young, has already resulted in a great variety of fascinating effects outlined here.  Moreover, the field of nonlinear transport and optical phenomena in graphene opens new prospects for further studies. Many of the effects addressed so far are not yet fully understood and await novel experimental and theoretical approaches and detailed studies. Some of the theoretical predictions discussed here demand an experimental verification. The new horizons appear related with tailoring of the nonlinear response of the material by external magnetic field, strain or artificial combinations of graphene layers with other materials. Similar effects await to be studied in details in the systems with akin atomic arrangement or band structure, like Boron nitride (BN), Molybdenum disulfide (MoS$_2$) and various kinds of topological insulators, for which first results on high frequency nonlinear transport have already been published~\cite{TI1,TI2,TI3}. Finally, we anticipate, that such effects in graphene will soon find their applications both for material characterization and development of graphene-based nonlinear devices.

\acknowledgements

We are grateful to S.A. Mikhailov, V.V. Bel'kov, L.E. Golub, E.L. Ivchenko, V.A. Shalygin, S.A. Tarasenko for valuable discussions.

This work was supported by DFG  (SPP~1459 and GRK~1570),
Linkage Grant of IB of BMBF at DLR,
RFBR and RF President Grant NSh-5442.2012.2.


\begin{thebibliography}{160}
\expandafter\ifx\csname natexlab\endcsname\relax\def\natexlab#1{#1}\fi
\expandafter\ifx\csname bibnamefont\endcsname\relax
  \def\bibnamefont#1{#1}\fi
\expandafter\ifx\csname bibfnamefont\endcsname\relax
  \def\bibfnamefont#1{#1}\fi
\expandafter\ifx\csname citenamefont\endcsname\relax
  \def\citenamefont#1{#1}\fi
\expandafter\ifx\csname url\endcsname\relax
  \def\url#1{\texttt{#1}}\fi
\expandafter\ifx\csname urlprefix\endcsname\relax\def\urlprefix{URL }\fi
\providecommand{\bibinfo}[2]{#2}
\providecommand{\eprint}[2][]{\url{#2}}

\bibitem[{\citenamefont{Novoselov et~al.}(2004)\citenamefont{Novoselov, Geim,
  Morozov, Jiang, Zhang, Dubonos, Grigorieva, and Firsov}}]{Novoselov04}
\bibinfo{author}{\bibfnamefont{K.~S.} \bibnamefont{Novoselov}},
  \bibinfo{author}{\bibfnamefont{A.~K.} \bibnamefont{Geim}},
  \bibinfo{author}{\bibfnamefont{S.~V.} \bibnamefont{Morozov}},
  \bibinfo{author}{\bibfnamefont{D.}~\bibnamefont{Jiang}},
  \bibinfo{author}{\bibfnamefont{Y.}~\bibnamefont{Zhang}},
  \bibinfo{author}{\bibfnamefont{S.~V.} \bibnamefont{Dubonos}},
  \bibinfo{author}{\bibfnamefont{I.~V.} \bibnamefont{Grigorieva}},
  \bibnamefont{and} \bibinfo{author}{\bibfnamefont{A.~A.}
  \bibnamefont{Firsov}}, {Electric field effect in atomically thin carbon films}, \bibinfo{journal}{Science}
  \textbf{\bibinfo{volume}{306}}, \bibinfo{pages}{666} (\bibinfo{year}{2004}).

\bibitem[{\citenamefont{Novoselov et~al.}(2005)\citenamefont{Novoselov, Geim,
  Morozov, Jiang, Katsnelson, Grigorieva, Dubonos, and Firsov}}]{Novoselov05}
\bibinfo{author}{\bibfnamefont{K.~S.} \bibnamefont{Novoselov}},
  \bibinfo{author}{\bibfnamefont{A.~K.} \bibnamefont{Geim}},
  \bibinfo{author}{\bibfnamefont{S.~V.} \bibnamefont{Morozov}},
  \bibinfo{author}{\bibfnamefont{D.}~\bibnamefont{Jiang}},
  \bibinfo{author}{\bibfnamefont{M.~I.} \bibnamefont{Katsnelson}},
  \bibinfo{author}{\bibfnamefont{I.~V.} \bibnamefont{Grigorieva}},
  \bibinfo{author}{\bibfnamefont{S.~V.} \bibnamefont{Dubonos}},
  \bibnamefont{and} \bibinfo{author}{\bibfnamefont{A.~A.}
  \bibnamefont{Firsov}}, {Two-dimensional gas of massless Dirac fermions in graphene}, \bibinfo{journal}{Nature}
  \textbf{\bibinfo{volume}{438}}, \bibinfo{pages}{197} (\bibinfo{year}{2005}).

\bibitem[{\citenamefont{Zhang et~al.}(2005)\citenamefont{Zhang, Tan, Stormer,
  and Kim}}]{Zhang05}
\bibinfo{author}{\bibfnamefont{Y.}~\bibnamefont{Zhang}},
  \bibinfo{author}{\bibfnamefont{Y.-W.} \bibnamefont{Tan}},
  \bibinfo{author}{\bibfnamefont{H.~L.} \bibnamefont{Stormer}},
  \bibnamefont{and} \bibinfo{author}{\bibfnamefont{P.}~\bibnamefont{Kim}}, {Experimental observation of the quantum Hall effect and Berry's phase in graphene},
  \bibinfo{journal}{Nature} \textbf{\bibinfo{volume}{438}},
  \bibinfo{pages}{201} (\bibinfo{year}{2005}).

\bibitem[{\citenamefont{Geim and Novoselov}(2007)}]{Geim07a}
\bibinfo{author}{\bibfnamefont{A.~K.} \bibnamefont{Geim}} \bibnamefont{and}
  \bibinfo{author}{\bibfnamefont{K.~S.} \bibnamefont{Novoselov}}, {The rise of graphene}, 
  \bibinfo{journal}{Nature Materials} \textbf{\bibinfo{volume}{6}},
  \bibinfo{pages}{183} (\bibinfo{year}{2007}).

\bibitem[{\citenamefont{Castro~Neto et~al.}(2009)\citenamefont{Castro~Neto,
  Guinea, Peres, Novoselov, and Geim}}]{Neto09}
\bibinfo{author}{\bibfnamefont{A.~H.} \bibnamefont{Castro~Neto}},
  \bibinfo{author}{\bibfnamefont{F.}~\bibnamefont{Guinea}},
  \bibinfo{author}{\bibfnamefont{N.~M.~R.} \bibnamefont{Peres}},
  \bibinfo{author}{\bibfnamefont{K.~S.} \bibnamefont{Novoselov}},
  \bibnamefont{and} \bibinfo{author}{\bibfnamefont{A.~K.} \bibnamefont{Geim}},
  The electronic properties of graphene, \bibinfo{journal}{Rev. Mod. Phys.} \textbf{\bibinfo{volume}{81}},
  \bibinfo{pages}{109} (\bibinfo{year}{2009}).
 

\bibitem[{\citenamefont{Wallace}(1947)}]{Wallace47}
\bibinfo{author}{\bibfnamefont{P.~R.} \bibnamefont{Wallace}}, The band theory of graphite, 
  \bibinfo{journal}{Phys. Rev.} \textbf{\bibinfo{volume}{71}},
  \bibinfo{pages}{622} (\bibinfo{year}{1947}).

\bibitem[{\citenamefont{McClure}(1956)}]{McClure56}
\bibinfo{author}{\bibfnamefont{J.~W.} \bibnamefont{McClure}}, Diamagnetism of graphite,
  \bibinfo{journal}{Phys. Rev.} \textbf{\bibinfo{volume}{104}},
  \bibinfo{pages}{666} (\bibinfo{year}{1956}).

\bibitem[{\citenamefont{Slonczewski and Weiss}(1958)}]{Slonczewski58}
\bibinfo{author}{\bibfnamefont{J.~C.} \bibnamefont{Slonczewski}}
  \bibnamefont{and} \bibinfo{author}{\bibfnamefont{P.~R.} \bibnamefont{Weiss}}, Band structure of graphite,
  \bibinfo{journal}{Phys. Rev.} \textbf{\bibinfo{volume}{109}},
  \bibinfo{pages}{272} (\bibinfo{year}{1958}).
  

\bibitem[{\citenamefont{Novoselov et~al.}(2007)\citenamefont{Novoselov, Jiang,
  Zhang, Morozov, Stormer, Zeitler, Maan, Boebinger, Kim, and
  Geim}}]{Novoselov07}
\bibinfo{author}{\bibfnamefont{K.~S.} \bibnamefont{Novoselov}},
  \bibinfo{author}{\bibfnamefont{Z.}~\bibnamefont{Jiang}},
  \bibinfo{author}{\bibfnamefont{Y.}~\bibnamefont{Zhang}},
  \bibinfo{author}{\bibfnamefont{S.~V.} \bibnamefont{Morozov}},
  \bibinfo{author}{\bibfnamefont{H.~L.} \bibnamefont{Stormer}},
  \bibinfo{author}{\bibfnamefont{U.}~\bibnamefont{Zeitler}},
  \bibinfo{author}{\bibfnamefont{J.~C.} \bibnamefont{Maan}},
  \bibinfo{author}{\bibfnamefont{G.~S.} \bibnamefont{Boebinger}},
  \bibinfo{author}{\bibfnamefont{P.}~\bibnamefont{Kim}}, \bibnamefont{and}
  \bibinfo{author}{\bibfnamefont{A.~K.} \bibnamefont{Geim}}, Room-temperature quantum Hall effect in graphene,
  \bibinfo{journal}{Science} \textbf{\bibinfo{volume}{315}},
  \bibinfo{pages}{1379} (\bibinfo{year}{2007}).  
  
\bibitem{Bib:McCann} E.~McCann, K.~Kechedzhi, V.~I.~Fal'ko, H.~Suzuura, T.~Ando, and B.~L.~Altshuler, Weak-localization magnetoresistance and valley symmetry in graphene,
{Phys. Rev. Lett.}~\textbf{97}, 146805 (2006).

\bibitem{Bib:Tikhonenko} F.~V.~ Tikhonenko, D.~W.~Horsell, R.~V.~Gorbachev, and A.~K.~Savchenko, Weak localization in graphene flakes,
{Phys. Rev. Lett.}~\textbf{100}, 056802 (2008).

\bibitem[{\citenamefont{Katsnelson}(2006)}]{Katsnelson06}
\bibinfo{author}{\bibfnamefont{M.~I.} \bibnamefont{Katsnelson}}, Zitterbewegung, chirality, and minimal conductivity in graphene,
  \bibinfo{journal}{Europ. Phys. J. B} \textbf{\bibinfo{volume}{51}},
  \bibinfo{pages}{157} (\bibinfo{year}{2006}).

\bibitem[{\citenamefont{Nomura and MacDonald}(2007)}]{Nomura07}
\bibinfo{author}{\bibfnamefont{K.}~\bibnamefont{Nomura}} \bibnamefont{and}
  \bibinfo{author}{\bibfnamefont{A.~H.} \bibnamefont{MacDonald}}, Quantum transport of massless Dirac fermions,
  \bibinfo{journal}{Phys. Rev. Lett.} \textbf{\bibinfo{volume}{98}},
  \bibinfo{pages}{076602} (\bibinfo{year}{2007}).

\bibitem[{\citenamefont{Tan et~al.}(2007)\citenamefont{Tan, Zhang, Bolotin,
  Zhao, Adam, Hwang, Das~Sarma, Stormer, and Kim}}]{Tan07}
\bibinfo{author}{\bibfnamefont{Y.-W.} \bibnamefont{Tan}},
  \bibinfo{author}{\bibfnamefont{Y.}~\bibnamefont{Zhang}},
  \bibinfo{author}{\bibfnamefont{K.}~\bibnamefont{Bolotin}},
  \bibinfo{author}{\bibfnamefont{Y.}~\bibnamefont{Zhao}},
  \bibinfo{author}{\bibfnamefont{S.}~\bibnamefont{Adam}},
  \bibinfo{author}{\bibfnamefont{E.~H.} \bibnamefont{Hwang}},
  \bibinfo{author}{\bibfnamefont{S.}~\bibnamefont{Das~Sarma}},
  \bibinfo{author}{\bibfnamefont{H.~L.} \bibnamefont{Stormer}},
  \bibnamefont{and} \bibinfo{author}{\bibfnamefont{P.}~\bibnamefont{Kim}}, Measurement of scattering rate and minimum conductivity in graphene,
  \bibinfo{journal}{Phys. Rev. Lett.} \textbf{\bibinfo{volume}{99}},
  \bibinfo{pages}{246803} (\bibinfo{year}{2007}).
  



\bibitem[{\citenamefont{Stander et~al.}(2009)\citenamefont{Stander, Huard, and
  Goldhaber-Gordon}}]{Stander09}
\bibinfo{author}{\bibfnamefont{N.}~\bibnamefont{Stander}},
  \bibinfo{author}{\bibfnamefont{B.}~\bibnamefont{Huard}}, \bibnamefont{and}
  \bibinfo{author}{\bibfnamefont{D.}~\bibnamefont{Goldhaber-Gordon}}, Evidence for Klein tunneling in graphene p-n junctions,
  \bibinfo{journal}{Phys. Rev. Lett.} \textbf{\bibinfo{volume}{102}},
  \bibinfo{pages}{026807} (\bibinfo{year}{2009}).

\bibitem[{\citenamefont{Young and Kim}(2009)}]{Young09}
\bibinfo{author}{\bibfnamefont{A.~F.} \bibnamefont{Young}} \bibnamefont{and}
  \bibinfo{author}{\bibfnamefont{P.}~\bibnamefont{Kim}}, Quantum interference and carrier collimation in graphene heterojunctions,
  \bibinfo{journal}{Nature Physics} \textbf{\bibinfo{volume}{5}},
  \bibinfo{pages}{222} (\bibinfo{year}{2009}).  

\bibitem{mccann2013} E. McCann and M. Koshino, The electronic properties of bilayer graphene (review), {\em Arxiv e-prints:} 1205.6953 (2012).  

\bibitem[{\citenamefont{Semenoff}(1984)}]{Semenoff84}
\bibinfo{author}{\bibfnamefont{G.~W.} \bibnamefont{Semenoff}}, Condensed-matter simulation of a three-dimensional anomaly,
  \bibinfo{journal}{Phys. Rev. Lett.} \textbf{\bibinfo{volume}{53}},
  \bibinfo{pages}{2449} (\bibinfo{year}{1984}).

\bibitem[{\citenamefont{Haldane}(1988)}]{Haldane88}
\bibinfo{author}{\bibfnamefont{F.~D.~M.} \bibnamefont{Haldane}}, Model for a quantum Hall effect without Landau levels: Condensed-matter realization of the ``parity anomaly'',
  \bibinfo{journal}{Phys. Rev. Lett.} \textbf{\bibinfo{volume}{61}},
  \bibinfo{pages}{2015} (\bibinfo{year}{1988}).



\bibitem{Morozov:2008eng}
S.~V. Morozov, K.~S. Novoselov, A.~K. Geim.
\newblock {Electronic transport in graphene}.
\newblock Physics-Uspekhi \textbf{51}, 744 (2008).

\bibitem{Lozovik:2008eng}
Y.~E. Lozovik, S.~P. Merkulova, A.~A. Sokolik.
\newblock {Collective electron phenomena in graphene}.
\newblock Physics-Uspekhi \textbf{51}, 727 (2008).







\bibitem{Bib:Valley}
A.~Rycerz, J.~Tworzydlo, C.~W.~J. Beenakker.
\newblock {Valley filter and valley valve in graphene}.
\newblock Nature Physics \textbf{3}, 172 (2007).



\bibitem{Falkovsky:2008eng}
L.~A. Falkovsky.
\newblock {Optical properties of graphene and IV - VI semiconductors}.
\newblock Physics-Uspekhi \textbf{51}, 887 (2008).

\bibitem{RevModPhys.82.2673}
N.~M.~R. Peres.
\newblock {Colloquium: The transport properties of graphene: An
  introduction}.
\newblock Rev. Mod. Phys. \textbf{82}, 2673 (2010).

\bibitem{RevModPhys.83.407}
S.~Das~Sarma, S.~Adam, E.~H. Hwang, E.~Rossi.
\newblock {Electronic transport in two-dimensional graphene}.
\newblock Rev. Mod. Phys. \textbf{83}, 407 (2011).



\bibitem[{\citenamefont{Bonaccorso et~al.}(2010)\citenamefont{Bonaccorso, Sun,
  Hasan, and Ferrari}}]{Bonaccorso10}
\bibinfo{author}{\bibfnamefont{F.}~\bibnamefont{Bonaccorso}},
  \bibinfo{author}{\bibfnamefont{Z.}~\bibnamefont{Sun}},
  \bibinfo{author}{\bibfnamefont{T.}~\bibnamefont{Hasan}}, \bibnamefont{and}
  \bibinfo{author}{\bibfnamefont{A.~C.} \bibnamefont{Ferrari}}, Graphene photonics and optoelectronics,
  \bibinfo{journal}{Nature Photonics} \textbf{\bibinfo{volume}{4}},
  \bibinfo{pages}{611} (\bibinfo{year}{2010}).


\bibitem{blombergen}
N.~Blombergen.
\newblock \emph{Nonlinear optics} (Benjamin, New York, 1965).

\bibitem{Boyd} R.W. Boyd, \emph{Nonlinear optics} (Academic Press, San Diego, 1993).

\bibitem{yariv} A. Yariv, P. Yeh, \emph{Optical waves in crystals: Propagation and control of laser radiation} (J. Wiley \& Sons, New York, 2003).

\bibitem{wegener}
M.~Wegener, \emph{Extreme nonlinear optics: An introduction} (Springer, Berlin, 2005).

\bibitem{shen}
Y.R.~Shen, \emph{The principles of nonlinear optics} (John Wiley \& Sons, New York,  2003).

  
\bibitem{sturmanBOOK}
B.~Sturman, V.~Fridkin.
\newblock \emph{The photovoltaic and photorefractive effects in
  non-centrosymmetric materials}  (Gordon \& Breach, Philadelphia, 1992).  

\bibitem{ivchenkopikus}
E.~L. Ivchenko, G.~E. Pikus.
\newblock \emph{Superlattices and other heterostructures} (Springer, 1997).

\bibitem{ivchenko05a}
E.~L. Ivchenko.
\newblock \emph{Optical Spectroscopy of Semiconductor Nanostructures} (Alpha
  Science, Harrow UK, 2005).

\bibitem{ganichev_book}
S.~Ganichev, W.~Prettl.
\newblock \emph{Intense Terahertz Excitation of Semiconductors} (Oxford University Press, 2006).




\bibitem{ivch_spi}
E.~L. Ivchenko, B.~Spivak.
\newblock {Chirality effects in carbon nanotubes}.
\newblock Phys. Rev. B \textbf{66}, 155404 (2002).

\bibitem{obraztsov:231112}
A.~N. Obraztsov, D.~A. Lyashenko, S.~Fang, R.~H. Baughman, P.~A. Obraztsov,
  S.~V. Garnov, Y.~P. Svirko.
\newblock {Photon drag effect in carbon nanotube yarns}.
\newblock Appl. Phys. Lett. \textbf{94}, 231112 (2009).

\bibitem{ISI:000072756100009}
V.~Margulis, T.~Sizikova.
\newblock {Theoretical study of third-order nonlinear optical response of
  semiconductor carbon nanotubes}.
\newblock Physica B \textbf{245}, 173 (1998).

\bibitem{PhysRevA.63.053808}
G.~Y. Slepyan, S.~A. Maksimenko, V.~P. Kalosha, A.~V. Gusakov, J.~Herrmann.
\newblock {High-order harmonic generation by conduction electrons in
  carbon nanotube ropes}.
\newblock Phys. Rev. A \textbf{63}, 053808 (2001).

\bibitem{Mele:2000oq}
E.~J. Mele, P.~Kr\'al, and D.~Tom\'anek.
{Coherent control of photocurrents in graphene and carbon nanotubes}.
\newblock { Phys. Rev. B} \textbf{61}, 7669 (2000).

\bibitem{Kral:2000nx}
P.~Kr\'al, E.~J. Mele, and D.~Tom\'anek.
\newblock {Photogalvanic effects in heteropolar nanotubes}.
\newblock {Phys. Rev. Lett.} {\bf 85}, 1512 (2000).

\bibitem{mikheev:4854}
G.~M. Mikheev, R.~G. Zonov, A.~N. Obraztsov, and Y.~P. Svirko,
{Giant optical rectification effect in nanocarbon films},
Appl. Phys. Lett. {\bf 84}, 4854 (2004).

\bibitem{doi:10.1021/nl203003p}
G.~M. Mikheev, A.~G. Nasibulin, R.~G. Zonov, A.~Kaskela, and E.~I. Kauppinen,  {Photon-drag effect in single-walled carbon nanotube films}, Nano Letters {\bf 12}, 77 (2012).





\bibitem{cnt:book} {A. Jorio, G. Dresselhaus, M. S. Dresselhaus, eds. \emph{Advanced Topics in the Synthesis, Structure, Properties and Applications}, (Springer, 2008).}

\bibitem{Liu}
Z.~Liu, X.~Zhang, X.~Yan, Y.~Chen, and J.~Tian.
\newblock {Nonlinear optical properties of graphene-based materials},
\newblock {Chinese Science Bulletin} {\bf 57}, 2971 (2012).
\bibitem{Wang} Wang Jun, Chen Yu, Li Rihong, Dong Hongxing, Zhang Long, Lotya Mustafa, N. Coleman Jonathan and J. Blau Werner, \emph{Nonlinear Optical Properties of Graphene and Carbon Nanotube Composites}, in Carbon Nanotubes - Synthesis, Characterization, Applications, Ed. Siva Yellampalli (2011).


\bibitem{0295-5075-79-2-27002}
S.~A. Mikhailov.
\newblock {Nonlinear electromagnetic response of graphene}.
\newblock EPL \textbf{79}, 27002 (2007).



\bibitem{10.1063/1.3483872}
M.~Dragoman, D.~Neculoiu, G.~Deligeorgis, G.~Konstantinidis, D.~Dragoman,
  A.~Cismaru, A.~A. Muller, R.~Plana.
\newblock {Millimeter-wave generation via frequency multiplication in
  graphene}, Appl. Phys. Lett. \textbf{97}, 093101 (2010).
  
  
\bibitem{dean:261910}
J.~J. Dean, H.~M. van Driel.
\newblock {Second harmonic generation from graphene and graphitic films}.
\newblock Appl. Phys. Lett. \textbf{95}, 261910 (2009).

\bibitem{PhysRevB.82.125411}
J.~J. Dean, H.~M. van Driel.
\newblock {Graphene and few-layer graphite probed by second-harmonic
  generation: Theory and experiment}.
\newblock Phys. Rev. B \textbf{82}, 125411 (2010).

\bibitem{murzina} A. Y. Bykov, T. V. Murzina, M. G. Rybin, and E. D. Obraztsova, {Second harmonic generation in multilayer graphene induced by direct electric current}, Phys. Rev. B {\bf 85}, 121413 (2012).

\bibitem[{\citenamefont{Hotopan et~al.}(2011)\citenamefont{Hotopan, {Ver
  Hoeye}, Vazquez, Camblor, Fern\'andez, Las~Heras, \'Alvarez, and
  Men\'endez}}]{Hotopan11}
\bibinfo{author}{\bibfnamefont{G.}~\bibnamefont{Hotopan}},
  \bibinfo{author}{\bibfnamefont{S.}~\bibnamefont{{Ver Hoeye}}},
  \bibinfo{author}{\bibfnamefont{C.}~\bibnamefont{Vazquez}},
  \bibinfo{author}{\bibfnamefont{R.}~\bibnamefont{Camblor}},
  \bibinfo{author}{\bibfnamefont{M.}~\bibnamefont{Fern\'andez}},
  \bibinfo{author}{\bibfnamefont{F.}~\bibnamefont{Las~Heras}},
  \bibinfo{author}{\bibfnamefont{P.}~\bibnamefont{\'Alvarez}},
  \bibnamefont{and}
  \bibinfo{author}{\bibfnamefont{R.}~\bibnamefont{Men\'endez}},
  Millimeter wave microstrip mixer based on graphene,
  \bibinfo{journal}{Progress In Electromagnetic Research}
  \textbf{\bibinfo{volume}{118}}, \bibinfo{pages}{57} (\bibinfo{year}{2011}).


\bibitem{2013arXiv1301.1042K}
N.~{Kumar}, J.~{Kumar}, C.~{Gerstenkorn}, R.~{Wang}, H.-Y. {Chiu}, A.~L.
  {Smirl}, and H.~{Zhao}.
\newblock {Third harmonic generation in graphene and few-layer graphite films}.
\newblock {\em ArXiv e-prints:} 1301.1042 (2013).

\bibitem{2013arXiv1301.1697H}
S.-Y. {Hong}, J.~I. {Dadap}, N.~{Petrone}, P.-C. {Yeh}, J.~{Hone}, and R.~M.
  {Osgood}, Jr.
\newblock {Optical Third-harmonic generation in graphene}.
\newblock {\em ArXiv e-prints:} 1301.1697 (2013).

\bibitem[{\citenamefont{Hendry et~al.}(2010)\citenamefont{Hendry, Hale, Moger,
  Savchenko, and Mikhailov}}]{Hendry10}
\bibinfo{author}{\bibfnamefont{E.}~\bibnamefont{Hendry}},
  \bibinfo{author}{\bibfnamefont{P.~J.} \bibnamefont{Hale}},
  \bibinfo{author}{\bibfnamefont{J.~J.} \bibnamefont{Moger}},
  \bibinfo{author}{\bibfnamefont{A.~K.} \bibnamefont{Savchenko}},
  \bibnamefont{and} \bibinfo{author}{\bibfnamefont{S.~A.}
  \bibnamefont{Mikhailov}}, 
  Coherent nonlinear optical response of graphene,
  \bibinfo{journal}{Phys. Rev. Lett.}
  \textbf{\bibinfo{volume}{105}}, \bibinfo{pages}{097401}
  (\bibinfo{year}{2010}).


\bibitem{2011rangel} N. L. Rangel, A. Gimenez, A. Sinitskii, and J. M. Seminario, {Graphene signal mixer for sensing applications}, J. Phys. Chem. C {\bf 115}, 12128 (2011). 

\bibitem{GuT.:2012fk}
T.~Gu, N.~Petrone, J.~F. McMillan, A.~van~der Zande, M.~Yu, G.~Q. Lo, D.~L.
  Kwong, J.~Hone, and C.~W. Wong.
\newblock Regenerative oscillation and four-wave mixing in graphene
  optoelectronics.
\newblock {Nature Photonics} {\bf 6}, 554 (2012).



\bibitem{karch2010}
J.~Karch, P.~Olbrich, M.~Schmalzbauer, C.~Zoth, C.~Brinsteiner,
  M.~Fehrenbacher, U.~Wurstbauer, M.~M. Glazov, S.~A. Tarasenko, E.~L.
  Ivchenko, D.~Weiss, J.~Eroms, R.~Yakimova, S.~Lara-Avila, S.~Kubatkin, S.~D.
  Ganichev.
\newblock {Dynamic Hall effect driven by circularly polarized light in a
  graphene layer}.
\newblock Phys. Rev. Lett. \textbf{105}, 227402 (2010).

\bibitem{2010arXiv1002.1047K}
J.~{Karch}, P.~{Olbrich}, M.~{Schmalzbauer}, C.~{Brinsteiner}, U.~{Wurstbauer},
  M.~M. {Glazov}, S.~A. {Tarasenko}, E.~L. {Ivchenko}, D.~{Weiss}, J.~{Eroms},
  S.~D. {Ganichev}.
\newblock {{Photon helicity driven electric currents in graphene}}.
\newblock \emph{ArXiv e-prints:} 1002.1047  (2010).

\bibitem{entin10}
M.~V. Entin, L.~I. Magarill, D.~L. Shepelyansky.
\newblock {Theory of resonant photon drag in monolayer graphene}.
\newblock Phys. Rev. B \textbf{81}, 165441 (2010).

\bibitem{edge}
J.~Karch, C.~Drexler, P.~Olbrich, M.~Fehrenbacher, M.~Hirmer, M.~M. Glazov,
  S.~A. Tarasenko, E.~L. Ivchenko, B.~Birkner, J.~Eroms, D.~Weiss, R.~Yakimova,
  S.~Lara-Avila, S.~Kubatkin, M.~Ostler, T.~Seyller, S.~D. Ganichev.
\newblock {Terahertz radiation driven chiral edge currents in graphene}.
\newblock Phys. Rev. Lett. \textbf{107}, 276601 (2011).

\bibitem{PRB2010}
C.~Jiang, V.~A. Shalygin, V.~Y. Panevin, S.~N. Danilov, M.~M. Glazov,
  R.~Yakimova, S.~Lara-Avila, S.~Kubatkin, S.~D. Ganichev.
\newblock {Helicity-dependent photocurrents in graphene layers excited by
  midinfrared radiation of a CO${}_{2}$ laser}.
\newblock Phys. Rev. B \textbf{84}, 125429 (2011).


\bibitem{sun2010}
D.~Sun, C.~Divin, J.~Rioux, J.~E. Sipe, C.~Berger, W.~A. de~Heer, P.~N. First,
  T.~B. Norris.
\newblock {Coherent control of ballistic photocurrents in multilayer
  epitaxial graphene using quantum interference}.
\newblock Nano Lett. \textbf{10}, 1293 (2010).



\bibitem{Sun:2012ys}
D.~Sun, J.~Rioux, J.~E. Sipe, Y.~Zou, M.~T. Mihnev, C.~Berger, W.~A. de~Heer,
  P.~N. First, and T.~B. Norris.
{Evidence for interlayer electronic coupling in multilayer epitaxial
  graphene from polarization-dependent coherently controlled photocurrent
  generation}, Phys. Rev. B {\bf 85}, 165427 (2012).

  
  \bibitem{2012winzent}
D.~Sun, C.~Divin, M.~Mihnev, T.~Winzer, E.~Malic, A.~Knorr, J.~E. Sipe,
  C.~Berger, W.~A. de~Heer, P.~N. First, and T.~B. Norris,
{Current relaxation due to hot carrier scattering in graphene},
\newblock {New Journal of Physics} {\bf 14}, 105012 (2012).

\bibitem{Prechtel:2012kx}
L.~Prechtel, L.~Song, D.~Schuh, P.~Ajayan, W.~Wegscheider, and A.~W.
  Holleitner.
\newblock Time-resolved ultrafast photocurrents and terahertz generation in
  freely suspended graphene.
\newblock {Nature Communications} {\bf 3}, 01 (2012).



\bibitem{Graham:2013uq}
M.~W. Graham, S.-F. Shi, D.~C. Ralph, J.~Park, and P.~L. McEuen.
\newblock Photocurrent measurements of supercollision cooling in graphene.
\newblock {Nature Physics} {\bf 9}, 103 (2013).


\bibitem{PhysRevB.78.045407}
S.~V. Syzranov, M.~V. Fistul, and K.~B. Efetov.
\newblock Effect of radiation on transport in graphene.
\newblock {Phys. Rev. B} {\bf 78}, 045407 (2008).
\bibitem{doi:10.1021/nl8033812}
F.~Xia, T.~Mueller, R.~Golizadeh-Mojarad, M.~Freitag, Y.-m. Lin, J.~Tsang,
  V.~Perebeinos, and P.~Avouris.
\newblock Photocurrent imaging and efficient photon detection in a graphene
  transistor.
\newblock {Nano Lett.} {\bf 9}, 1039 (2009).
\bibitem{Mai:2011oq}
S.~Mai, S.~V. Syzranov, and K.~B. Efetov.
\newblock Photocurrent in a visible-light graphene photodiode.
\newblock {Phys. Rev. B} {\bf 83}, 033402 (2011).


\bibitem{doi:10.1021/nl2023405}
R.~Wu, Y.~Zhang, S.~Yan, F.~Bian, W.~Wang, X.~Bai, X.~Lu, J.~Zhao, E.~Wang.
\newblock {Purely Coherent Nonlinear Optical Response in Solution
  Dispersions of Graphene Sheets}.
\newblock Nano Lett. \textbf{11}, 5159 (2011).


\bibitem{2012chu} S. Chu, S. Wang, Q. Gong, {Ultrafast third-order nonlinear optical properties of graphene in aqueous solution and polyvinyl alcohol film}, Chem. Phys. Lett. {\bf 523}, 104 (2012).

\bibitem{Yan} Xiao-Qing Yan, Zhi-Bo Liu, Jun Yao, Xin Zhao, Xu-Dong Chen, Fei Xing, Yongsheng Chen, Jian-Guo Tian, {Experimental observation of polarization-dependent ultrafast carrier dynamics in multi-layer graphene}, \emph{Arxiv e-prints:} 1301.1743  (2013).




\bibitem{PhysRevB.84.045432}
S.~A. Mikhailov.
\newblock {Theory of the giant plasmon-enhanced second-harmonic generation
  in graphene and semiconductor two-dimensional electron systems}.
\newblock Phys. Rev. B \textbf{84}, 045432 (2011).

\bibitem{novoselov_pl1} T. J. Echtermeyer, L. Britnell, P. K. Jasnos, A. Lombardo, R. V. Gorbachev, A. N. Grigorenko, A. K. Geim, A. C. Ferrari, and K. S. Novoselov, Strong plasmonic enhancement of photovoltage in graphene. Nature Communications {\bf 2}, 458, 08 (2011).

\bibitem{novoselov_pl2} A. N. Grigorenko, M. Polini, and K. S. Novoselov, Graphene plasmonics, Nature Photonics {\bf 6}, 749 (2012).

\bibitem[{\citenamefont{L\'{o}pez-Rodr\'{i}guez and Naumis}(2008)}]{Lopez08}
\bibinfo{author}{\bibfnamefont{F.~J.} \bibnamefont{L\'{o}pez-Rodr\'{i}guez}}
  \bibnamefont{and} \bibinfo{author}{\bibfnamefont{G.~G.}
  \bibnamefont{Naumis}}, {Analytic solution for electrons and holes in graphene under electromagnetic waves: Gap appearance and nonlinear effects} \bibinfo{journal}{Phys. Rev. B}
  \textbf{\bibinfo{volume}{78}}, \bibinfo{pages}{201406(R)}
  (\bibinfo{year}{2008}).

\bibitem{Saleh} B. E. A. Saleh, M. C. Teich, \emph{Fundamentals of Photonics} (John Wiley \& Sons, New York,  2003).

\bibitem{BornWolf} M. Born, E. Wolf, \emph{Principles of Optics: Electromagnetic Theory of Propagation, Interference and Diffraction of Light} (Cambridge University Press,  1999).




\bibitem{reviewG}
S.~D. Ganichev, W.~Prettl.
\newblock {Spin photocurrents in quantum wells}.
\newblock J. Phys.: Condens. Matter \textbf{15}, R935 (2003).





\bibitem{ivchenko_ganichev}
E.~Ivchenko, S.~Ganichev in 
\newblock \emph{Spin physics in semiconductors}, ed. M. Dyakonov (Springer, 2008).
  
  
\bibitem{Ch7Yaroshetskii80p173}
I.~D. Yaroshetskii, S.~M. Ryvkin, in
\newblock \emph{Semiconductor Physics}, (Cons. Bureau, New York,
  1986).

\bibitem{Ch7Gibson80p182}
A.~F. Gibson, M.~F. Kimmitt.
\newblock \emph{Infrared and Millimeter Waves, Vol. 3, Detection of Radiation},
  181--217 (Academic Press, New York, 1980).
  

\bibitem{BARLOW1954} H. M. Barlow, {Application of the Hall effect in a semi-conductor to the measurement of power in an electromagnetic field}, Nature {\bf 173}, 41 (1954).

\bibitem{Ivchenko1980}
E.~L. Ivchenko, G.~E. Pikus, in 
\newblock \emph{Semiconductor Physics} (Cons. Bureau, New York,
  1986).

\bibitem{belinicher_cpde}
V.~I. Belinicher.
\newblock {On the mechanisms underlying the circular drag effect}.
\newblock Sov. Phys. Solid State \textbf{23}, 2012 (1981).

\bibitem{Shalygin2007}
V.~Shalygin, H.~Diehl, C.~Hoffmann, S.~Danilov, T.~Herrle, S.~Tarasenko,
  D.~Schuh, C.~Gerl, W.~Wegscheider, W.~Prettl, S.~Ganichev.
\newblock {Spin photocurrents and the circular photon drag effect in
  (110)-grown quantum well structures}.
\newblock JETP Letters \textbf{84}, 570 (2007).

\bibitem{PhysRevLett.103.103906}
T.~Hatano, T.~Ishihara, S.~G. Tikhodeev, N.~A. Gippius,
\newblock {Transverse photovoltage induced by circularly polarized light},
\newblock Phys. Rev. Lett. \textbf{103}, 103906 (2009).

\bibitem{weber:GaN} W. Weber, L. E. Golub, S. N. Danilov, J. Karch, C. Reitmaier, B. Wittmann, V. V. Bel'kov, E. L. Ivchenko, Z. D. Kvon, N. Q. Vinh, A. F. G. van der Meer, B. Murdin, and S. D. Ganichev, Quantum ratchet effects induced by terahertz radiation in GaN-based two-dimensional structures, Phys. Rev. B {\bf 77}, 245304 (2008).

\bibitem{Kiselev11}
Y.~Y. Kiselev, L.~E. Golub.
\newblock {Optical and photogalvanic properties of graphene superlattices
  formed by periodic strain}.
\newblock Phys. Rev. B \textbf{84}, 235440 (2011).


\bibitem{Nalitov:2012vn}
{A.~V. Nalitov, L.~E. Golub, and E.~L. Ivchenko.
\newblock Ratchet effects in two-dimensional systems with a lateral periodic
  potential.
\newblock {Phys. Rev. B} {\bf 86} 115301 (2012).}




\bibitem{TaisukeOhta08182006}
T.~Ohta, A.~Bostwick, T.~Seyller, K.~Horn, and E.~Rotenberg.
\newblock {Controlling the electronic structure of bilayer graphene}.
\newblock {Science} {\bf 313}, 951 (2006).

\bibitem{castro:216802}
E.~V. Castro, K.~S. Novoselov, S.~V. Morozov, N.~M.~R. Peres, J.~M. B.~L. dos
  Santos, J.~Nilsson, F.~Guinea, A.~K. Geim, and A.~H.~C. Neto.
\newblock Biased bilayer graphene: Semiconductor with a gap tunable by the
  electric field effect.
\newblock {Phys. Rev. Lett.} {\bf 99}, 216802 (2007).

\bibitem{FengWang04112008}
F.~Wang, Y.~Zhang, C.~Tian, C.~Girit, A.~Zettl, M.~Crommie, and Y.~R. Shen.
\newblock {Gate-Variable Optical Transitions in Graphene}.
\newblock {Science} {\bf 320}, 206 (2008).

\bibitem{Mayorov12082011}
A.~S. Mayorov, D.~C. Elias, M.~Mucha-Kruczynski, R.~V. Gorbachev,
  T.~Tudorovskiy, A.~Zhukov, S.~V. Morozov, M.~I. Katsnelson, V.~I. Fal'ko,
  A.~K. Geim, and K.~S. Novoselov.
\newblock Interaction-driven spectrum reconstruction in bilayer graphene.
\newblock {Science} {\bf 333}, 860 (2011).

\bibitem{Bao2011}
W.~Bao, L.~Jing, J.~Velasco, Y.~Lee, G.~Liu, D.~Tran, B.~Standley, M.~Aykol,
  S.~B. Cronin, D.~Smirnov, M.~Koshino, E.~McCann, M.~Bockrath, and C.~N. Lau.
\newblock Stacking-dependent band gap and quantum transport in trilayer
  graphene.
\newblock {Nature Physics} {\bf 7}, 948 (2011).

\bibitem{Zhang2011}
L.~Zhang, Y.~Zhang, J.~Camacho, M.~Khodas, and I.~Zaliznyak.
\newblock The experimental observation of quantum {H}all effect of $l=3$ chiral
  quasiparticles in trilayer graphene.
\newblock {Nature Physics} {\bf 7}, 953 (2011).

\bibitem{Lui2011}
C.~H. Lui, Z.~Li, K.~F. Mak, E.~Cappelluti, and T.~F. Heinz.
\newblock Observation of an electrically tunable band gap in trilayer graphene.
\newblock {Nature Physics} {\bf 7}, 944 (2011).

\bibitem{PhysRevB.75.155424}
J.~L. Ma\~nes, F.~Guinea, M.~A.~H. Vozmediano.
\newblock {Existence and topological stability of Fermi points in
  multilayered graphene}.
\newblock Phys. Rev. B \textbf{75}, 155424 (2007).

\bibitem{PhysRevB.79.125426}
L.~M. Malard, M.~H.~D. Guimar\~aes, D.~L. Mafra, M.~S.~C. Mazzoni, A.~Jorio.
\newblock {Group-theory analysis of electrons and phonons in $N$ -layer
  graphene systems}.
\newblock Phys. Rev. B \textbf{79}, 125426 (2009).


\bibitem{valleyLPGE}
L.~E. Golub, S.~A. Tarasenko, M.~V. Entin, L.~I. Magarill.
\newblock {Valley separation in graphene by polarized light}.
\newblock Phys. Rev. B \textbf{84}, 195408 (2011).

\bibitem{portnoi_book}
R.~R. Hartmann, M.~E. Portnoi.
\newblock \emph{Optoelectronic Properties of Carbon-based Nanostructures:
  Steering electrons in graphene by electromagnetic fields} (LAP LAMBERT
  Academic Publishing, Saarbrucken, 2011).

\bibitem{glazov:shg}
M.~Glazov.
\newblock {Second harmonic generation in graphene}.
\newblock JETP Letters \textbf{93}, 366 (2011).

  \bibitem{PhysRev.138.A534}
M.~Bass, P.~A. Franken, J.~F. Ward.
\newblock {Optical Rectification}.
\newblock Phys. Rev. \textbf{138}, A534 (1965).



\bibitem{cote}
D.~C\^{o}t\'{e}, N.~Laman, H.~M. van Driel.
\newblock {Rectification and shift currents in GaAs}.
\newblock Applied Physics Letters \textbf{80}, 905 (2002).


\bibitem{legurevich67}
L.E.~Gurevich and A.A.~Rumyantsev.
\newblock {Theory of the photoelectric effect in finite crystals at high
  frequencies and in the presence of an external magnetic field},
\newblock {Sov. Phys. Solid State} {\bf 9}, 55 (1967).



\bibitem{perelpinskii73}
V.~I. Perel' and Ya.~M. Pinskii, Constant current in conducting media due to a high-frequency electron electromagnetic field,
\newblock {Sov. Phys. Solid State}, {\bf 15}, 688 (1973).



\bibitem{Tom:1983dq}
H.~W.~K. Tom, T.~F. Heinz, and Y.~R. Shen,
{Second-harmonic reflection from silicon surfaces and its relation to
  structural symmetry}, Phys. Rev. Lett. {\bf 51}, 1983--1986 (1983).



\bibitem{Nair06062008}
R.~R. Nair, P.~Blake, A.~N. Grigorenko, K.~S. Novoselov, T.~J. Booth,
  T.~Stauber, N.~M.~R. Peres, A.~K. Geim.
\newblock {Fine structure constant defines visual transparency of
  graphene}.
\newblock Science \textbf{320}, 1308 (2008).

\bibitem{PhysRevB.38.87}
A.~A.~Grinberg, S.~Luryi.
\newblock {Theory of the photon-drag effect in a two-dimensional electron
  gas}.
\newblock Phys. Rev. B \textbf{38}, 87 (1988).


\bibitem{suppllara2011} S.~Lara-Avila, K. Moth-Poulsen, R. Yakimova, T. Bjornholm, V. Fal'ko, A. Tzalenchuk, S. Kubatkin,
Non-volatile photochemical gating of an epitaxial graphene.
\textit{Advanced Materials} \textbf{23}, 878 (2011).



\bibitem{DrexlerC.:2013uq} C. Drexler,
S. A. Tarasenko, P. Olbrich, J. Karch, M. Hirmer, F. Muller, M. Gmitra, J. Fabian, R. Yakimova, S. Lara-Avila, S. Kubatkin, M. Wang, R. Vajtai, P.M. Ajayan, J. Kono, and S. D. Ganichev
 {Magnetic quantum ratchet effect in graphene},
 {Nature Nanotechnology} {\bf 8}, 104 (2013).




\bibitem{Bassani} F. Bassani and G. Pastori-Parravicini, \emph{Electronic states and optical transitions in solids} (Oxford, New York, Pergamon Press 1975).


\bibitem{PhysRevB.17.626}
A.~Zunger.
\newblock {Self-consistent LCAO calculation of the electronic properties
  of graphite. I. The regular graphite lattice}.
\newblock Phys. Rev. B \textbf{17}, 626 (1978).

\bibitem{Tarasenko2007}
S.~Tarasenko.
\newblock {Orbital mechanism of the circular photogalvanic effect in
  quantum wells}.
\newblock JETP Letters \textbf{85}, 182 (2007).

\bibitem{PhysRevB.79.121302}
P.~Olbrich, S.~A. Tarasenko, C.~Reitmaier, J.~Karch, D.~Plohmann, Z.~D. Kvon,
  S.~D. Ganichev.
\newblock {Observation of the orbital circular photogalvanic effect}.
\newblock Phys. Rev. B \textbf{79}, 121302 (2009).

\bibitem{tarasenko11}
S.~A. Tarasenko.
\newblock {Direct current driven by ac electric field in quantum wells}.
\newblock Phys. Rev. B \textbf{83}, 035313 (2011).





\bibitem{PhysRevB.48.8307}
V.~L. Gurevich and R.~Laiho.
\newblock Photomagnetism of metals: Microscopic theory of the photoinduced
  surface current.
\newblock {Phys. Rev. B} {\bf 48}, 8307 (1993).

\bibitem{Gurevich2000}
V.~Gurevich and R.~Laiho.
\newblock Photomagnetism of metals. First observation of dependence on
  polarization of light.
\newblock {Physics of the Solid State} {\bf 42}, 1807 (2000).



\bibitem{magarill81}
L.~Magarill, M.~Entin.
\newblock {Surface photogalvanic effect in metals}.
\newblock JETP \textbf{54}, 531 (1981).

\bibitem{alperovich80eng}
V.~L. Al'perovich, V.~I. Belinicher, V.~N. Novikov, A.~S. Terekhov.
\newblock {Surface photovoltaic effect in gallium arsenide}.
\newblock JETP Lett. \textbf{31}, 546 (1980).

\bibitem{Bolotin2008351}
K.~Bolotin, K.~Sikes, Z.~Jiang, M.~Klima, G.~Fudenberg, J.~Hone, P.~Kim, and
  H.~Stormer.
\newblock {Ultrahigh electron mobility in suspended graphene}.
\newblock {Solid State Communications} {\bf 146}, 351 (2008).



\bibitem{doi:10.1021/nl8032697}
C.~Casiraghi, A.~Hartschuh, H.~Qian, S.~Piscanec, C.~Georgi, A.~Fasoli, K.~S.
  Novoselov, D.~M. Basko, and A.~C. Ferrari.
\newblock Raman spectroscopy of graphene edges.
\newblock {Nano Letters} {\bf 9}, 1433 (2009).

\bibitem{10.1063/1.3474613}
S.~Heydrich, M.~Hirmer, C.~Preis, T.~Korn, J.~Eroms, D.~Weiss, and
  C.~Schueller.
\newblock Scanning {R}aman spectroscopy of graphene antidot lattices: Evidence
  for systematic $p$-type doping.
\newblock {Appl. Phys. Lett.} {\bf 97}, 043113 (2010).





\bibitem{Tzalenchuk2010}
A.~Tzalenchuk, S.~Lara-Avila, A.~Kalaboukhov, S.~Paolillo, M.~Syvajarvi,
  R.~Yakimova, O.~Kazakova, J.~J.~B. M., V.~Fal'ko, and S.~Kubatkin.
\newblock Towards a quantum resistance standard based on epitaxial graphene.
\newblock {Nature Nanotechnology} {\bf 5}, 186 (2010).

\bibitem{Emtsev2009}
K.~V. Emtsev, A.~Bostwick, K.~Horn, J.~Jobst, G.~L. Kellogg, L.~Ley, J.~L.
  McChesney, T.~Ohta, S.~A. Reshanov, J.~Rohrl, E.~Rotenberg, A.~K. Schmid,
  D.~Waldmann, H.~B. Weber, and T.~Seyller.
\newblock Towards wafer-size graphene layers by atmospheric pressure
  graphitization of silicon carbide.
\newblock {Nature Materials} {\bf 8}, 203 (2009).

\bibitem{erl}
M.~Ostler, F.~Speck, M.~Gick, and T.~Seyller.
\newblock Automated preparation of high-quality epitaxial graphene on
  6{H}-{SiC}(0001).
\newblock {physica status solidi (b)} {\bf 247}, 2924 (2010).




\bibitem{ch1Ziemann2000p3843} E.~Ziemann,
S.~D.~Ganichev, I.~N.~Yassievich, V.~I.~Perel, and W.~Prettl,
Characterization of deep impurities in semiconductors
by terahertz tunneling ionization, J. Appl. Phys. {\bf 87}, 3843 (2000). 

\bibitem{EJHLee} E.J.H. Lee, K. Balasubramanian, R.T. Weitz, M. Burghard,  K. Kern, Contact and edge effects in graphene devices, Nature Nanotechnology {\bf 3}, 486 (2008).

\bibitem{2010arXiv1011.4841V}
F.~T. {Vasko}.
\newblock {{Carrier heating and high-order harmonics generation in doped
  graphene by a strong ac electric field}}.
\newblock \emph{ArXiv e-prints:} 1011.4841  (2010).

\bibitem{doi:10.1021/nl300084j} 
S.~Wu, L.~Mao, A.~M. Jones, W.~Yao, C.~Zhang, and X.~Xu.
\newblock Quantum-enhanced tunable second-order optical nonlinearity in bilayer
  graphene.
\newblock {Nano Letters} {\bf 12}, 2032 (2012).

\bibitem{ISI:000275777700012}
W.~S. Bao, S.~Y. Liu, X.~L. Lei.
\newblock {Hot-electron transport in graphene driven by intense terahertz
  fields}.
\newblock Phys.  Lett. A \textbf{374}, 1266 (2010).

\bibitem{PhysRevB.77.195433}
F.~T. Vasko, V.~Ryzhii.
\newblock {Photoconductivity of intrinsic graphene}.
\newblock Phys. Rev. B \textbf{77}, 195433 (2008).

\bibitem{0295-5075-96-3-37006}
M.~Trushin, J.~Schliemann.
\newblock {Anisotropic photoconductivity in graphene}.
\newblock EPL \textbf{96}, 37006 (2011).

\bibitem{0953-8984-21-44-445802}
N.~M. Vildanov.
\newblock {Optical conductivity and electron-hole pair creation in
  graphene}.
\newblock Journal of Physics: Condensed Matter \textbf{21}, 445802 (2009).

\bibitem{doi:10.1021/nl202318u}
J.~C.~W. Song, M.~S. Rudner, C.~M. Marcus, L.~S. Levitov.
\newblock {Hot Carrier Transport and Photocurrent Response in Graphene}.
\newblock Nano Letters \textbf{11}, 4688 (2011).

\bibitem{Sun2012}
 D.~Sun, G.~Aivazian, A.~M. Jones, J.~S. Ross, W.~Yao, D.~Cobden, X.~Xu.
\newblock {Ultrafast hot-carrier-dominated photocurrent in graphene}.
\newblock  Nature Nanotechnology {\bf 7}, 114 (2012).

\bibitem{PhysRevB.79.081406}
T.~Oka, H.~Aoki.
\newblock {Photovoltaic Hall effect in graphene}.
\newblock Phys. Rev. B \textbf{79}, 081406 (2009).

\bibitem{KibisPRB}
O.V. Kibis, Metal-insulator transition in graphene induced by circularly polarized photons, Phys. Rev. B {\bf 81}, 165433 (2010).

\bibitem{PhysRevLett.107.216601}
Z.~Gu, H.~A. Fertig, D.~P. Arovas, A.~Auerbach.
\newblock {Floquet spectrum and transport through an irradiated graphene
  ribbon}.
\newblock Phys. Rev. Lett. \textbf{107}, 216601 (2011).

\bibitem{2011PhRvB..84w5108K}
T.~{Kitagawa}, T.~{Oka}, A.~{Brataas}, L.~{Fu}, E.~{Demler}.
\newblock {{Transport properties of nonequilibrium systems under the
  application of light: Photoinduced quantum Hall insulators without Landau
  levels}}.
\newblock Phys. Rev. B \textbf{84}, 235108 (2011).

\bibitem{torres1} Hernan L. Calvo, Horacio M. Pastawski, Stephan Roche, and Luis E. F. Foa Torres, Tuning laser-induced band gaps in graphene, Appl. Phys. Lett. {\bf 98}, 232103 (2011).

\bibitem{torres2} Hernan L. Calvo, Pablo M. Perez-Piskunow, Stephan Roche, and Luis E. F. Foa Torres, Laser-induced effects on the electronic features of graphene nanoribbons, Appl. Phys. Lett. {\bf 101}, 253506 (2012).

\bibitem{torres3} Eric Suarez Morell, and Luis E. F. Foa Torres, Radiation effects on the electronic properties of bilayer graphene, Phys. Rev. B {\bf 86}, 125449 (2012)


\bibitem{opt_or_book}
M. I. Dyakonov, Ed. \emph{Spin physics in semiconductors} (Springer-Verlag, Berlin, Heidelberg, 2008).

\bibitem{shmelev}
G.~M. Shmelev, N.~H. Shon, G.~I. Tsurkan.
\newblock {Photostimulated even acousto-electric effect}.
\newblock Izv. Vyssh. Uchebn. Zaved. Fiz. \textbf{28}, 84 (1985).

\bibitem{entin89}
M.~V. Entin.
\newblock {Theory of coherent photogalvanic effect}.
\newblock Sov. Phys. Semicond. \textbf{23}, 664 (1989).

\bibitem{0953-8984-20-38-384204}
S.~A. Mikhailov, K.~Ziegler.
\newblock {Nonlinear electromagnetic response of graphene: frequency
  multiplication and the self-consistent-field effects}.
\newblock Journal of Physics: Condensed Matter \textbf{20}, 384204 (2008).

\bibitem{Bass1986237}
F.~Bass, A.~Tetervov.
\newblock {High-frequency phenomena in semiconductor superlattices}.
\newblock Physics Reports \textbf{140}, 237  (1986).

\bibitem{AlekseevErement}
K.~N. Alekseev, M.~V. Erementchouk, F.~V. Kusmartsev.
\newblock {Direct-current generation due to wave mixing in
  semiconductors}.
\newblock EPL \textbf{47}, 595 (1999).

\bibitem{doi:10.1021/nl204283q}
A.~O'Hare, F.~V. Kusmartsev, and K.~I. Kugel.
\newblock A stable ``flat" form of two-dimensional crystals: Could graphene,
  silicene, germanene be minigap semiconductors?
\newblock {Nano Letters} {\bf 12}, 1045 (2012).



\bibitem[{\citenamefont{Wright et~al.}(2009)\citenamefont{Wright, Xu, Cao, and
  Zhang}}]{Wright09}
\bibinfo{author}{\bibfnamefont{A.~R.} \bibnamefont{Wright}},
  \bibinfo{author}{\bibfnamefont{X.~G.} \bibnamefont{Xu}},
  \bibinfo{author}{\bibfnamefont{J.~C.} \bibnamefont{Cao}}, \bibnamefont{and}
  \bibinfo{author}{\bibfnamefont{C.}~\bibnamefont{Zhang}}, Strong nonlinear optical response of graphene in the terahertz regime,
  \bibinfo{journal}{Appl. Phys. Lett.} \textbf{\bibinfo{volume}{95}},
  \bibinfo{pages}{072101} (\bibinfo{year}{2009}).


\bibitem{PhysRevB.83.195406}
J.~Rioux, G.~Burkard, J.~E. Sipe.
\newblock {Current injection by coherent one- and two-photon excitation in
  graphene and its bilayer}.
\newblock Phys. Rev. B \textbf{83}, 195406 (2011).


\bibitem{doi:10.1163/156939311798072090}
R.~Camblor, S.~V. Hoeye, G.~Hotopan, C.~V{\'a}zquez, M.~Fern{\'a}ndez, F.~L.
  Heras, P.~{\'A}lvarez, and R.~Men{\'e}ndez.
{Microwave frequency tripler based on a microstrip gap with graphene}, {Journal of Electromagnetic Waves and Applications},
 {\bf 25}, 1921--1929 (2011).


\bibitem{Shareef:12}
S.~Shareef, Y.~S. Ang, and C.~Zhang.
 {Room-temperature strong terahertz photon mixing in graphene},
 {J. Opt. Soc. Am. B} {\bf 29}, 274 (2012).

\bibitem{13} M. Orlita, C. Faugeras, P. Plochocka, P. Neugebauer, G. Martinez, D. K. Maude, A.-L. Barra, M. Sprinkle, C. Berger, W. A. de Heer, and M. Potemski, {Approaching the Dirac Point in High-Mobility
Multilayer Epitaxial Graphene}, Phys. Rev. Lett. {\bf 101}, 267601 (2008).
\bibitem{24} C. Faugeras, A. Nerriere, M. Potemski, A. Mahmood, E. Dujardin, C. Berger, and W. A. de Heer, {Few-layer graphene on SiC, pyrolitic graphite, and graphene: A Raman scattering study}, Appl. Phys. Lett. {\bf 92}, 011914 (2008).
\bibitem{25} J. Hass, F. Varchon, J. E. Millan-Otoya, M. Sprinkle, N. Sharma, W. A. de Heer, C. Berger, P. N. First, L. Magaud, and E. H. Conrad, {Why multilayer graphene on 4H-SiC$(000{\bar1})$ behaves
like a single sheet of graphene}, Phys. Rev. Lett. {\bf 100},
125504 (2008).
\bibitem{26} D.L. Miller, K.D. Kubista, G.M. Rutter, M. Ruan, W.A. de Heer, P.N. First, J.A. Stroscio, {Observing the quantization of zero mass
carriers in graphene}, Science {\bf 324}, 924 (2009).
\bibitem{27} M.L. Sadowski, G. Martinez, M. Potemski, C. Berger, W.A. De Heer, {Landau level spectroscopy of ultrathin graphite layers}, 
Phys. Rev. Lett. {\bf 97}, 266405 (2006).
\bibitem{28} X. Wu, X. Li, Z. Song, C. Berger, W.A De Heer, Weak
{Antilocalization in epitaxial graphene: evidence for chiral electrons},
Phys. Rev. Lett. {\bf 98}, 136801 (2007).
\bibitem{kop} I.A. Luk'yanchuk, Y. Kopelevich. Phase analysis of quantum oscillations in graphite. Phys. Rev. Lett. {\bf 93}, 166402 (2004).
\bibitem{15} P.R. Smith, D.H. Auston, and M.C. Nuss, {Subpicosecond photoconductive
dipole antennas}, IEEE J. Quant. Electron. {\bf QE-24}, 255 (1988).
\bibitem{17} X.-C. Zhang, B.B. Hu, J.T. Darrow, and D.H. Auston, {Generation of femtosecond
electromagnetic pulses from semiconductor surfaces}, Appl. Phys.
Lett. {\bf 56}, 1011 (1990).
\bibitem{Sakai} Kiyomi Sakai \textit{Terahertz Optoelectronics (Topics in Applied Physics)} (Springer 2005).
\bibitem{Lee} Yun-Shik Lee,  \textit{Principles of Terahertz Science and Technology}  (Springer 2009).

\bibitem{TI1} 
D. Hsieh, J.W. McIver, D.H. Torchinsky, D.R. Gardner, Y.S. Lee, and N. Gedik, Nonlinear Optical Probe of Tunable Surface Electrons on a Topological Insulator, Phys. Rev. Lett. {\bf 106}, 057401 (2011).
\bibitem{TI2} J.C.W. Song, and L.S. Levitov, 
System-Wide Photocurrent Response in Gapless Materials, {\em ArXiv e-prints:} 1112.5654 (2011).
\bibitem{TI3} P. Hosur, Circular photogalvanic effect on topological insulator surfaces: Berry-curvature-dependent response, Phys. Rev. B {\bf 83}, 035309 (2011) 


\end{thebibliography}
\end{document}